\documentclass[dvips,12pt,a4paper]{article}
\usepackage{epsfig}
\usepackage{a4}
\usepackage{amsmath}
\usepackage{latexsym}

\renewcommand{\baselinestretch}{1.5}
\def\lsim{\mathrel{\rlap{\raise 2.5pt \hbox{$<$}}\lower 2.5pt
\hbox{$\sim$}}}
\def\gsim{\mathrel{\rlap{\raise 2.5pt \hbox{$>$}}\lower 2.5pt
\hbox{$\sim$}}}

\def\thW{\theta_{\rm W}}
\def\GeV{{\rm GeV}}
\def\TeV{{\rm TeV}}
\def\dd{{\rm d}}
\def \sup{^{\vphantom{2}}}
\newcommand\half{\textstyle\frac{1}{2}}
\renewcommand\Re{{\rm Re}}

\begin{document}
\thispagestyle{empty}
\phantom{AAA}
\vspace*{-20mm}
\begin{flushright}
\tt
CERN-TH/98-189 \\[-4mm]
hep-ph/9806351
\end{flushright}
\vspace{-2mm}
\begin{center}
{\bf{\Large Measuring the Trilinear Couplings of MSSM \\
Neutral Higgs Bosons at High-Energy $e^+  e^-$ Colliders}}
\vskip 0.5cm
P. Osland$^{a,b,c}$ and P. N. Pandita$^{a,d}$\\
$^a$ Department of Physics, University of Bergen, 
N-5007 Bergen, Norway$^{*}$\\
$^b$ Deutsches Elektronen-Synchrotron DESY, D-22603 Hamburg, Germany\\
$^c$ Theoretical Physics Division, CERN, CH 1211 Geneva 23, Switzerland\\
$^{d}$ Department of Physics, North Eastern Hill University,
Shillong 793 022, India$^{*}$\\[-2mm]
\end{center}

\begin{abstract}
We present a detailed analysis of multiple production of the lightest
$CP$-even Higgs boson ($h$) of the Minimal Supersymmetric Standard Model 
(MSSM) at high-energy $e^+  e^-$ colliders. 
We consider the production of 
the heavier $CP$-even Higgs boson ($H$) via
Higgs-strahlung $e^+e^- \rightarrow ZH$, in association with 
the $CP$-odd Higgs boson ($A$) in $e^+e^- \rightarrow AH$, 
or via the fusion mechanism $e^+e^- \rightarrow \nu_e \bar\nu_e H$, 
with $H$ subsequently decaying through $H \rightarrow hh$, thereby 
resulting in a pair of lighter Higgs bosons ($h$) in the final state.
These processes can enable one to measure the trilinear Higgs couplings 
$\lambda_{Hhh}$ and
$\lambda_{hhh}$, which can be used to theoretically reconstruct the
Higgs potential. 
We delineate the regions of the MSSM parameter space in which these
trilinear Higgs couplings could be measured at a future $e^+ e^-$
collider. In our calculations, we include in detail 
the radiative corrections to the Higgs sector of the MSSM, 
especially the mixing in the squark sector.

\end{abstract}

\noindent{\underline{\hspace{11.6cm}}}\\[-2mm]
* Permanent addresses
\begin{flushleft}
CERN-TH/98-189 \\[-4mm]
June 1998
\end{flushleft}
\newpage
\section{Introduction}
The Higgs potential of the Standard Model (SM), which 
is crucial in implementing
the mechanism of spontaneous symmetry breaking, contains the 
unknown quartic coupling of the Higgs field. 
As a consequence, the mass of the only Higgs boson in the SM, 
which is determined by this quartic coupling, is not known~\cite{GHKD}. 
If a Higgs boson is discovered and its mass measured, 
the Higgs potential of the Standard Model can be uniquely determined. 

On the other hand, supersymmetry is at present the only known 
framework in which the Higgs
sector of the Standard Model (SM), so crucial for its internal
consistency, is natural~\cite{HPN}.
The minimal version of the Supersymmetric 
Standard Model (MSSM) contains two Higgs doublets $(H_1, H_2)$ 
with opposite hypercharges: $Y(H_1) = -1$, $Y(H_2) = +1$, so as to 
generate masses for up- and down-type
quarks (and leptons), and to cancel gauge anomalies. 
After spontaneous symmetry breaking induced by the neutral 
components of $H_1$ and $H_2$ obtaining vacuum 
expectation values, $\langle H_1\rangle = v_1$, 
$\langle H_2\rangle = v_2$, $\tan\beta = v_2/v_1$, 
the MSSM contains two neutral $CP$-even\footnote{When unambiguous, 
we shall denote the $CP$-even Higgs particles as $h$ and $H$.}
($h$, $H$), one neutral 
$CP$-odd ($A$), and two charged ($H^{\pm}$) Higgs bosons \cite{GHKD}. 
Although gauge invariance and supersymmetry fix the quartic couplings 
of the Higgs bosons in the MSSM in terms of $SU(2)_L$ and $U(1)_Y$ 
gauge couplings, $g$ and $g^{\prime}$, respectively, there still 
remain two independent parameters that describe the Higgs sector
of the MSSM. These are usually chosen to be $\tan\beta$ and 
$m_A$, the mass of the $CP$-odd Higgs boson. All the Higgs masses
and the Higgs couplings in the MSSM can be described (at tree level) 
in terms of these two parameters.

In particular, all the trilinear self-couplings of the physical
Higgs particles can be predicted theoretically (at the tree level)
in terms of $m_A$ and $\tan\beta$. Once a light Higgs boson is 
discovered, the measurement of these trilinear couplings can be used to 
reconstruct the Higgs potential of the MSSM. This will go a long way
toward establishing the Higgs mechanism as the basic mechanism
of spontaneous symmetry breaking in gauge theories. Although the 
measurement of all the Higgs couplings in the MSSM is a difficult task,
preliminary theoretical investigations by Plehn, Spira and Zerwas
\cite{PSZ}, and by Djouadi, Haber and Zerwas
\cite{DHZ} (referred to as `DHZ' in the following), 
of the measurement of these couplings at the LHC and
at a high-energy $e^+ e^-$ linear collider, respectively, are encouraging.

In this paper we consider in detail the question of possible measurements 
of the trilinear Higgs couplings of the MSSM at a high-energy $e^+ e^-$ 
linear collider. We assume that such a facility will operate at
an energy of 500~GeV with an integrated luminosity per year of  
${\mathcal L}_{\rm int} = 500~\mbox{fb}^{-1}$ \cite{NLC}.
(This is a factor of 10 more than the earlier estimate.)
In a later phase one may envisage an upgrade to an energy of 1.5~TeV.
Since the `interesting' cross sections fall off like $1/E^2$, the
luminosity should increase by a corresponding factor.
An earlier estimated luminosity of $500~\mbox{fb}^{-1}$
at 1.5~TeV may turn out to be too conservative.

The trilinear Higgs couplings that are of interest are 
$\lambda_{Hhh}$, $\lambda_{hhh}$, and $\lambda_{hAA}$, 
involving both the $CP$-even 
($h$, $H$) and $CP$-odd ($A$) Higgs bosons.\footnote{These are not 
the only couplings that occur in the Higgs potential.
However, these are the only ones which could possibly be measured
at future colliders.}
The couplings $\lambda_{Hhh}$ and $\lambda_{hhh}$
are rather small with respect to the corresponding trilinear coupling
$\lambda_{hhh}^{\rm SM}$ in the SM (for a given mass 
of the lightest Higgs boson $m_h$), 
unless $m_h$ is close to the upper value (decoupling limit).
The coupling $\lambda_{hAA}$ remains small for all parameters.

Throughout, we include one-loop radiative corrections
\cite{ERZ1} to the Higgs sector in the effective potential
approximation. In particular, we take into account 
the parameters $A$ and $\mu$, the soft supersymmetry
breaking trilinear parameter and the bilinear Higgs(ino)  
parameter in the superpotential, respectively, and as a consequence
the left--right mixing in the squark sector, in our calculations. 
We thus include all the relevant parameters of the MSSM in our study, 
which is more detailed than the preliminary one of DHZ.

For a given value of $m_h$, the values of these
couplings significantly depend on the soft supersymmetry-breaking 
trilinear parameter $A$, as well as on $\mu$, 
and thus on the resulting mixing in the squark sector.
Since the trilinear couplings tend to be small,
and depend on several parameters, their effects are somewhat difficult
to estimate.

The plan of the paper is as follows. 
In Section 2 we review the Higgs sector of the MSSM, including
the radiative corrections to the masses.
The trilinear couplings are presented in Section 3.
In Section 4 we review the possible production mechanisms 
for the multiple production of Higgs bosons through 
which the trilinear Higgs couplings can be measured at an $e^+ e^-$ 
linear collider. 
In Section 5 we consider the dominant source of the multiple production
of the Higgs ($h$) boson through Higgs-strahlung of $H$, and 
through production of $H$ in association  with the $CP$-odd Higgs 
boson ($A$), and the background to these
processes.  This source of multiple production can be used to 
extract the trilinear Higgs coupling $\lambda_{Hhh}$.

Section 6 deals with a detailed calculation of the cross section for
the double Higgs-strahlung process $e^+ e^- \to Zhh$. This process
involves the trilinear couplings $\lambda_{Hhh}$ and  $\lambda_{hhh}$
of the $CP$-even Higgs bosons ($h$, $H$).
In Section 7 we consider the different fusion mechanisms for multiple
$h$ production, especially the non-resonant 
process $e^+e^-\to \nu_e\bar\nu_e hh$,
for which we present a detailed calculation of the cross section
in the `effective $WW$ approximation'. This process also involves 
the two trilinear Higgs couplings, $\lambda_{Hhh}$ and $\lambda_{hhh}$,
and is the most useful one for extracting the coupling $\lambda_{hhh}$.
In Section 8 we present, based on our calculations,
the regions of the MSSM parameter space in which the
trilinear couplings $\lambda_{Hhh}$ and $\lambda_{hhh}$ could 
be measured; finally, in
Section 9 we present a summary of our results and conclusions.

\section{The Higgs Sector of the MSSM}
In this section we review the Higgs sector of the Minimal 
Supersymmetric Standard Model in order to set the notation and
to describe the approximations we use in our calculations.
As mentioned in the introduction, we shall include the dependence
on the parameters $A$ and $\mu$ through mixing in the squark sector.
Where there is an overlap, our notation and approach closely follow 
those of Ref.~\cite{DHZ}.

At the tree level, the Higgs sector of the MSSM 
is described by two parameters, which can be conveniently chosen as
$m_A$ and $\tan\beta$ \cite{GHKD}. There are, however, substantial
radiative corrections to the $CP$-even neutral Higgs masses and
couplings~\cite{ERZ1}. In the one-loop effective potential approximation,
the radiatively corrected squared-mass matrix for the 
$CP$-even Higgs bosons can be written as \cite{ERZ2}
\smallskip
\begin{eqnarray}
\label{Eq:M2}
{\mathcal M}^2   
&=&  \left[ \begin{array}{cc}
m_A^2 \sin^2 \beta + m_Z^2 \cos^2\beta & 
-(m_Z^2 + m_A^2) \sin\beta \cos \beta\\
-(m_Z^2 + m_A^2) \sin\beta \cos \beta & 
m_A^2 \cos^2 \beta + m_Z^2 \sin^2\beta
\end{array} \right] \nonumber \\
& &
+ \frac{3 g^2}{16 \pi^{2} m_W^2}
\left[ \begin{array}{cc}
\Delta_{11} & \Delta_{12}\\
\Delta_{12} & \Delta_{22}
\end{array} \right], 
\end{eqnarray}
where the second matrix represents the radiative 
corrections.\footnote{We note that two-loop corrections to the Higgs 
masses in the MSSM are sizable, especially for large mixing in the stop 
sector. 
For the dominant two-loop radiative corrections to the Higgs sector
of the MSSM, see, e.g.\ \cite{Carena}.
In this paper we restrict ourselves to one-loop corrections only.}

The functions $\Delta_{ij}$ depend, 
besides the top- and bottom-quark masses, on the Higgs 
bilinear parameter $\mu$ in the superpotential, 
the soft supersymmetry-breaking trilinear couplings
($A_t$, $A_b$) and soft scalar masses ($m_Q$, $m_U$, $m_D$), 
as well as on $\tan\beta$.
We shall ignore the $b$-quark mass effects in 
$\Delta_{ij}$ in our calculations, which is a reasonable 
approximation for moderate values of $\tan\beta \lsim 20$--30.
Furthermore, we shall assume, as is often done,
\begin{eqnarray}
A &\equiv& A_t = A_b, \nonumber \\
\tilde m &\equiv& m_Q = m_U = m_D. \label{2}
\end{eqnarray}

With these approximations we can write ($m_t$ is the top quark mass) 
\cite{ERZ2}:
\begin{eqnarray}
\Delta_{11} & = & \frac{m_t^4}{\sin^2\beta}
                \left(\frac{\mu(A+\mu\cot\beta)}
                {m_{\tilde t_1}^2 - m_{\tilde t_2}^2} \right)^2 
                g(m_{\tilde t_1}^2, m_{\tilde t_2}^2), \label{3} \\
\Delta_{22} & = & \frac{m_t^4}{\sin^2\beta}
 \left(\log\frac{m_{\tilde t_1}^2 m_{\tilde t_2}^2}{m_t^4}
+  \frac{2A(A+\mu\cot\beta)}{m_{\tilde t_1}^2 - m_{\tilde t_2}^2}
   \log\frac{m_{\tilde t_1}^2}{m_{\tilde t_2}^2} \right ) \nonumber \\ 
 & + & \frac{m_t^4}{\sin^2\beta}
       \left(\frac{\mu(A+\mu\cot\beta)}
      {m_{\tilde t_1}^2 - m_{\tilde t_2}^2} \right)^2
     g(m_{\tilde t_1}^2, m_{\tilde t_2}^2), \label{4}\\
\Delta_{12} & = & \frac{m_t^4}{\sin^2\beta} \frac{\mu(A+\mu\cot\beta)}
     {m_{\tilde t_1}^2 - m_{\tilde t_2}^2}
      \left(\log\frac{m_{\tilde t_1}^2}{m_{\tilde t_2}^2}
       +  \frac{A(A+\mu\cot\beta)}{m_{\tilde t_1}^2 - m_{\tilde t_2}^2}
       g(m_{\tilde t_1}^2, m_{\tilde t_2}^2) \right), \label{5} 
\end{eqnarray}
where $m_{\tilde t_1}^2$ and $m_{\tilde t_2}^2$ are squared stop masses
given by  
\begin{equation}
m_{\tilde t_{1,2}}^2   =  m_t^2 + \tilde m^2 \pm 
m_t(A + \mu\cot\beta) \label{6}
\end{equation}
(we have ignored the small $D$-term contributions to the stop masses)
and
\begin{equation}
g(m_{\tilde t_1}^2, m_{\tilde t_2}^2) =  2 - 
\frac{m_{\tilde t_1}^2 + m_{\tilde t_2}^2}{m_{\tilde t_1}^2 - 
m_{\tilde t_2}^2} \log\frac{m_{\tilde t_1}^2}{m_{\tilde t_2}^2}. 
\label{7}
\end{equation}

The one-loop radiatively corrected masses ($m_{h}$, $m_{H}$;
$m_{h} <  m_{H}$) of the $CP$-even Higgs bosons ($h$, $H$)
can be obtained by diagonalizing the $2 \times 2$ mass matrix in
Eq.~(\ref{Eq:M2}). The radiative corrections are, in general, positive,
and they shift the mass of the lightest Higgs boson upwards from 
its tree-level value.
We show in Fig.~\ref{Fig:masses} the resulting mass of 
the lightest Higgs boson, 
$m_h$, as a function of $\mu$ and $\tan\beta$, for two 
values of $A$ and two values of $m_A$, and for $\tilde m=1$~TeV.
With a wider range of parameter values,
or when the squark mass scale 
is taken to be smaller, 
the dependence on $\mu$ and $\tan\beta$ can be more dramatic \cite{KOP}.

The Higgs mass falls rapidly at small values of $\tan\beta$.
Since the LEP experiments are obtaining 
lower bounds on the mass of the lightest Higgs boson, they are
beginning to rule out significant parts of the small-$\tan\beta$ 
parameter space, depending on the model assumptions.
For $\tan\beta>1$, ALEPH finds $m_h>62.5~\GeV$ at 95\% C.L.\ 
\cite{ALEPH}.\footnote{Recently, a new study has been presented,
with a lower limit of $m_h>72.2$~GeV, irrespective of $\tan\beta$,
and a limit of $\sim 88$~GeV for $1<\tan\beta\lsim2$ \cite{ALEPH98}.}
In our calculations, we shall therefore take $\tan\beta=2$ to be 
a representative value.
[For a recent discussion on how the lower allowed value of $\tan\beta$
depends on some of the model parameters, see Ref.~\cite{CCPW}.]

\setcounter{equation}{0}
\section{Trilinear Higgs couplings}

The trilinear Higgs couplings that are of interest
can be written~\cite{BBSP} as a sum of the tree-level coupling and 
one-loop radiative corrections: 
\begin{eqnarray}
\lambda_{Hhh} & = & \lambda_{Hhh}^0 + \Delta \lambda_{Hhh}, 
\label{Eq:lambda-Hhh} \\
\lambda_{hhh} & = & \lambda_{hhh}^0 + \Delta \lambda_{hhh}, 
\label{Eq:lambda-hhh} \\
\lambda_{hAA} & = & \lambda_{hAA}^0 + \Delta\lambda_{hAA}. 
\label{Eq:lambda-hAA}
\end{eqnarray}  
In units of $gm_Z/(2\cos\thW)=(\sqrt{2}G_F)^{1/2}m_Z^2$,
the tree-level couplings are given by
\begin{eqnarray} 
\lambda_{Hhh}^0 & = & 2\sin2\alpha \sin(\beta + \alpha) - \cos 2\alpha
\cos(\beta + \alpha), \\
\label{Eq:lambda-Hhh0}
\lambda_{hhh}^0 & = & 3 \cos2\alpha \sin(\beta + \alpha), \\
\label{Eq:lambda-hhh0}
\lambda_{hAA}^0 & = & \cos2\beta \sin(\beta + \alpha),      
\label{Eq:lambda-hAA0}
\end{eqnarray}
with $\alpha$ the mixing angle in the $CP$-even Higgs sector, which can
be calculated in terms of the parameters appearing in the $CP$-even
Higgs mass matrix (\ref{Eq:M2}). The one-loop radiative corrections 
in (\ref{Eq:lambda-Hhh})--(\ref{Eq:lambda-hAA}) are 
(in the above units):
\begin{eqnarray}
\Delta \lambda_{Hhh} & = & \left( \frac{3g^2 \cos^2\theta_W}{16 \pi^2}
\frac{m_t^4}{m_W^4} \frac{\sin\alpha\cos^2\alpha}{\sin^3\beta} \right) 
\nonumber\\
& \times & \left[ 3\log\frac{m_{\tilde t_1}^2 m_{\tilde t_2}^2}{m_t^4}
+ (m_{\tilde t_1}^2 - m_{\tilde t_2}^2)C_t(E_t + 2F_t)
   \log\frac{m_{\tilde t_1}^2}{m_{\tilde t_2}^2} \right. \nonumber\\
& & + 2\left(\frac{m_t^2}{m_{\tilde t_1}^2}
\left[1+(m_{\tilde t_1}^2 - m_{\tilde t_2}^2)C_t E_t\right]
\left[1+(m_{\tilde t_1}^2 - m_{\tilde t_2}^2)C_t F_t\right]^2 \right. 
\nonumber \\
& & + \left. \left.\frac{m_t^2}{m_{\tilde t_2}^2}
\left[1-(m_{\tilde t_1}^2 - m_{\tilde t_2}^2)C_t E_t\right]
\left[1-(m_{\tilde t_1}^2 - m_{\tilde t_2}^2)C_t F_t\right]^2 - 2 \right) 
\right], \label{17}  \\
\Delta \lambda_{hhh} & = & \left( \frac{3g^2 \cos^2\theta_W}{16 \pi^2}
\frac{m_t^4}{m_W^4} \frac{\cos^3\alpha}{\sin^3\beta} \right) 
\nonumber \\
& \times & \left[ 3\log\frac{m_{\tilde t_1}^2 m_{\tilde t_2}^2}{m_t^4}
+ 3 (m_{\tilde t_1}^2 - m_{\tilde t_2}^2)C_t F_t
   \log\frac{m_{\tilde t_1}^2}{m_{\tilde t_2}^2} \right. \nonumber\\ 
& & + \left. 2\left(\frac{m_t^2}{m_{\tilde t_1}^2}
\left[1+(m_{\tilde t_1}^2 - m_{\tilde t_2}^2)C_t F_t\right]^3
+\frac{m_t^2}{m_{\tilde t_2}^2}
\left[1-(m_{\tilde t_1}^2 - m_{\tilde t_2}^2)C_t F_t\right]^3 
-2 \right) \right], \label{16} 
\end{eqnarray}
\begin{eqnarray}
\Delta\lambda_{hAA} & = & \left( \frac{3g^2 \cos^2\theta_W}{16 \pi^2}
\frac{m_t^4}{m_W^4} \frac{\cos\alpha \cos^2\beta}{\sin^3\beta} \right)
\nonumber\\ 
& \times & \biggl[ 
\log\frac{m_{\tilde t_1}^2 m_{\tilde t_2}^2}{m_t^4}
+  (m_{\tilde t_1}^2 - m_{\tilde t_2}^2)(D_t^2 + C_t F_t)
   \log\frac{m_{\tilde t_1}^2}{m_{\tilde t_2}^2}  \nonumber\\
& & + (m_{\tilde t_1}^2 - m_{\tilde t_2}^2)^2 
C_tD_t^2F_t\, g(m_{\tilde t_1}^2,m_{\tilde t_2}^2) \biggr],      \label{21}
\end{eqnarray}
where
\begin{eqnarray}
C_t & = & (A + \mu\cot\beta)/(m_{\tilde t_1}^2 - m_{\tilde t_2}^2), 
\nonumber\\
D_t & = & (A - \mu\tan\beta)/(m_{\tilde t_1}^2 - m_{\tilde t_2}^2),
\nonumber\\
E_t & = & (A + \mu\cot\alpha)/(m_{\tilde t_1}^2 - m_{\tilde t_2}^2), 
\nonumber\\
F_t & = & (A - \mu\tan\alpha)/(m_{\tilde t_1}^2 - m_{\tilde t_2}^2), 
\label{Eq:CEF}
\end{eqnarray}
and we have ignored the contributions from $b$-quarks and $b$-squarks, 
which are in general small with respect to those arising from $t$-quarks
and $t$-squarks. 
We have also adopted the simplification
described in Eq.~(\ref{2}) in writing the above results. 
We shall make these approximations throughout this paper.

We show in Figs.~\ref{Fig:lamHhh2}, \ref{Fig:lamhhh2} and
\ref{Fig:lamhAA2} the couplings $\lambda_{Hhh}$, $\lambda_{hhh}$ 
and $\lambda_{hAA}$ as functions of $\mu$ and $\tan\beta$, 
for two values of $A$ and two values of $m_A$, 
all for $\tilde m=1$~TeV.
The explicit dependence on $A$ and $\mu$ is not dramatic,
but it should be kept in mind that unless $m_A$ is rather small,
$m_h$ may change considerably with $A$.

The trilinear couplings change significantly with $m_A$,
and thus also with $m_h$. This is shown more explicitly
in Fig.~\ref{Fig:lam-mh},
where we compare $\lambda_{Hhh}$, $\lambda_{hhh}$ and $\lambda_{hAA}$
for three different values of $\tan\beta$,
and the SM quartic coupling $\lambda^{\rm SM}$.
The SM quartic coupling includes one-loop radiative corrections 
\cite{SirZuc}, and its normalization is such that at the tree-level,
it coincides with the trilinear coupling.

At low values of $m_h$, the MSSM trilinear couplings are rather small.
For some value of $m_h$ the couplings $\lambda_{Hhh}$ and $\lambda_{hhh}$
start to increase in magnitude, whereas $\lambda_{hAA}$ remains small.
The values of $m_h$ at which they start becoming significant
depend crucially on $\tan\beta$.
For $\tan\beta=2$ (Fig.~\ref{Fig:lam-mh}a) this transition
takes place around $m_h\sim 90$--100~GeV, whereas
for $\tan\beta=5$ and 15, the critical values of $m_h$ increase
to 100--110 and 120~GeV, respectively (see Figs.~\ref{Fig:lam-mh}b and c).
In this region, the actual values of $\lambda_{Hhh}$ and $\lambda_{hhh}$
(for a given value of $m_h$) change significantly if $A$ becomes
large and positive. 
A non-vanishing squark-mixing parameter $A$ is thus seen to be quite 
important. Also, we note that for special values of the parameters,
the couplings may vanish \cite{DKZ1}. See also Fig.~1 of Ref.~\cite{PSZ}.

To sum up the behaviour of the trilinear couplings, we note that
$\lambda_{Hhh}$ and  $\lambda_{hhh}$ are small ($\le1$) for 
$m_h \lsim 100$--120~GeV, depending on the value of $\tan\beta$. 
However, as $m_h$ approaches its maximum value, 
which is reached rapidly as $m_A$ becomes large,
$m_A \gsim 200$~GeV, these trilinear couplings become large ($\sim 2-4$).
Thus, as functions of $m_A$, the trilinear couplings
$\lambda_{Hhh}$ and $\lambda_{hhh}$ are large
for most of the parameter space.
We also note that, for large values of $\tan\beta$, $\lambda_{Hhh}$
tends to be relatively small, whereas $\lambda_{hhh}$ becomes large,
if also $m_A$ (or, equivalently, $m_h$) is large.

We note that for a given Higgs boson mass 
$m_h$, the tree level SM trilinear Higgs coupling is 
given by
\begin{equation}
\lambda^{\rm SM}_{hhh} = 3(m_h/m_Z)^2. 
\end{equation}
On the other hand, for large values of $m_A$ (the decoupling limit) the
corresponding MSSM trilinear coupling, Eq.~(\ref{Eq:lambda-Hhh0}), becomes
\begin{equation}
\lambda^0_{hhh}=3\cos(2\alpha)\sin(\beta+\alpha)\to 
                  3 (m_h/m_Z)^2,
\end{equation}
i.e., it approaches the SM trilinear coupling.

\setcounter{equation}{0}
\section{Production mechanisms}

The different mechanisms for the multiple production of the MSSM 
Higgs bosons in $e^+ e^-$ collisions have been discussed by DHZ.
The dominant mechanism for the production of multiple  
$CP$-even light Higgs bosons ($h$) is through the 
production of the heavy $CP$-even Higgs boson $H$, which then decays
by $H \rightarrow hh$. The heavy Higgs boson $H$ can be produced
by $H$-strahlung, in association with $A$, 
and by the resonant $WW$ fusion mechanism. These mechanisms
for multiple production of $h$
\begin{eqnarray}
\left. \begin{array}{ccc}
e^+e^- & \rightarrow & ZH,AH \\ 
e^+e^- & \rightarrow & \nu_e \bar \nu_e H
\end{array}
\right\}, \qquad H \rightarrow hh, \label{Eq:res-Hhh} 
\end{eqnarray}
are shown in Fig.~\ref{Fig:Feynman-resonant}. We note that all
the diagrams of Fig.~\ref{Fig:Feynman-resonant} involve the
trilinear coupling $\lambda_{Hhh}$.


A background to (\ref{Eq:res-Hhh}) comes from the production of the 
pseudoscalar $A$ in association with $h$ and its subsequent decay to $hZ$
\begin{equation}
e^+e^- \rightarrow hA, \qquad A \rightarrow hZ, \label{Eq:bck-hA}
\end{equation}
leading to $Zhh$ final states [see Fig.~\ref{Fig:Feynman-nonres-Zhh}d].

A second mechanism for $hh$ production is double Higgs-strahlung 
in the continuum with a $Z$ boson in the final state
[see Fig.~\ref{Fig:Feynman-nonres-Zhh}a--d],
\begin{equation}
e^+e^-  \rightarrow Z^* \rightarrow Zhh. \label{Eq:Zstar}
\end{equation} 
We note that the Feynman diagram of Fig.~\ref{Fig:Feynman-nonres-Zhh}c
involves, apart from the coupling $\lambda_{Hhh}$,  
the trilinear Higgs coupling $\lambda_{hhh}$ as well, 
whereas the other diagrams
do not involve any of the trilinear Higgs couplings.

A third way of generating multiple Higgs bosons in $e^+ e^-$ collisions
is through associated production of ($hh$) with the pseudoscalar $A$ 
in the continuum [see Fig.~\ref{Fig:Feynman-nonres-Ahh}]:
\begin{equation}
e^+e^-  \rightarrow Z^* \rightarrow hhA. \label{Eq:Zstar-hhA}
\end{equation}
This process will be briefly discussed in Section~6.
It involves, besides $\lambda_{Hhh}$ and $\lambda_{hhh}$, the
trilinear coupling $\lambda_{hAA}$ as well.
It is, however, difficult~\cite{DHZ}  
to measure this coupling $\lambda_{hAA}$ through the process 
(\ref{Eq:Zstar-hhA}). 

Finally, there is a mechanism of multiple production of the lightest Higgs
boson through non-resonant $WW$ ($ZZ$) fusion in the continuum
[see Fig.~\ref{Fig:Feynman-nonres-WW}]:
\begin{equation}
e^+e^-  \rightarrow \bar \nu_e \nu_e W^* W^* \rightarrow 
\bar \nu_e \nu_e hh, \label{Eq:WW-fusion}
\end{equation}
which will be discussed in Section~7.

It is important to note that all the diagrams of  
Fig.~\ref{Fig:Feynman-resonant}
involve the trilinear coupling $\lambda_{Hhh}$ only. On the other hand,
Fig.~\ref{Fig:Feynman-nonres-Zhh}c, Fig.~\ref{Fig:Feynman-nonres-Ahh}b
and  Fig.~\ref{Fig:Feynman-nonres-WW}c all involve both
the trilinear Higgs couplings $\lambda_{Hhh}$ and $\lambda_{hhh}$.

\setcounter{equation}{0}
\section{Higgs-strahlung and Associated Production of $H$}
As stated in Section~4, the dominant source for the production of
multiple Higgs bosons ($h$) in $e^+ e^-$ collisions is through 
the production of the heavier $CP$-even Higgs boson $H$ either via
Higgs-strahlung or in association with $A$ \cite{GHKD},  
followed, if kinematically allowed, by the cascade decay 
$H \rightarrow hh$.
In terms of the $Z$-electron couplings $v_e = -1 + 4\sin^2\theta_W$,  
$a_e = -1$,
the cross sections for these processes can be written as
\cite{PocZsi,GETAL}
\begin{eqnarray}
\sigma (e^+e^- \rightarrow ZH) & = & \frac{G_F^2m_Z^4}{96\pi s}
(v_e^2 + a_e^2)\cos^2(\beta - \alpha) 
\frac{\lambda_Z^{1/2} \left [\lambda_Z + 12m_Z^2/s
\right ]}{(1-m_Z^2/s)^2}, \label{Eq:sigZH}\\
\sigma (e^+e^- \rightarrow AH) & = & \frac{G_F^2m_Z^4}{96\pi s}
(v_e^2 + a_e^2)\sin^2(\beta - \alpha) 
\frac{\lambda_A^{3/2}}{(1-m_Z^2/s)^2}, 
\label{Eq:sigAH}
\end{eqnarray}
where $\lambda_j$ refers to $\lambda(m_j^2, m_H^2; s)$, 
the two-body phase-space function, and is given as
\begin{equation}
\lambda(m_a^2, m_b^2; m_c^2) = \left(1 - \frac{m_a^2}{m_c^2} - 
\frac{m_b ^2}{m_c^2}\right)^2 - \frac{4 m_a^2m_b^2}{m_c^4}.
\label{27}
\end{equation}

In Fig.~\ref{Fig:sigma-500-1500} we plot the cross sections 
(\ref{Eq:sigZH}) and (\ref{Eq:sigAH}) for the 
$e^+e^-$ centre-of-mass energies $\sqrt s = 500~\GeV$ and $1.5$ TeV, 
as functions of the Higgs mass $m_H$ and for $\tan\beta = 2.0$. 
For large values of the mass $m_A$ of the pseudoscalar 
Higgs boson, all the Higgs bosons, 
except the lightest one ($h$), become heavy and  decouple~\cite{HABER1} 
from the rest of the spectrum. In this case
\begin{equation}
\cos^2(\beta - \alpha) \simeq\frac{m_Z^4 \sin^2 4\beta}{4 m_A^4}, 
\label{Eq:decouple}
\end{equation}
and the associated $AH$ production (\ref{Eq:sigAH}) becomes 
the dominant production mechanism for $H$.

At values of $\tan\beta$ that are not too large, the trilinear
$Hhh$ coupling $\lambda_{Hhh}$ can be measured by the decay process
$H \rightarrow hh$, which has a width 
\begin{equation}
\Gamma(H \rightarrow hh)  =  
\frac{G_F \lambda_{Hhh}^2}{16\pi\sqrt 2}
\frac{m_Z^4}{m_H} \left( 1 - \frac{4m_h^2}{m_H^2} \right)^{1/2}.
\label{Eq:GamHhh}
\end{equation}
However, this is possible only if the decay is kinematically
allowed, and the branching ratio is sizeable.
In Fig.~\ref{Fig:BR-H-A-2} we show the branching ratios (at $\tan\beta=2$)
for the main decay modes of the heavy $CP$-even Higgs boson 
as a function of the $H$ mass.
Apart from the $hh$ decay mode, the other important decay modes 
are $H \rightarrow WW^*$, $ZZ^*$. 
(We have here disregarded decays 
to supersymmetric particles: charginos, stops, etc. 
If such particles are kinematically accessible, the $H\to hh$
and $A\to Zh$ rates could be much smaller \cite{Bartl}.)
We note that the couplings of $H$ 
to gauge bosons can be measured through the production cross sections
for $e^+e^- \rightarrow \nu_e\bar\nu_eH$; therefore the branching
ratio $BR(H \rightarrow hh)$ can be used to measure the triple Higgs
coupling $\lambda_{Hhh}$. 

The Higgs-strahlung process [Fig.~\ref{Fig:Feynman-resonant}a, 
Eq.~(\ref{Eq:sigZH})] gives rise to 
resonant two-Higgs $[hh]$ final states. 
This is to be contrasted with the associated 
production process [Fig.~\ref{Fig:Feynman-resonant}b, 
Eq.~(\ref{Eq:sigAH})], which typically yields 
three Higgs $h[hh]$
final states, since the channel $A \rightarrow hZ$ is the dominant 
decay mode of $A$ in the mass range of interest. The decay width for
$A \rightarrow hZ$ can be written as \cite{DKZ2} 
\begin{equation}
\Gamma(A \rightarrow hZ) =  \frac{G_F}{8\pi\sqrt 2}
\cos^2(\beta - \alpha) \frac{m_Z^4}{m_A}
\lambda^{1/2}(m_Z^2, m_h^2; m_A^2) \lambda(m_A^2, m_h^2; m_Z^2),
\label{26}
\end{equation}
where the $\lambda$ are phase-space factors
given by Eq.~(\ref{27}).
In Fig.~\ref{Fig:BR-H-A-2} we show the 
branching ratios for the pseudoscalar $A$ for $\tan\beta = 2.0$.

A background to the multiple production of lighter Higgs bosons $h$
comes from $hh$ states generated in the sequential reaction $e^+e^-
\rightarrow  Ah \rightarrow [Zh]h$ 
[see Fig.~\ref{Fig:Feynman-nonres-Zhh}d].
This is a genuine background in the sense that no Higgs
self-couplings are involved.
But these background events are expected to be topologically
very different from the signal events, since the 
two $h$ bosons do not form a resonance,
whereas the $[Zh]$ does. The cross section for the 
process $e^+e^- \rightarrow Ah$ can be written as 
\cite{GETAL} 
\begin{equation}
\sigma(e^+e^- \rightarrow Ah)  =  \frac{G_F^2 m_Z^4}{96\pi s}
(v_e^2 + a_e^2) \cos^2(\beta - \alpha)
\frac{\lambda^{3/2}(m_h^2, m_A^2; s)}{(1 - m_Z^2/ s)^2}, 
\label{28}
\end{equation}
and is shown in Fig.~\ref{Fig:sigma-500-1500} together with 
the signal cross sections
(\ref{Eq:sigZH}) and (\ref{Eq:sigAH}). As a consequence of the decoupling
theorem~\cite{HABER1}, the cross section becomes small for large
values of $m_H$.

For increasing values of $\tan\beta$, the $Hhh$ coupling gradually 
gets weaker (see Figs.~\ref{Fig:lamHhh2} and \ref{Fig:lam-mh}),
and hence the prospects for measuring $\lambda_{Hhh}$ diminish.
This is indicated by Fig.~\ref{Fig:BR-H-A-5}, where we show the
$H$ and $A$ branching ratios for $\tan\beta=5$.

There is in fact a sizeable region in the $m_A$--$\tan\beta$ plane
where the decay $H\to hh$ is kinematically forbidden.
This is indicated in Fig.~\ref{Fig:hole}.
In this figure we also display the regions where the $H\to hh$
branching ratio is in the range 0.1--0.9.
Clearly, in the forbidden region, the $\lambda_{Hhh}$ cannot be
determined from resonant production.

\setcounter{equation}{0}
\section{Double Higgs-strahlung and Triple $h$ Production}
For small and moderate values of $\tan\beta$, the study of decays
of the heavy $CP$-even Higgs boson $H$ provides a means of determining
the triple-Higgs coupling $\lambda_{Hhh}$.
In order to extract the coupling $\lambda_{hhh}$, other processes
involving two-Higgs ($h$) final states must be considered.
The $Zhh$ final states, which can be produced in the double 
Higgs-strahlung $e^+e^- \rightarrow Zhh$ 
of Fig.~\ref{Fig:Feynman-nonres-Zhh}, could provide one possible
opportunity, since it involves the coupling $\lambda_{hhh}$
through the mechanism of Fig.~\ref{Fig:Feynman-nonres-Zhh}c.
In this section we shall study these non-resonant processes in detail.

\subsection{The Double Higgs-strahlung $e^+e^- \rightarrow Zhh$}
The doubly differential cross section for the process 
$e^+e^- \rightarrow Zhh$ shown in Fig.~\ref{Fig:Feynman-nonres-Zhh} 
can be written as
\cite{DHZ}
\begin{equation} 
\frac{d\sigma(e^+e^- \rightarrow Zhh)}{dx_1 dx_2}
= \frac{G_F^3 m_Z^6}{384\sqrt 2 \pi^3s}(v_e^2 + a_e^2) 
\frac{\mathcal A}{(1-\mu_Z)^2},
\label{Eq:sigZhh}
\end{equation}
\noindent
where the couplings $v_e$ and $a_e$ have been defined 
at the beginning of Section 5.
Because of some misprints in the formulas given in \cite{DHZ} for 
the coefficient $\mathcal A$, we have recalculated it.
Following \cite{DHZ}, we introduce $x_{1,2} = {{2E_{1,2}}/{\sqrt s}}$
for the scaled energies of the Higgs particles, $x_3 = 2 - x_1 -x_2$
for the scaled energy of the $Z$ boson, and $y_k = 1 - x_k$. 
Also, we denote by $\mu_i = m_i^2/s$ the scaled squared masses of various 
particles:
\begin{equation}
\mu_h = m_h^2/s, \qquad  \mu_H = m_H^2/s, \qquad \mu_W = m_W^2/s.
\label{Eq:mu}
\end{equation}
We can express our result in a compact form as follows:
\begin{equation}
{\mathcal A}
 = \mu_Z\, \left\{ \half|a|^2\, f_a
+|b(y_1)|^2\, f_b +2\, \Re[a b^*(y_1)]\, g_{ab}
+\Re[b(y_1)b^*(y_2)]\,g_{bb}\right\} +\{x_1\leftrightarrow x_2\}.
\label{Eq:calA}
\end{equation}
Here,
\begin{equation}
a = \frac{1}{2}
\biggl[\frac{\sin(\beta-\alpha)\lambda_{hhh}}{y_3+\mu_Z-\tilde\mu_h}
+\frac{\cos(\beta-\alpha)\lambda_{Hhh}}{y_3+\mu_Z-\tilde\mu_H}\biggr]
+\biggl[\frac{\sin^2(\beta-\alpha)}{y_1+\mu_h-\tilde\mu_Z}
+\frac{\sin^2(\beta-\alpha)}{y_2+\mu_h-\tilde\mu_Z}\biggr]
+\frac{1}{2\mu_Z}
\label{Eq:lca}
\end{equation}
represents a contribution from diagram  
\ref{Fig:Feynman-nonres-Zhh}a, where the lepton tensor couples directly to 
the final-state $Z$ polarization tensor,
as well as the contributions of diagrams \ref{Fig:Feynman-nonres-Zhh}b 
and \ref{Fig:Feynman-nonres-Zhh}c.
Similarly,
\begin{equation}
b(y) = \frac{1}{2\mu_Z}\left(
 \frac{\sin^2(\beta-\alpha)}{y+\mu_h-\tilde\mu_Z}
+\frac{\cos^2(\beta-\alpha)}{y+\mu_h-\tilde\mu_A}
\right)
\label{Eq:lcb}
\end{equation}
represents the part of diagram \ref{Fig:Feynman-nonres-Zhh}a where 
the lepton tensor couples 
to the final-state $Z$ polarization tensor indirectly via the Higgs
momenta $q_1$ and $q_2$, as well as diagram \ref{Fig:Feynman-nonres-Zhh}d.
The tildes on $\mu_i$ keep track of the widths, e.g.\
$\tilde\mu_Z=(m_Z^2+im_Z\Gamma_Z)/s$.

The Higgs self-couplings $\lambda_{Hhh}$ and $\lambda_{hhh}$ occur only
in the function $a$, Eq.~(\ref{Eq:lca}). The coefficients $f$ and $g$,
which do not involve any Higgs couplings, can be expressed rather 
compactly as
\begin{eqnarray}
f_a &=&  x_3^2+8\mu_Z, \nonumber \\
f_b &=&  (x_1^2-4\mu_h)[(y_1-\mu_Z)^2-4\mu_Z\mu_h], \nonumber \\
g_{ab} &=&  \mu_Z[2(\mu_Z-4\mu_h)+x_1^2+x_2(x_2+x_3)]
-y_1(2y_2-x_1x_3),\nonumber \\
g_{bb} &=& \mu_Z^2(4\mu_h +6 -x_1x_2) +2\mu_Z(\mu_Z^2 +y_3 -4\mu_h) 
\nonumber \\
& & +(y_3-x_1x_2-x_3\mu_Z-4\mu_h\mu_Z)(2y_3-x_1x_2-4\mu_h+4\mu_Z).
\end{eqnarray}
These coefficients (we use a mixed notation, 
which involves both $x$ and $y$)
correspond to those of \cite{DHZ}
as follows: $(f_a,f_b,g_{ab},g_{bb})=(f_0,f_1,f_3,f_2)$.
With this identification, we agree with the result given in 
the Erratum to \cite{DHZ}.

In the limit of large $m_A$, 
$\sin(\beta - \alpha) \rightarrow 1$, the cross 
section reduces to the Standard Model cross section
with
\begin{eqnarray}
a &=&\frac{1}{2}\,
\frac{\lambda_{hhh}}{y_3+\mu_Z-\tilde\mu_h}
+\biggl[\frac{1}{y_1+\mu_h-\tilde\mu_Z}
+\frac{1}{y_2+\mu_h-\tilde\mu_Z}\biggr]
+\frac{1}{2\mu_Z} \\
b(y)&=&
\frac{1}{2\mu_Z}\, \frac{1}{y+\mu_h-\tilde\mu_Z},
\end{eqnarray}
where at the tree level, $\lambda_{hhh}\to\lambda^{\rm SM}_{hhh}$,
as discussed in Sec.~3.

We show in Fig.~\ref{Fig:sig-Zll-2}a the $Zhh$ cross section, as given by 
Eqs.~(\ref{Eq:sigZH}), (\ref{Eq:GamHhh}) and (\ref{Eq:sigZhh}),
in the limit of no squark mixing, and with $\tilde m = 1~\TeV$.
The structure around $m_h=70~\GeV$ is due to the vanishing
and near-vanishing of the trilinear coupling.
In Fig.~\ref{Fig:sig-Zll-2}b--d we have introduced squark mixing: 
$A=1~\TeV$, $\mu=0, \pm1~\TeV$.
(For the decoupling-limit cross section, which is also shown,
we use the MSSM coupling, instead of the SM coupling, for the
reason given in Sec.~3.)

In the case of no mixing, there is a broad minimum from $m_h\simeq78$
to 90~GeV, followed by an enhancement around $m_h\sim90$--100~GeV.
This structure is due to the vanishing of the branching ratio 
for $H\to hh$, which is kinematically forbidden in the region 
$m_h\simeq78$--90~GeV, see Fig.~\ref{Fig:hole} (this coincides with 
the opening up of the channel $H\to WW$), 
followed by an increase of the trilinear couplings.
This particular structure depends considerably on the exact mass values
$m_H$ and $m_h$. Thus, it depends on details of the radiative corrections 
and the mixing parameters $A$ and $\mu$. 

The $A\to hZ$ channel contributes of the order of 20\% in the region
of the maximum at $m_h=90$--100~GeV.
\subsection{Triple-$h$ production}
The resonant and non-resonant production of $Ahh$ 
[Fig.~\ref{Fig:Feynman-nonres-Ahh}] can lead to three-$h$
final states in the region of $m_A$, where $A$ has a significant
branching ratio for decaying to $Zh$, i.e.\ for $m_A$ below
the $t\bar t$ threshold, and relatively low values of $\tan\beta$
[cf. Figs.~\ref{Fig:BR-H-A-2} and \ref{Fig:BR-H-A-5}].

In principle, this channel allows for a study of the coupling
$\lambda_{hAA}$ [cf.\ Fig.~\ref{Fig:Feynman-nonres-Ahh}a].
However, the prospects for measuring this coupling, which is
rather small [see Fig.~\ref{Fig:lamhAA2}], was studied in
Ref.~\cite{DHZ} and found not to be very encouraging.
\setcounter{equation}{0}
\section{Fusion Mechanism for Multiple-$h$ Production}
As mentioned in Section 4, a double Higgs ($hh$) final state
in $e^+ e^-$ collisions can also result from the $WW$ fusion mechanism, 
which can either be a resonant process as in (\ref{Eq:res-Hhh}), 
or a non-resonant one like (\ref{Eq:WW-fusion}). 
Since the neutral-current couplings are smaller than 
the charged-current ones, 
the cross section for the $ZZ$ fusion mechanism in (\ref{Eq:res-Hhh}) 
and (\ref{Eq:WW-fusion}) is an order of magnitude smaller than the 
$WW$ fusion mechanism. We shall thus, in the following,
ignore the $ZZ$ fusion mechanism, 
and concentrate instead on the $WW$ mechanism.

\subsection{Resonant $WW$ fusion}

The $WW$ fusion mechanism provides another large 
cross section for the multiple production of $h$ bosons. 
The cross section for 
$e^+e^- \rightarrow H\bar\nu_e\nu_e$ can be written
as~\cite{DHKMZ} 
\begin{equation}
\sigma(e^+e^- \rightarrow H\bar\nu_e\nu_e) 
= \frac{G_F^3 m_W^4}{64 \sqrt 2\pi^3 }
\left[\int_{\mu_H}^1 dx\int_{x}^1 \frac{dy}
{\left[1 + (y-x)/\mu_W\right]^2}\ {\cal F}(x,y)\right]
\cos^2(\beta - \alpha), 
\label{Eq:fusion-exact}
\end{equation}
where 
\begin{eqnarray}
{\cal F}(x,y)
& = & 16[F(x,y)+G(x,y)], \\
F(x,y) & = & \left[ \frac{2x}{y^3} - \frac{1 + 2x}{y^2}
+\frac{2 + x}{2y} - \frac{1}{2}\right]
\left[\frac{z}{1 + z} - \log(1 + z)\right]
+ \frac{x}{y^3}\frac{z^2(1 - y)}{(1 + z)}, \\ 
G(x,y) & = & \left[ -\frac{x}{y^2} 
+\frac{2 + x}{2y} - \frac{1}{2}\right]
\left[\frac{z}{1 + z} - \log(1 + z)\right],
\end{eqnarray}
with $\mu_i$ defined by Eq.~(\ref{Eq:mu}) and
\begin{equation}
z = \frac{y(x - \mu_H)}{\mu_W x}.
\end{equation}
For $\sqrt s$, $m_H$ $\gg$ $m_W$, and
in the effective longitudinal $W$ approximation,
the cross section (\ref{Eq:fusion-exact}) for 
$e^+e^- \rightarrow H\bar\nu_e\nu_e$ can be written in  
the following simple form \cite{CDCG}
\begin{equation}
\sigma(e^+e^- \rightarrow H\bar\nu_e\nu_e) 
= \frac{G_F^3 m_W^4}
{4\sqrt 2\ \pi^3}\left[\left(1 + \frac{m_H^2}{s}\right)
\log\frac{s}{m_H^2}
-2\left(1 - \frac{m_H^2}{s}\right)\right]\cos^2(\beta - \alpha).
\label{32}
\end{equation}
However, in this approximation the cross section may be overestimated
by a factor of $2$ for small values of masses and/or small 
centre-of-mass energies. 
For example, at $\sqrt s = 500$~GeV the 
equivalent $W$ approximation gives a result that is 
twice as large as the exact cross section. 
Therefore, we use the exact cross section (\ref{Eq:fusion-exact}) in 
our calculations.

The cross section (\ref{Eq:fusion-exact}) is plotted in 
Fig.~\ref{Fig:sigma-500-1500} for centre-of-mass energies, 
$\sqrt s = 500$ GeV and $1.5$ TeV, and for $\tan\beta = 2.0$,
as a function of $m_H$.  
The resonant fusion mechanism, which leads to $[hh]$ + [missing energy]
final states is competitive with the process
$e^+e^- \rightarrow HZ \rightarrow [hh]$ + [missing energy], 
particularly at high energies.
Since the dominant decay of $h$ will be into
$b\bar b$ pairs, the $H$-strahlung and the fusion mechanism will give 
rise to final states that will predominantly include four $b$-quarks.
On the other hand, the process $e^+e^- \rightarrow AH$ will give rise to
six $b$-quarks in the final state, since the $AH$ final state typically 
yields three-Higgs $h[hh]$ final states.
\subsection{Non-resonant fusion $WW \rightarrow hh$}
Besides the resonant $WW$ fusion mechanism for the multiple
production of $h$ bosons, there is also a non-resonant $WW$ 
fusion mechanism:
\begin{equation}
\label{Eq:WW-nonres}
e^+e^-\to\nu_e\bar\nu_e hh,
\end{equation}
through which the same final state of two $h$ bosons can be produced.
The cross section for this process, which arises
through $WW$ exchange as indicated in 
Fig.~\ref{Fig:Feynman-nonres-WW}, can be written in the
`effective $WW$ approximation' as\footnote{There could be sizable 
deviations of the effective $WW$ approximation from the exact result.}
\begin{equation}
\label{Eq:sigWW-nonres}
\sigma(e^+e^-\to\nu_e\bar\nu_e hh)
=\int_\tau^1\dd x\, \frac{\dd L}{\dd x}\, \hat\sigma\sup_{WW}(x),
\end{equation}
where $\tau=4m_h^2/s$. 
In the above, the cross section is written as a $WW$ cross section,
at invariant energy squared $\hat s=xs$, 
folded with the $WW$ `luminosity' \cite{CDCG}:
\begin{equation}
\frac{\dd L(x)}{\dd x}
=\frac{G_{\rm F}^2m_W^4}{2}\,\left(\frac{v^2+a^2}{4\pi^2}\right)^2
\frac{1}{x}\biggl\{(1+x)\log\frac{1}{x}-2(1-x)\biggr\},
\end{equation}
where $v^2+a^2=2$.

The $WW$ cross section receives contributions from several amplitudes,
according to the diagrams (a)--(d) \footnote{For each of the diagrams (a) 
and (d) there are two contributions, corresponding to the interchange
of the two Higgs particles.}
in Fig.~\ref{Fig:Feynman-nonres-WW}.
We have evaluated\footnote{There are some misprints in Ref.~\cite{DHZ},
so we present here results of an independent calculation.} 
these contributions and express the result in a form analogous to 
that of Ref.~\cite{DHZ}\footnote{In Ref.~\cite{DHZ}, the factor
in front of the term involving $g_1$ and $g_2$ reads
$\beta_W^2/(\beta_W\beta_h)$; it should be
$(1+\beta_W^2)/(\beta_W\beta_h)$. With $\beta_W=1$,
the prefactors in their Eq.~(16) reduce to ours.}:
\begin{eqnarray}
\label{Eq:sighat}
\hat\sigma\sup_{WW}(x)
&=&\frac{G^2_{\rm F}\hat s}{64\pi}\beta_h
\biggl\{4\biggl[
 \frac{\hat\mu_Z\sin(\beta-\alpha)}{1-\hat\mu_h}\,\lambda_{hhh}
+\frac{\hat\mu_Z\cos(\beta-\alpha)}{1-\hat\mu_H}\,\lambda_{Hhh}
+1\biggr]^2\,g_0 \nonumber \\
&& \phantom{\frac{2}{\beta_h}}
+\frac{2}{\beta_h}\biggl[
 \frac{\hat\mu_Z\sin(\beta-\alpha)}{1-\hat\mu_h}\,\lambda_{hhh}
+\frac{\hat\mu_Z\cos(\beta-\alpha)}{1-\hat\mu_H}\,\lambda_{Hhh}
+1\biggr]\nonumber \\
&& \phantom{\frac{2}{\beta_h}AAAA}
\times[\sin^2(\beta-\alpha)\,g_1 +\cos^2(\beta-\alpha)\,g_2] \nonumber \\
&& \phantom{\frac{2}{\beta_h}}
+\frac{1}{\beta_h^2}
\{\sin^4(\beta-\alpha)\,g_3+\cos^4(\beta-\alpha)\,g_4
+\sin^2[2(\beta-\alpha)]\,g_5\}\biggr\},
\end{eqnarray}
where we have introduced `reduced squared masses'
\begin{equation}
\hat\mu_Z=m_Z^2/\hat s, \qquad
\hat\mu_W=m_W^2/\hat s, \qquad
\hat\mu_h=m_h^2/\hat s, \qquad
\hat\mu_H=m_H^2/\hat s, 
\end{equation}
and the Higgs velocity is $\beta_h=\left(1-\hat\mu_h\right)^{1/2}$.

Our approach differs from that of DHZ in that we do not project out 
the longitudinal degrees of freedom of the intermediate $W$ bosons.
Instead, we follow the approach of Ref.~\cite{AMP}, where transverse
momenta are ignored everywhere except in the $W$ propagators,
the integrations over which are approximated as (here $p_1$ and
$p_1'$ denote electron and neutrino momenta, respectively,
in the process (\ref{Eq:WW-nonres})):
\begin{equation}
\int\dd^2\pmb{p}_{1\perp}\,\frac{1}{[(p_1-p_1')^2-m_W^2]^2}
\simeq\frac{\pi(1-x_1)}{m_W^2},
\end{equation}
where $x_1$ [and $x_2$] represents the energy of the $W$.
The energy squared of the subprocess is given as
$\hat s=(p_1x_1+p_2x_2)^2=x_1x_2s=xs$.

The contributions of diagrams (b)+(c), (a) and (d) are given
by the terms $g_0$, $g_3$ and $g_4$, respectively,
with $g_0(x)=1$, and 
\begin{eqnarray}
g_3(x)&=&8\beta_h[2\hat\mu_W+(\hat\mu_h-\hat\mu_W)^2]
[2\hat\mu_W+1-3(\hat\mu_h-\hat\mu_W)^2]\frac{l_W}{a_W}
\nonumber \\
&&+16[2\hat\mu_W+(\hat\mu_h-\hat\mu_W)^2]^2y_W
+16\beta_h^2(1+a_W)^2, \nonumber \\
g_4(x)&=&8\beta_h(\hat\mu_h-\hat\mu_C)^2[1-3(\hat\mu_h-\hat\mu_C)^2]
\frac{l_C}{a_C} \nonumber \\
&&+16(\hat\mu_h-\hat\mu_C)^4y_C
+16\beta_h^2(1+a_C)^2, 
\end{eqnarray}
where
\begin{eqnarray}
l_W&=&\log\frac{1-2\hat\mu_h+2\hat\mu_W-\beta_h}
               {1-2\hat\mu_h+2\hat\mu_W+\beta_h}, \\
y_W&=&\frac{2\beta_h^2}{(1-2\hat\mu_h+2\hat\mu_W)^2-\beta_h^2}, \\
a_W&=&-\half+\hat\mu_h-\hat\mu_W,
\end{eqnarray}
and similarly $l_C$, $y_C$ and $a_C$, with $\hat\mu_W$ replaced
by $\hat\mu_C$, the latter being defined in terms of the charged
Higgs mass $m_{H^+}$.

The interference between diagrams (b)+(c) and (a) is given by
the term $g_1$, whereas the interferences between diagrams (b)+(c) 
and (d), and between (a) and (d) are given by $g_2$ and $g_5$, 
respectively.
For these interference terms, we find
\begin{eqnarray}
g_1(x)&=&8[2\hat\mu_W+(\hat\mu_h-\hat\mu_W)^2]l_W
-4\beta_h(1+2\hat\mu_h-2\hat\mu_W),
\nonumber \\
g_2(x)&=&8(\hat\mu_h-\hat\mu_C)^2l_C
-4\beta_h(1+2\hat\mu_h-2\hat\mu_C),
\nonumber \\
g_5(x)&=&\frac{\beta_h}{4}(Z_Wl_W+Z_Cl_C)+8\beta^2_h(1+a_W)(1+a_C),
\end{eqnarray}
with
\begin{eqnarray}
Z_W&=&\frac{(1+2a_W)^2}{a_C-a_W}[8\hat\mu_W+(1+2a_W)^2] 
     +\frac{(1-2a_W)^2}{a_C+a_W}[8\hat\mu_W+(1+2a_W)^2], \\
Z_C&=&-\frac{(1+2a_C)^2}{a_C-a_W}[8\hat\mu_W+(1+2a_C)^2] 
      +\frac{(1+2a_C)^2}{a_C+a_W}[8\hat\mu_W+(1-2a_C)^2].
\end{eqnarray}

Our functions $g_1$--$g_5$ correspond to those of DHZ, cf.\ our
Eq.~(\ref{Eq:sighat}) and their Eq.~(16) in \cite{DHZ}.
At small $m_h$, the cross section is sensitive to small $x$, 
where the `effective $WW$ approximation'
is not well defined, and our results differ from those of DHZ.
However, apart from the contributions from small $x$, 
our results agree with those of DHZ to a precision of 1--5\%.

We show in Fig.~\ref{Fig:sig-WW-2} the $WW$ fusion cross section, 
at $\sqrt{s}=1.5~\TeV$,
as given by Eqs.~(\ref{Eq:fusion-exact}) and (\ref{Eq:sigWW-nonres}),
in the limit of no squark mixing, 
as well as with mixing (as indicated), and with $\tilde m = 1~\TeV$.
The structure is very reminiscent of that of Fig.~\ref{Fig:sig-Zll-2},
and for the same reasons. However, the scale is different.

\setcounter{equation}{0}
\section{Sensitivity to $\lambda_{Hhh}$ and $\lambda_{hhh}$}
Following \cite{DHZ}, we have indicated in the $m_A$--$\tan\beta$
plane the regions where $\lambda_{Hhh}$ and $\lambda_{hhh}$
might be measurable, according to criteria analogous
to those given there.
In Fig.~\ref{Fig:sensi-500}, we consider $\sqrt{s}=500~\GeV$,
and identify regions according to the following criteria:
\begin{itemize}
\item[(i)]
Regions where $\lambda_{Hhh}$ might become measurable are identified
as those where 
$\sigma(H)\times\mbox{BR}(H\to hh)> 0.1\mbox{ fb}$ (solid),
with the simultaneous requirement of 
$0.1 < \mbox{BR}(H\to hh) < 0.9$
[see Figs.~\ref{Fig:BR-H-A-2}--\ref{Fig:hole}].
In view of the recent, more optimistic, view on the
luminosity that might become available, 
we also give the corresponding contours for 0.05~fb (dashed) 
and 0.01~fb (dotted). For $\sigma(H)$ we take the sum of
(\ref{Eq:sigZH}), (\ref{Eq:sigAH}) and (\ref{Eq:fusion-exact}).
\item[(ii)]
Regions where $\lambda_{hhh}$ might become measurable
are those where the {\it continuum} $WW\to hh$
cross section [Eq.~(\ref{Eq:sigWW-nonres})] is larger than 
0.1~fb (solid).
Also included are contours at 0.05 (dashed) and 0.01~fb (dotted).
\end{itemize}
Such regions are given for four cases of the mixing parameters
$A$ and $\mu$, as indicated.
We have excluded from the plots the region where $m_h<62.5~\GeV$,
according to the LEP lower bound \cite{ALEPH}.
This corresponds to low values of $m_A$.

We note that with an integrated luminosity of 500~fb$^{-1}$,
the contours at 0.1~fb correspond to 50 events per year.
This will of course be reduced by efficiencies, but should indicate
the order of magnitude that can be reached.

At $\sqrt{s}=500~\GeV$, with a luminosity of 500~fb$^{-1}$ per year,
the trilinear coupling $\lambda_{Hhh}$ is accessible in a considerable
part of the $m_A$--$\tan\beta$ parameter space: at $m_A$ of the order
of 200--300~GeV and $\tan\beta$ up to the order of 5.
With increasing luminosity, the region extends somewhat 
to higher values of $m_A$.

At values of $m_A$ below 100~GeV, there is also a narrow band where 
$\lambda_{Hhh}$ is accessible. 

The `steep' edge around $m_A\simeq200~\GeV$ (where increased luminosity
does not help) is determined by the vanishing of $\mbox{BR}(H\to hh)$,
see Fig.~\ref{Fig:hole}.

The coupling $\lambda_{hhh}$ is accessible in a much larger part
of this parameter space, but with a moderate luminosity,
`large' values of $\tan\beta$ are accessible only if $A$ is small.

In Fig.~\ref{Fig:sensi-1500}, we consider $\sqrt{s}=1.5$~TeV,
and present the analogous contours.
Here, for the case of $\lambda_{Hhh}$ we demand
$\sigma(H)\times\mbox{BR}(H\to hh)> 0.5\mbox{ fb}$ (solid)
and 0.1~fb (dashed), and for the case of $\lambda_{hhh}$ we require
the corresponding cross section [Eq.~(\ref{Eq:sigWW-nonres})] 
to be larger than 0.5~fb (solid) and 0.1~fb (dashed).
If a luminosity corresponding to these cross sections becomes available
at $\sqrt{s}=1.5$~TeV, a somewhat larger region 
than at $\sqrt{s}=500$~GeV is accessible
in the $m_A$--$\tan\beta$ plane.

It should be stressed that the requirements discussed here
are necessary, but not sufficient conditions for the trilinear
couplings to be measurable. We also note that there might be 
sizable corrections to the $WW$ approximation, and that 
it would be desirable to incorporate the dominant two-loop
corrections to the trilinear couplings in the calculations.
\setcounter{equation}{0}
\section{Conclusions}
We have carried out a detailed investigation of the
possibility of measuring
the MSSM trilinear couplings $\lambda_{Hhh}$ and $\lambda_{hhh}$
at an $e^+ e^-$ collider.
Where there is an overlap, we  
have confirmed the results of Ref.~\cite{DHZ}.
Our emphasis has been on taking into account
all the parameters of the MSSM Higgs sector.
We have studied the importance of mixing in the squark sector,
as induced by the trilinear coupling $A$ and the bilinear coupling $\mu$.

At moderate energies ($\sqrt{s}=500~\GeV$) the range in 
the $m_A$--$\tan\beta$ plane that is accessible for studying 
$\lambda_{Hhh}$ changes quantitatively for non-zero values of
the parameters $A$ and $\mu$.
As far as the coupling $\lambda_{hhh}$
is concerned, however, there is a qualitative change from the case of
no mixing in the squark sector.
If $A$ is large, then high luminosity is required to reach 
`high' values of $\tan\beta$.
At higher energies ($\sqrt{s}=1.5~\TeV$), the mixing parameters
$A$ and $\mu$ change the accessible region of the
parameter space only in a quantitative manner.

\section*{Acknowledgements}
P.~O. would like to thank the DESY Theory Group and the
CERN Theory Division, 
whereas P.~N.~P. would like to thank the University of Bergen, 
for kind hospitality while parts of this work were finished.
It is also a pleasure to thank Abdel Djouadi, Wolfgang Hollik, 
Bernd Kniehl, Conrad Newton and Peter Zerwas 
for valuable discussions and advice.
This research was supported by the Research Council of Norway,
and (PNP) by the University Grants Commission, India under project 
number 10-26/98(SR-I).

\clearpage

\begin{figure}[htb]
\refstepcounter{figure}
\label{Fig:masses}
\addtocounter{figure}{-1}
\begin{center}
\setlength{\unitlength}{1cm}
\begin{picture}(16,16.4)
\put(0,-1)
{\mbox{\epsfxsize=15cm\epsffile{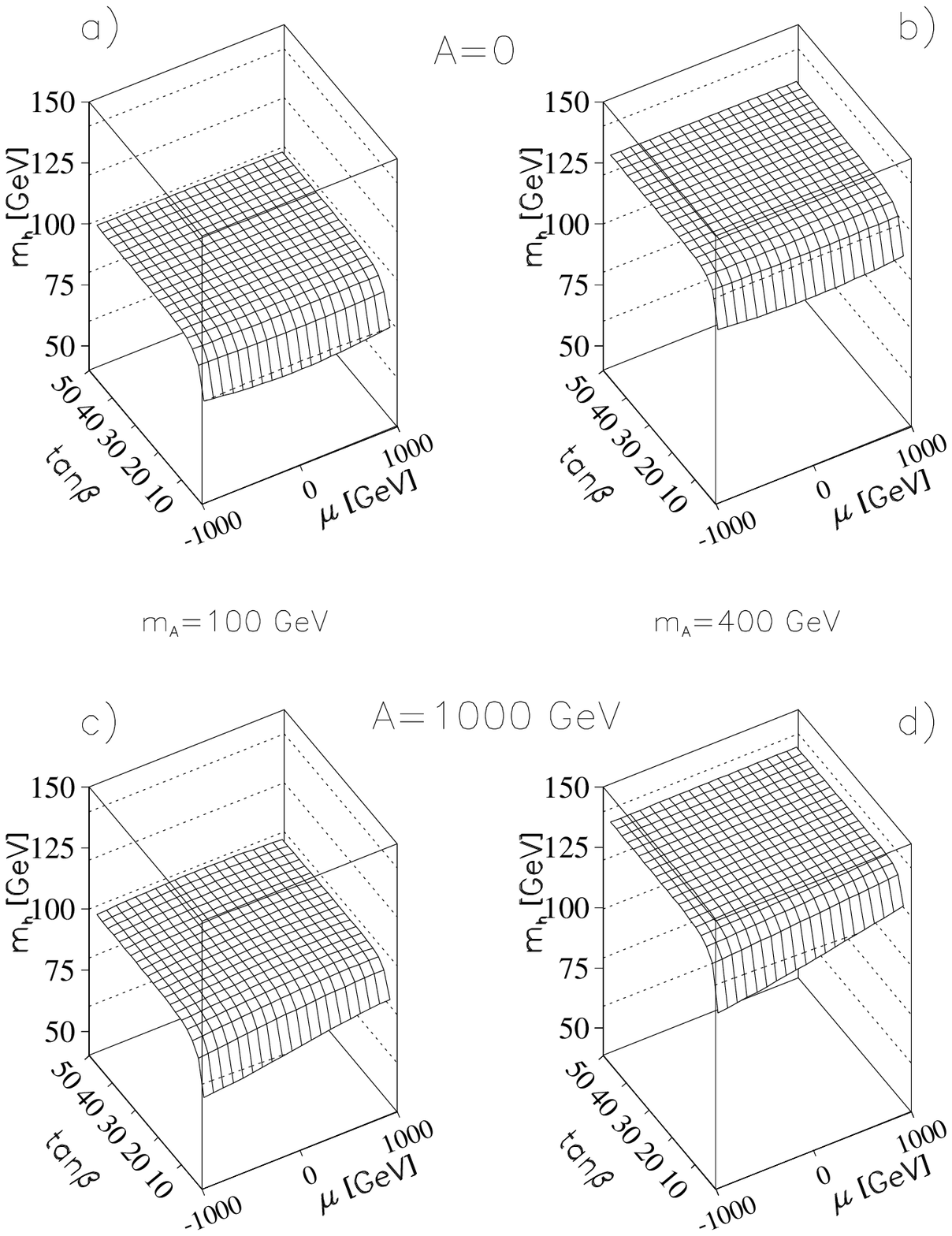}}}
\end{picture}
\vspace*{0mm}
\caption{Mass of the lightest Higgs boson $m_h$ as a function of 
$\mu$ and $\tan\beta$.
Two values of $m_A$ and two values of $A$ are considered: 
a)~$m_A=100~\GeV$, $A=0$,
b)~$m_A=400~\GeV$, $A=0~\GeV$, 
c)~$m_A=100~\GeV$, $A=1~\TeV$,
d)~$m_A=400~\GeV$, $A=1~\TeV$. 
We have taken $\tilde m = 1~\TeV$.}
\end{center}
\end{figure}

\begin{figure}[htb]
\refstepcounter{figure}
\label{Fig:lamHhh2}
\addtocounter{figure}{-1}
\begin{center}
\setlength{\unitlength}{1cm}
\begin{picture}(16,16.4)
\put(0,-1)
{\mbox{\epsfxsize=15cm\epsffile{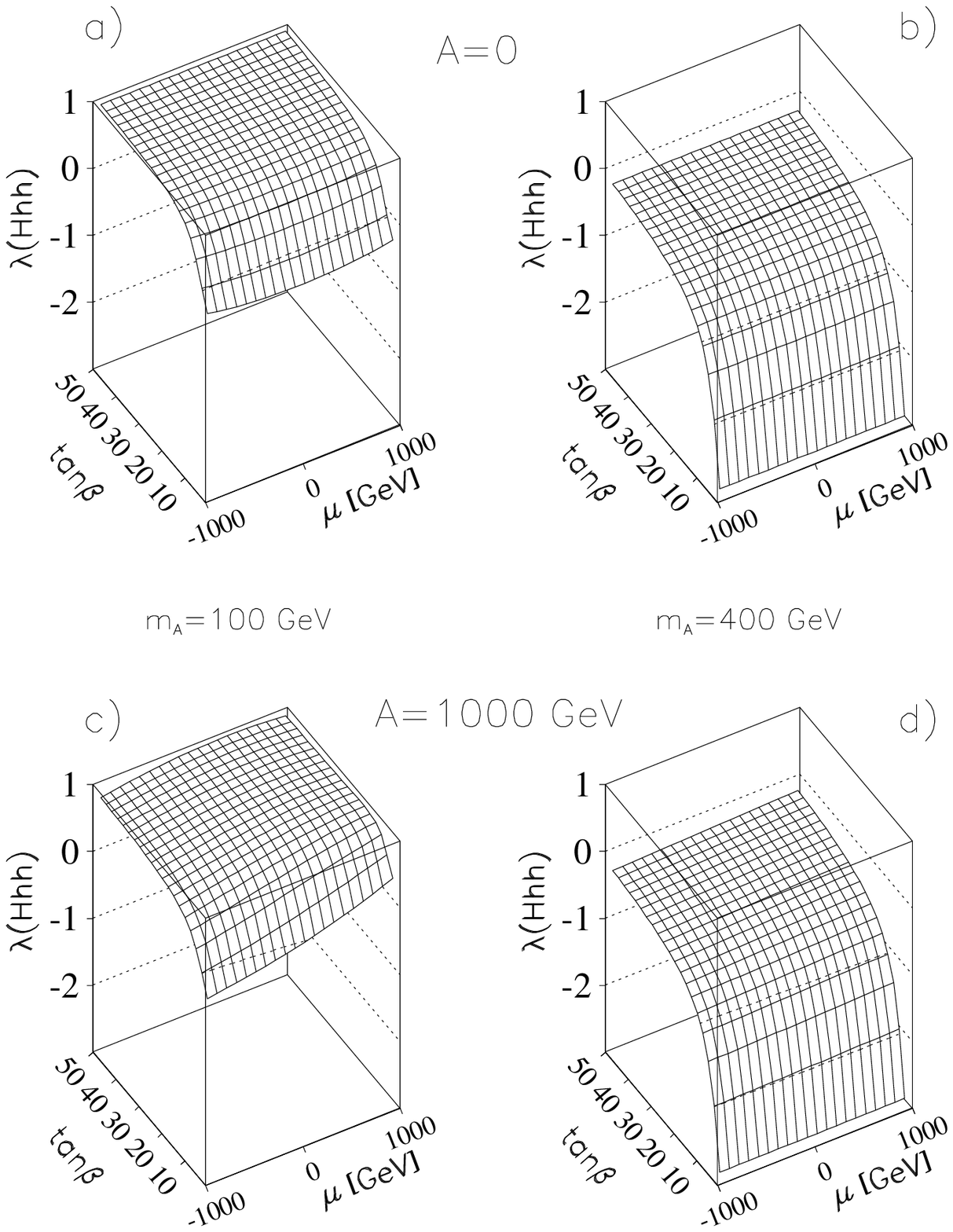}}}
\end{picture}
\vspace*{0mm}
\caption{Trilinear Higgs coupling $\lambda_{Hhh}$ as a function of $\mu$ 
and $\tan\beta$.
The values of the parameters are the same as in Fig.~\ref{Fig:masses}.}
\end{center}
\end{figure}

\begin{figure}[htb]
\refstepcounter{figure}
\label{Fig:lamhhh2}
\addtocounter{figure}{-1}
\begin{center}
\setlength{\unitlength}{1cm}
\begin{picture}(16,16.4)
\put(0,-1)
{\mbox{\epsfxsize=15cm\epsffile{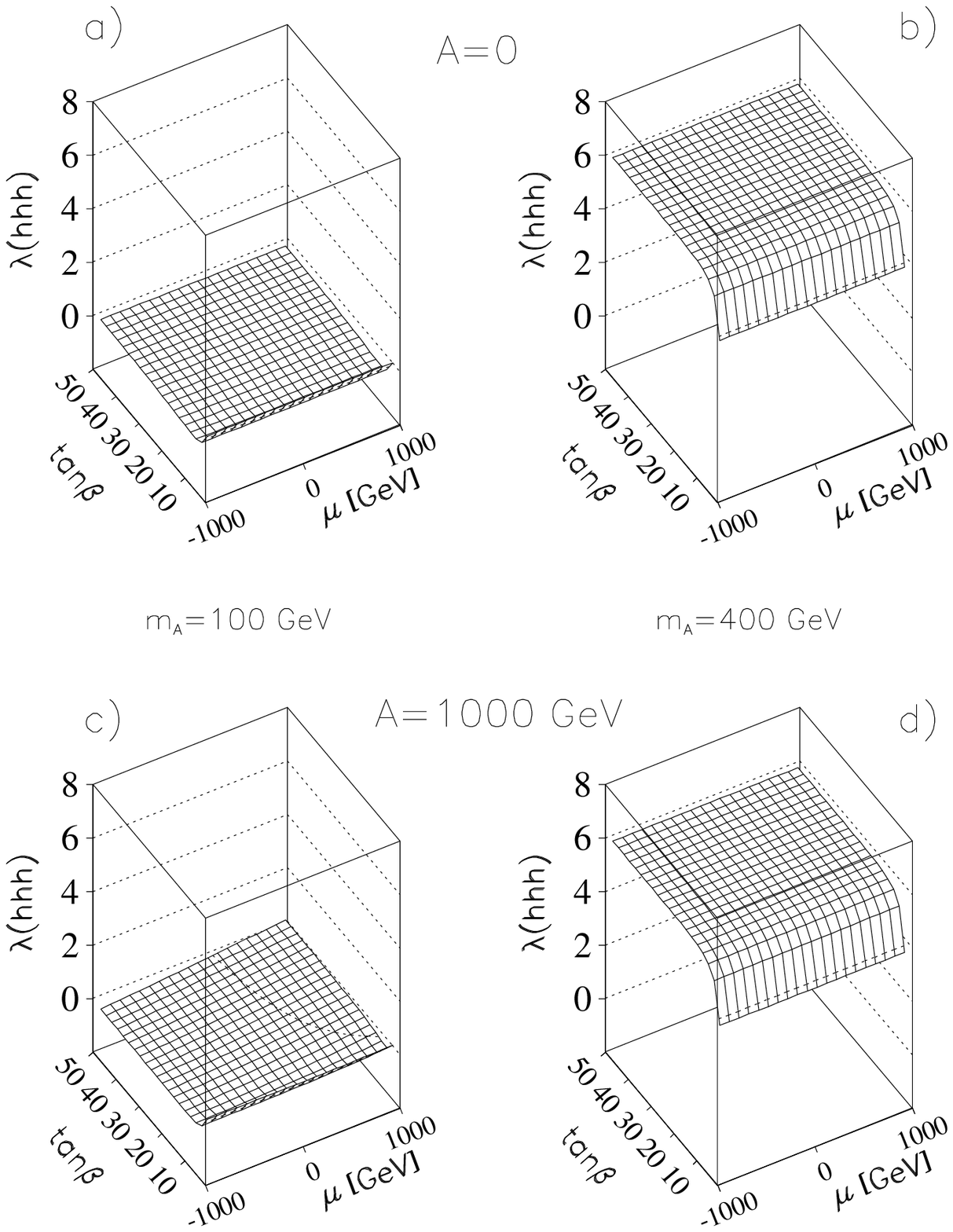}}}
\end{picture}
\vspace*{0mm}
\caption{Trilinear Higgs coupling $\lambda_{hhh}$ as a function of $\mu$ 
and $\tan\beta$.
The values of the parameters are the same as in Fig.~\ref{Fig:masses}.}
\end{center}
\end{figure}

\begin{figure}[htb]
\refstepcounter{figure}
\label{Fig:lamhAA2}
\addtocounter{figure}{-1}
\begin{center}
\setlength{\unitlength}{1cm}
\begin{picture}(16,16.4)
\put(0,-1)
{\mbox{\epsfxsize=15cm\epsffile{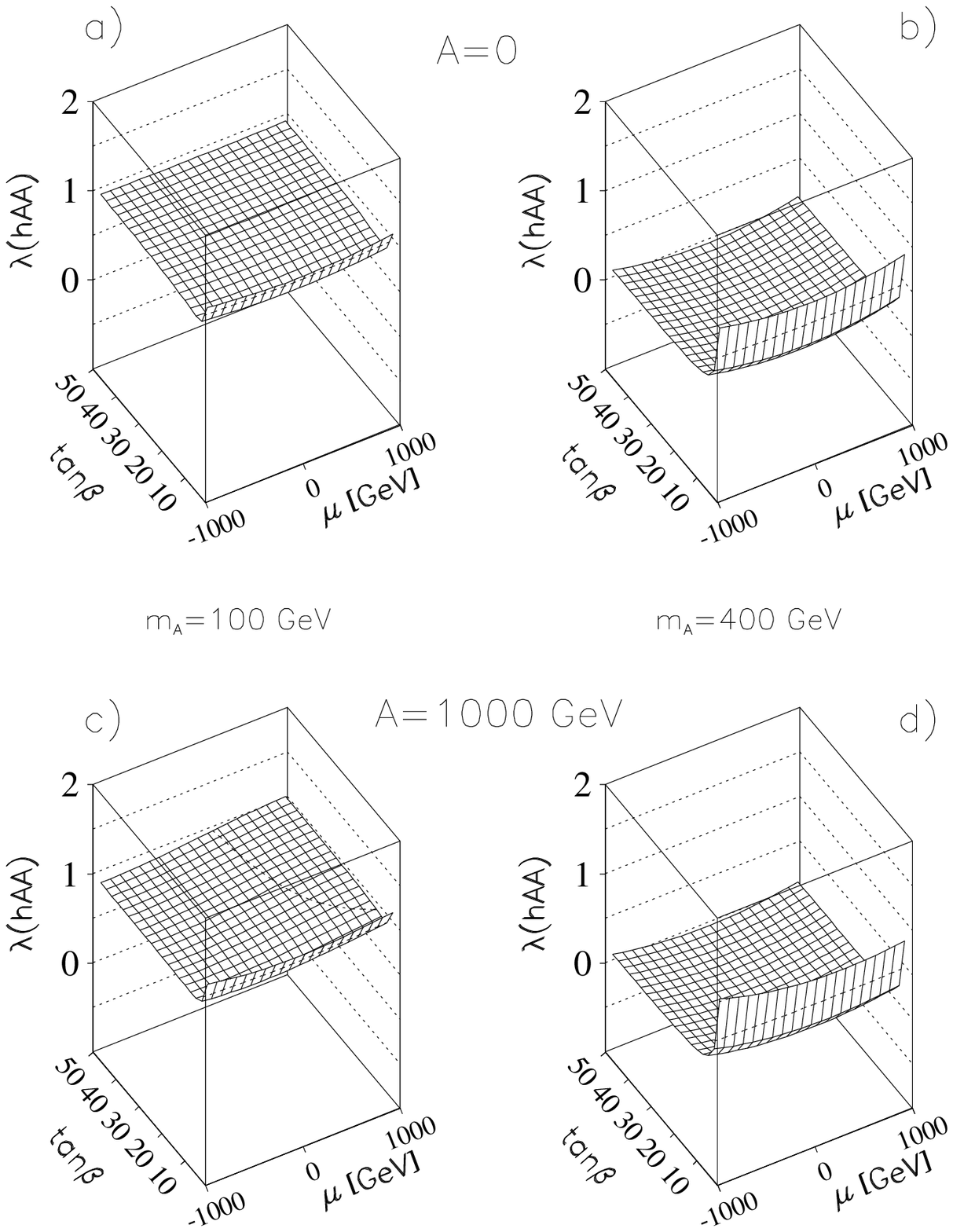}}}
\end{picture}
\vspace*{0mm}
\caption{Trilinear Higgs coupling $\lambda_{hAA}$ as a function of $\mu$ 
and $\tan\beta$.
The values of the parameters are the same as in Fig.~\ref{Fig:masses}.}
\end{center}
\end{figure}

\begin{figure}[htb]
\refstepcounter{figure}
\label{Fig:lam-mh}
\addtocounter{figure}{-1}
\begin{center}
\setlength{\unitlength}{1cm}
\begin{picture}(16,9)
\put(-1,1)
{\mbox{\epsfysize=9cm\epsffile{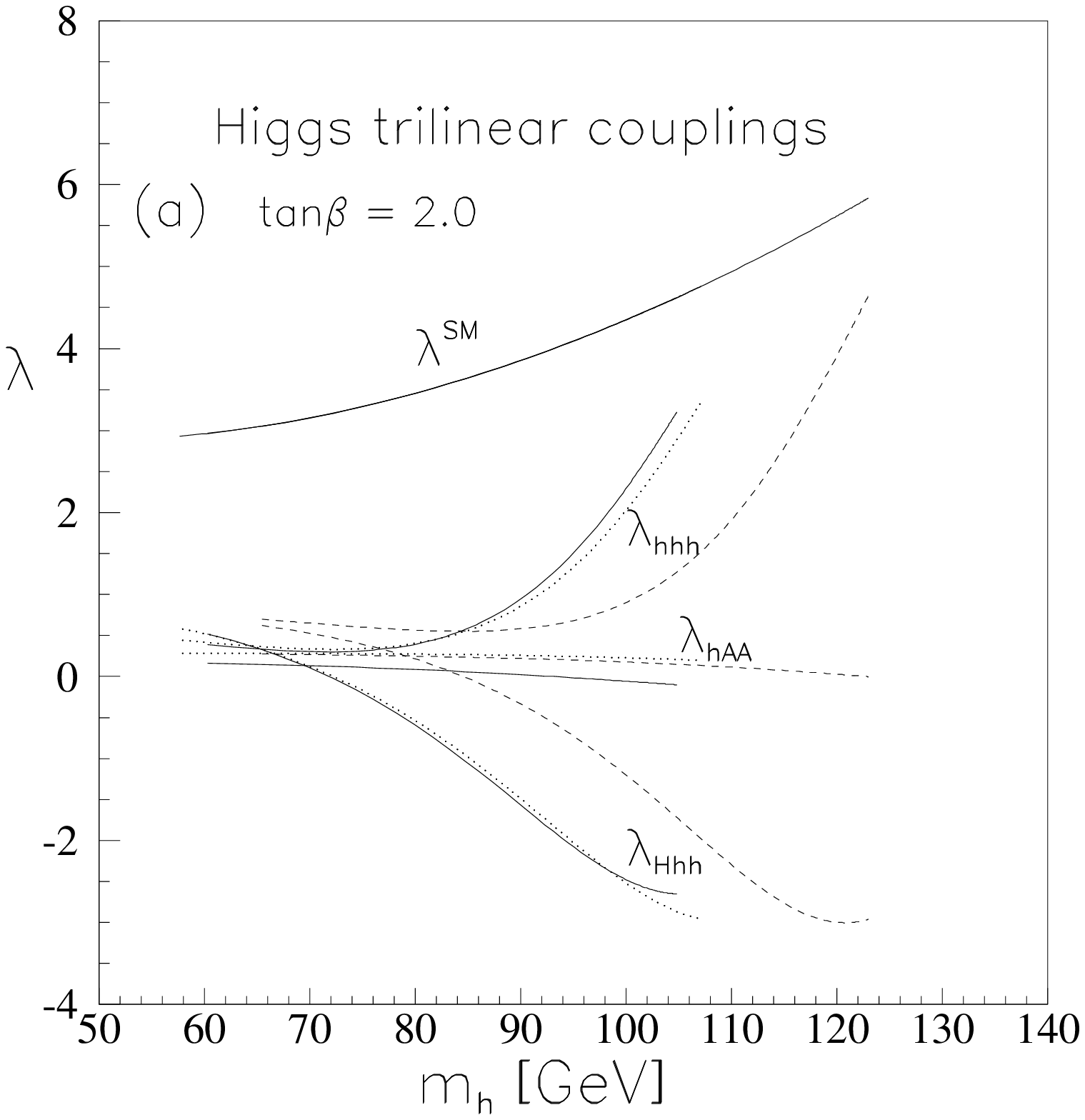}}
 \mbox{\epsfysize=9cm\epsffile{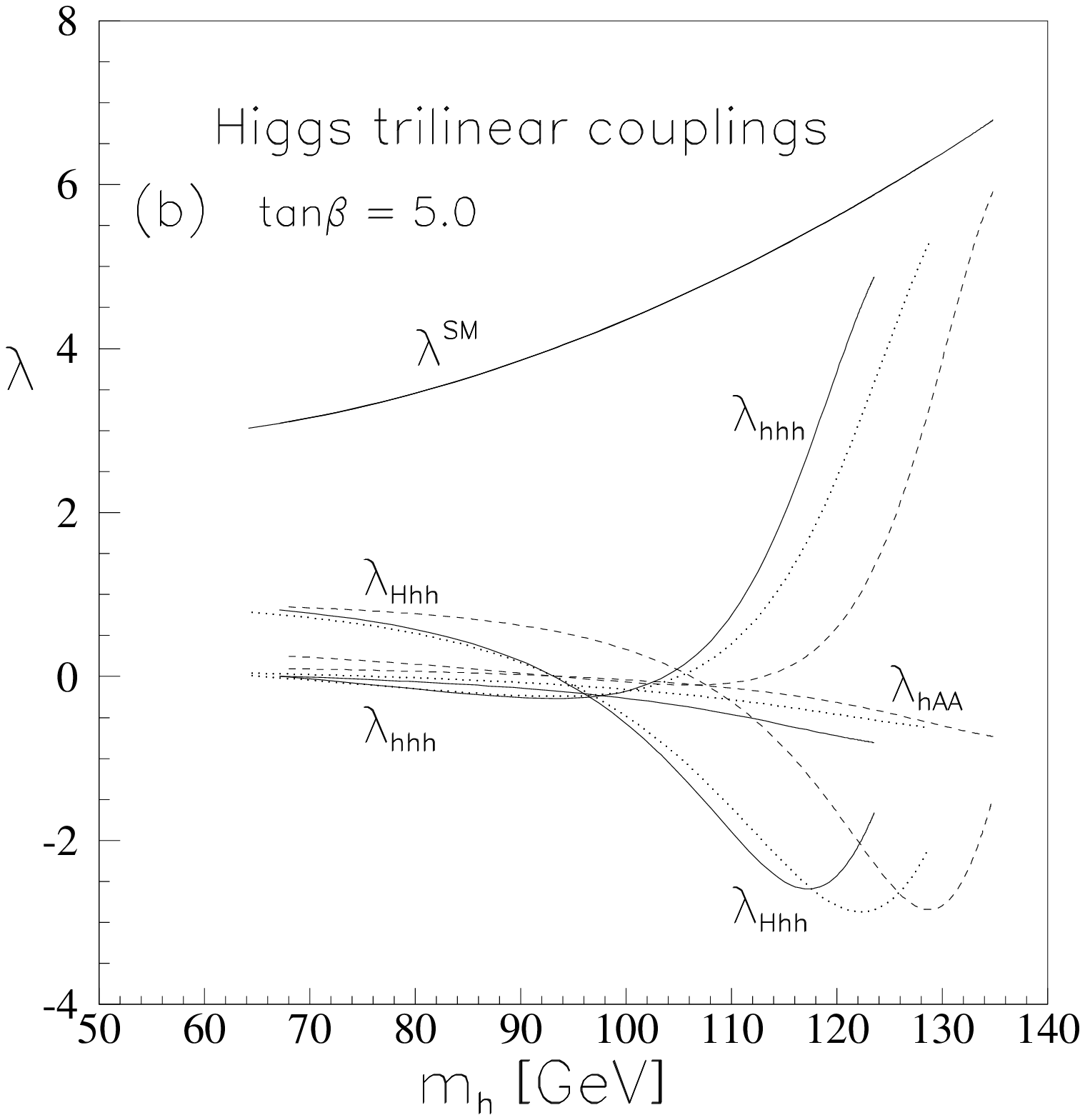}}}
\put(3,-8)
{\mbox{\epsfysize=9cm\epsffile{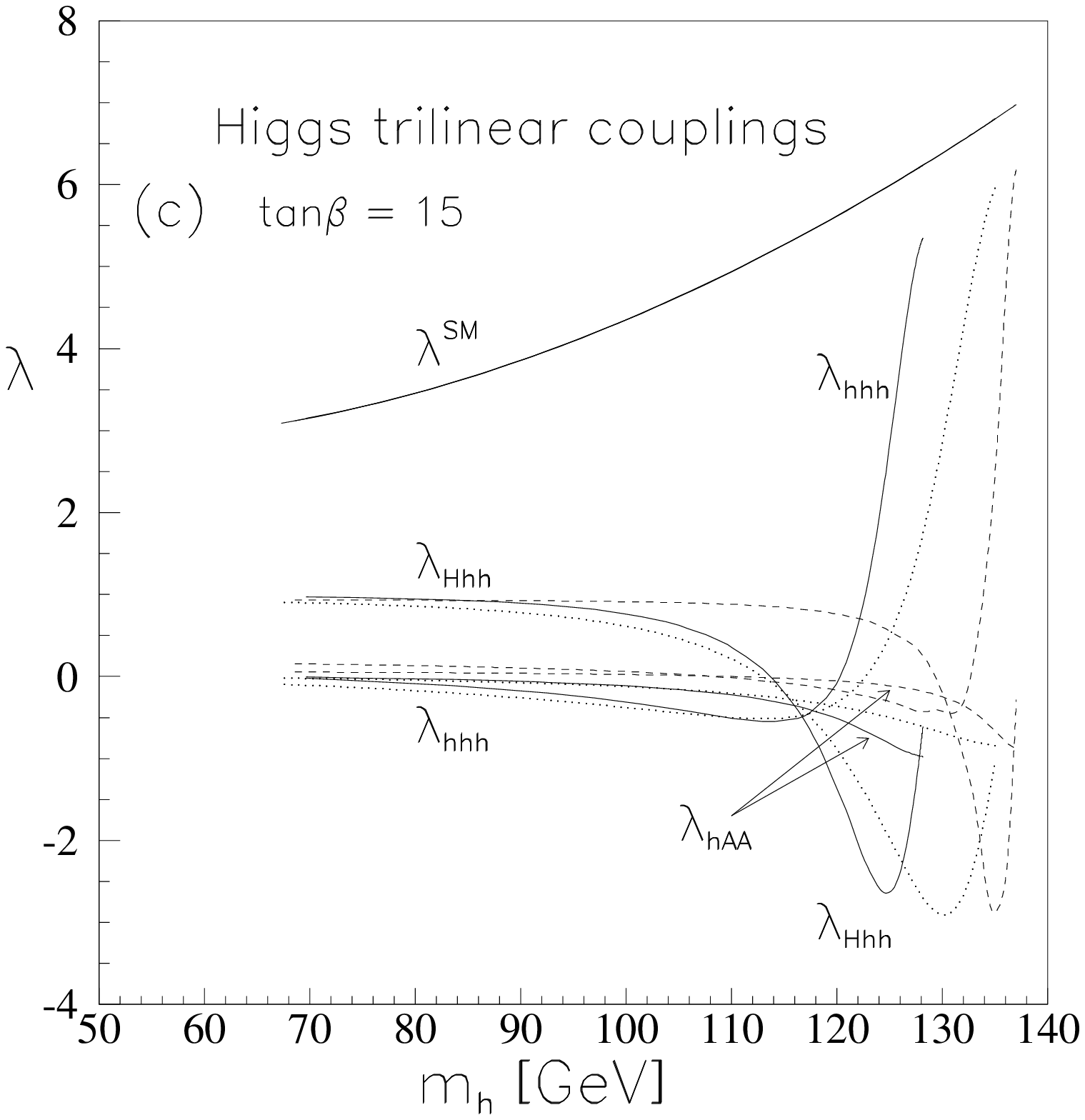}}}
\end{picture}
\vspace*{68mm}
\caption{Trilinear Higgs couplings $\lambda_{Hhh}$, $\lambda_{hhh}$ and
$\lambda_{hAA}$ as functions of $m_h$ for three values of $\tan\beta$:
(a) $\tan\beta=2.0$,
(b) $\tan\beta=5.0$, 
(c) $\tan\beta=15$.
Each coupling is shown for three cases of the
mixing parameters:
no mixing ($A=0$, $\mu=0$, solid),
mixing with $A=1$~TeV and $\mu=-1$~TeV (dotted),
as well as 
$A=1$~TeV and $\mu=1$~TeV (dashed).
For comparison the SM quartic coupling $\lambda^{\rm SM}$ 
is also shown.}
\end{center}
\end{figure}

\begin{figure}[htb]
\refstepcounter{figure}
\label{Fig:Feynman-resonant}
\addtocounter{figure}{-1}
\begin{center}
\setlength{\unitlength}{1cm}
\begin{picture}(16,6)
\put(2,-12)
{\mbox{\epsfxsize=16cm\epsffile{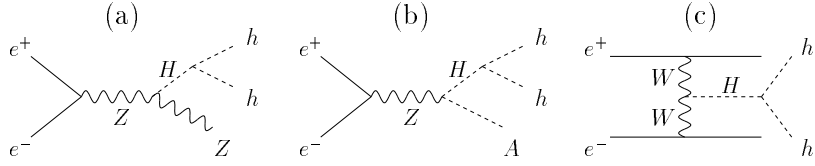}}}
\end{picture}
\vspace*{-60mm}
\caption{Feynman diagrams for the resonant production
of $hh$ final states in $e^+ e^-$ collisions. Diagrams (a) and (b)
represent the production of $H$ in association with
$Z$ and $A$, respectively, whereas diagram (c) is the 
$WW$ fusion  mechanism for the production of $H$. The Higgs boson
$H$ decays via $H \rightarrow hh$ to produce the two-Higgs 
final state.}
\end{center}
\end{figure}
\vspace*{-5mm}

\begin{figure}[htb]
\refstepcounter{figure}
\label{Fig:Feynman-nonres-Zhh}
\addtocounter{figure}{-1}
\begin{center}
\setlength{\unitlength}{1cm}
\begin{picture}(16,7)
\put(1,-12)
{\mbox{\epsfxsize=16cm\epsffile{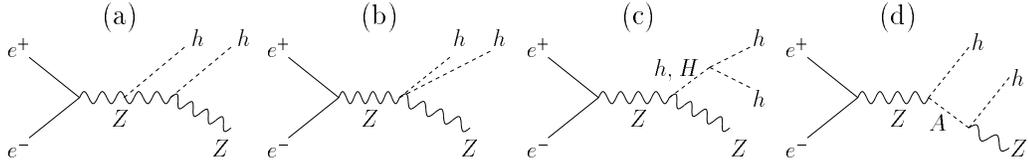}}}
\end{picture}
\vspace*{-60mm}
\caption{Feynman diagrams for the non-resonant 
production of $hh$ final states in association with $Z$.
The diagram (d), with $A$ produced on the mass shell, 
which subsequently decays via $A \rightarrow hZ$,
is a background to the resonance process, 
Fig.~\ref{Fig:Feynman-resonant}a.}
\end{center}
\end{figure}
\vspace*{-5mm}

\begin{figure}[htb]
\refstepcounter{figure}
\label{Fig:Feynman-nonres-Ahh}
\addtocounter{figure}{-1}
\begin{center}
\setlength{\unitlength}{1cm}
\begin{picture}(16,7)
\put(1,-12)
{\mbox{\epsfxsize=16cm\epsffile{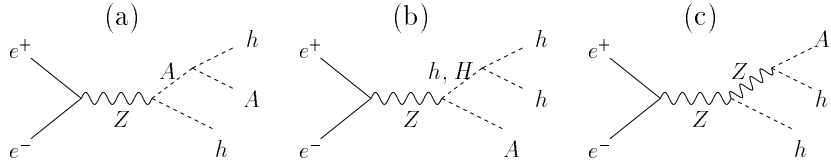}}}
\end{picture}
\vspace*{-60mm}
\caption{Feynman diagrams for the associated 
production of $hh$ with the pseudoscalar $A$ in the continuum.}
\end{center}
\end{figure}
\vspace*{-5mm}

\begin{figure}[htb]
\refstepcounter{figure}
\label{Fig:Feynman-nonres-WW}
\addtocounter{figure}{-1}
\begin{center}
\setlength{\unitlength}{1cm}
\begin{picture}(16,7)
\put(1,-12)
{\mbox{\epsfxsize=16cm\epsffile{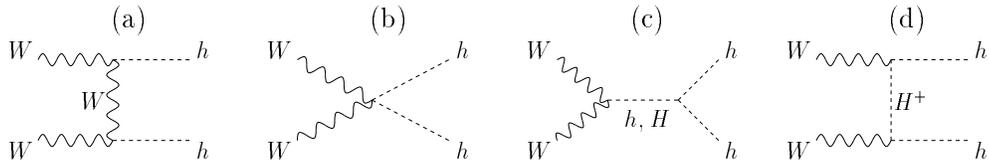}}}
\end{picture}
\vspace*{-60mm}
\caption{Feynman diagrams for the non-resonant $WW$ fusion
mechanism for the production of $hh$ states in $e^+ e^-$ collisions.}
\end{center}
\end{figure}
\vspace*{-5mm}

\begin{figure}[htb]
\refstepcounter{figure}
\label{Fig:sigma-500-1500}
\addtocounter{figure}{-1}
\begin{center}
\setlength{\unitlength}{1cm}
\begin{picture}(16,9)
\put(-1,4)
{\mbox{\epsfysize=9cm\epsffile{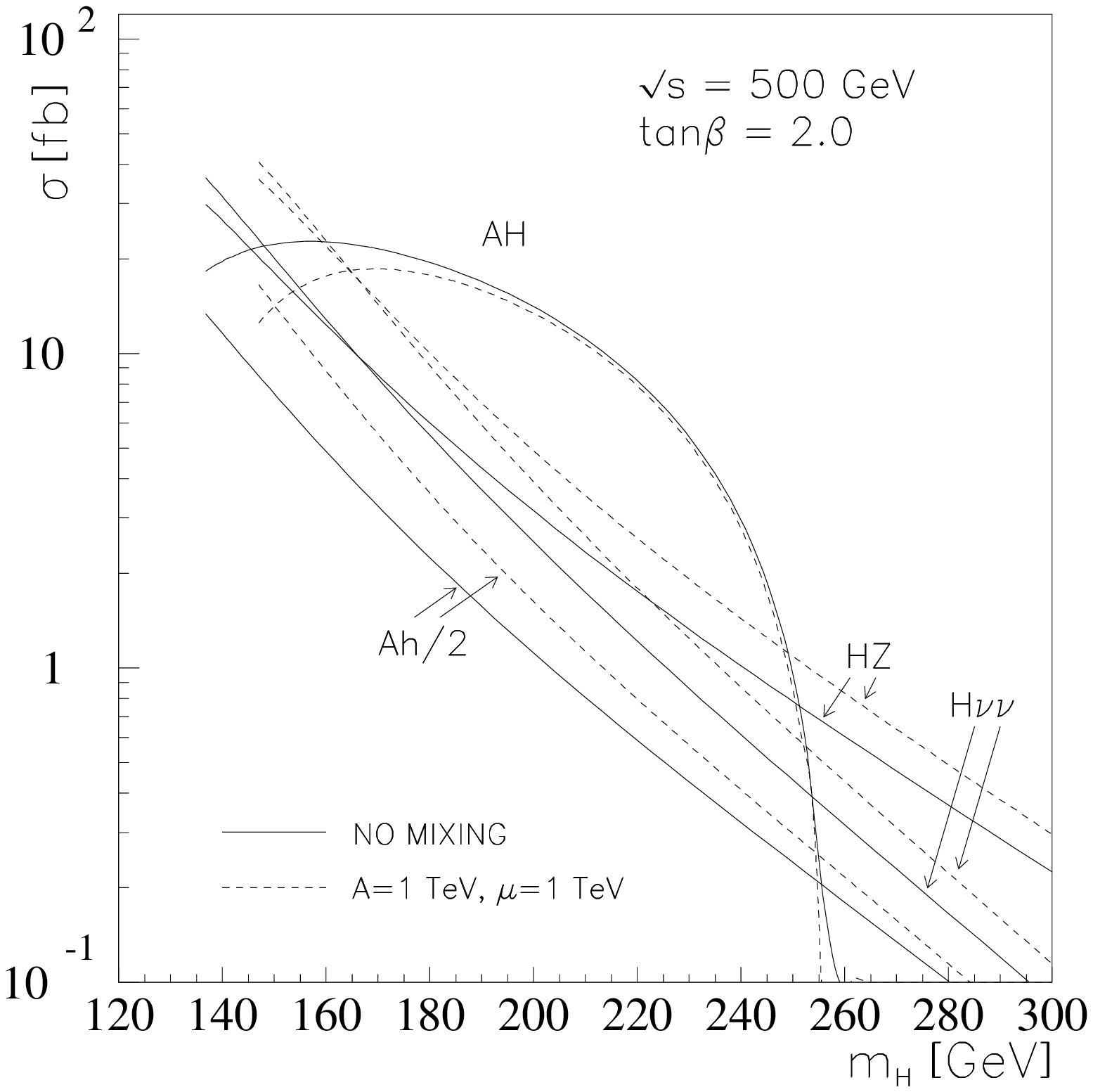}}
 \mbox{\epsfysize=9cm\epsffile{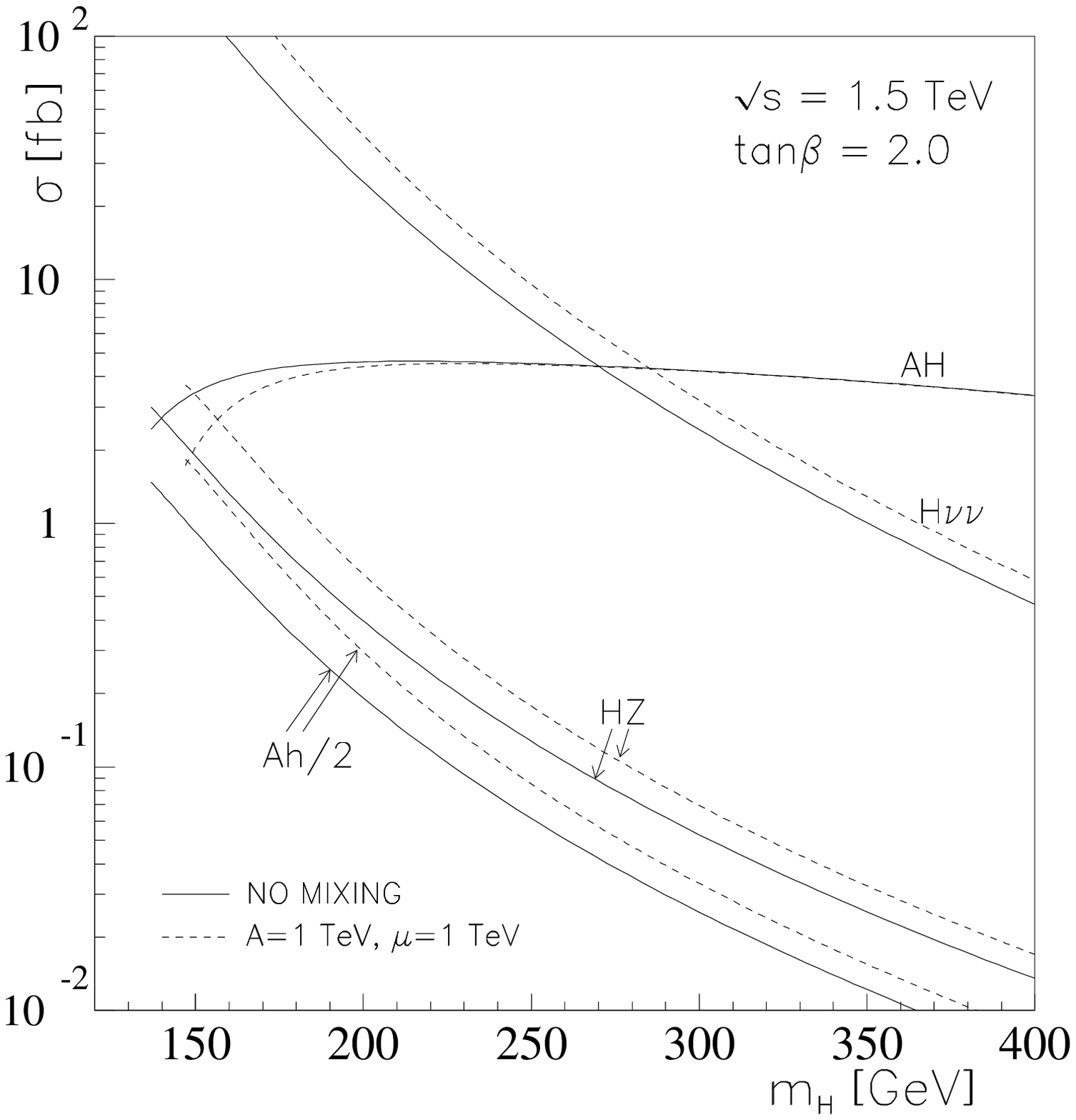}}}
\put(-1,-5)
{\mbox{\epsfysize=9cm\epsffile{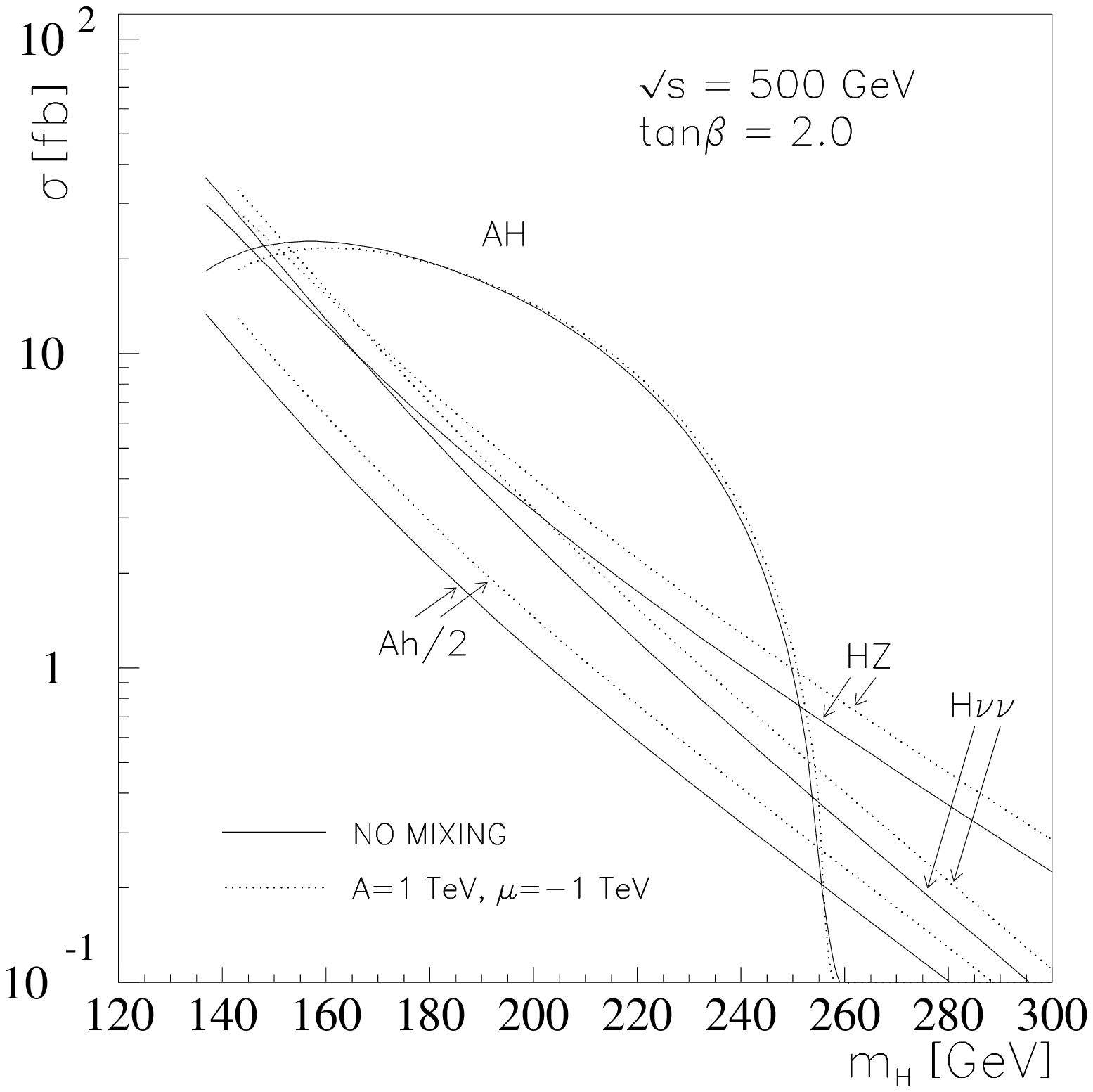}}
 \mbox{\epsfysize=9cm\epsffile{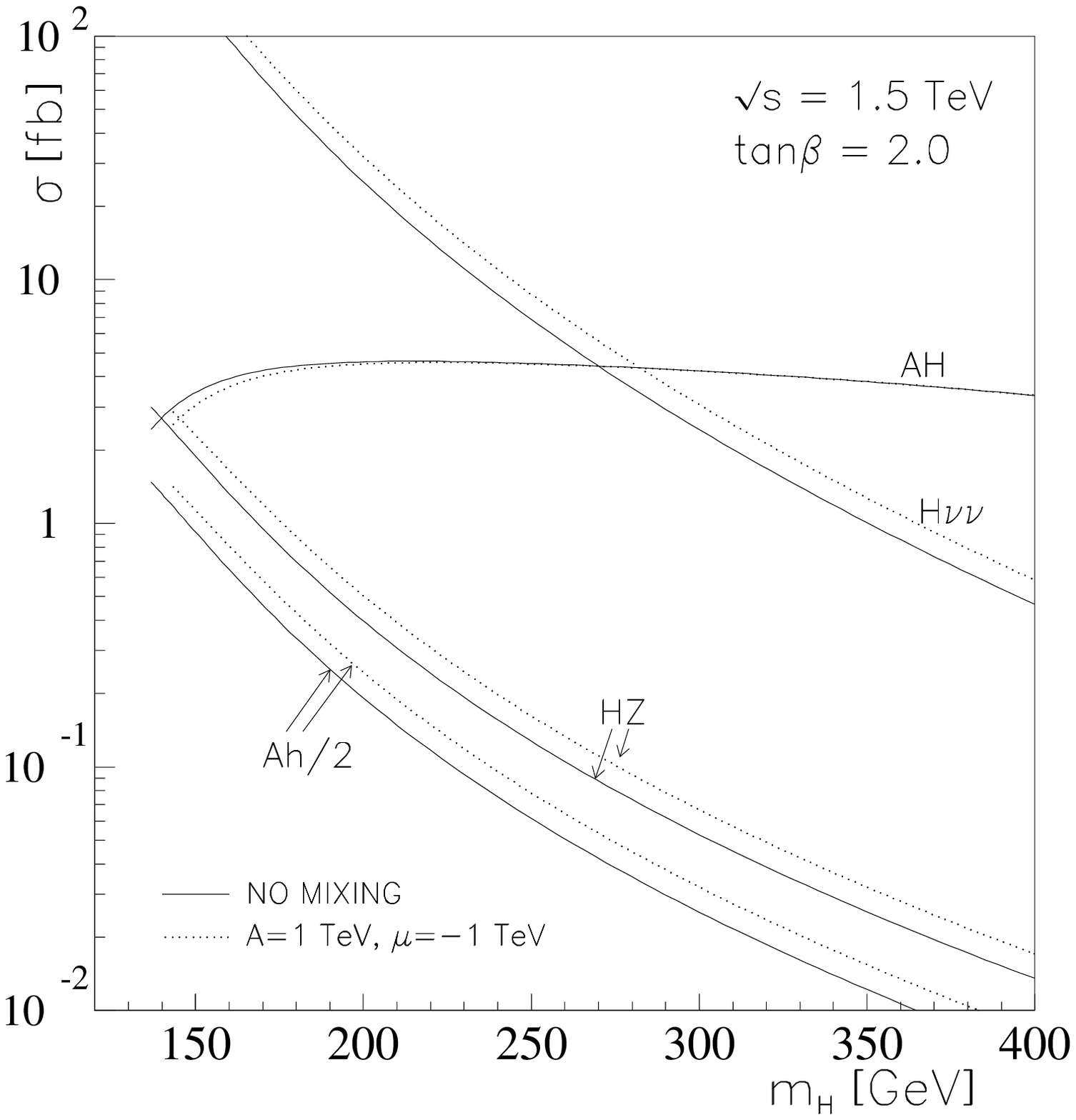}}}
\end{picture}
\vspace*{38mm}
\caption{Cross sections for the production of the heavy Higgs
boson $H$ in $e^+ e^-$ collisions. Also shown is the
cross section for the background process in which $Ah$ is produced
in the final state. We have taken $\sqrt s = 500$ GeV and 1.5 TeV.
Solid curves are for no mixing, $A=0$, $\mu=0$. 
Dashed and dotted curves refer to mixing:
$A=1.0$~TeV, $\mu=1.0$~TeV (dashed) and 
$A=1.0$~TeV, $\mu=-1.0$~TeV (dotted).}
\end{center}
\end{figure}
\clearpage
\begin{figure}[htb]
\refstepcounter{figure}
\label{Fig:BR-H-A-2}
\addtocounter{figure}{-1}
\begin{center}
\setlength{\unitlength}{1cm}
\begin{picture}(16,9)
\put(-1,4)
{\mbox{\epsfysize=9cm\epsffile{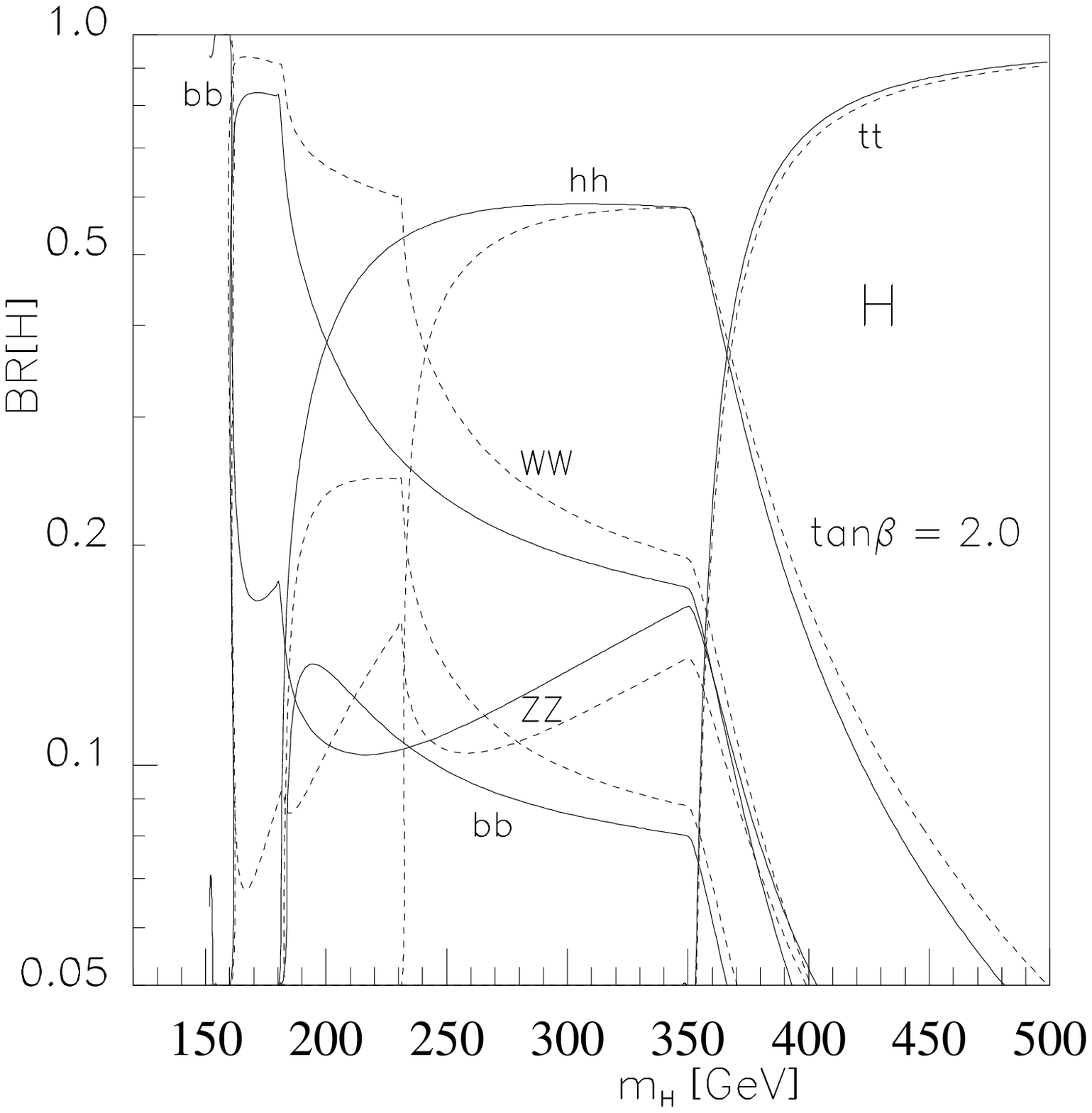}}
 \mbox{\epsfysize=9cm\epsffile{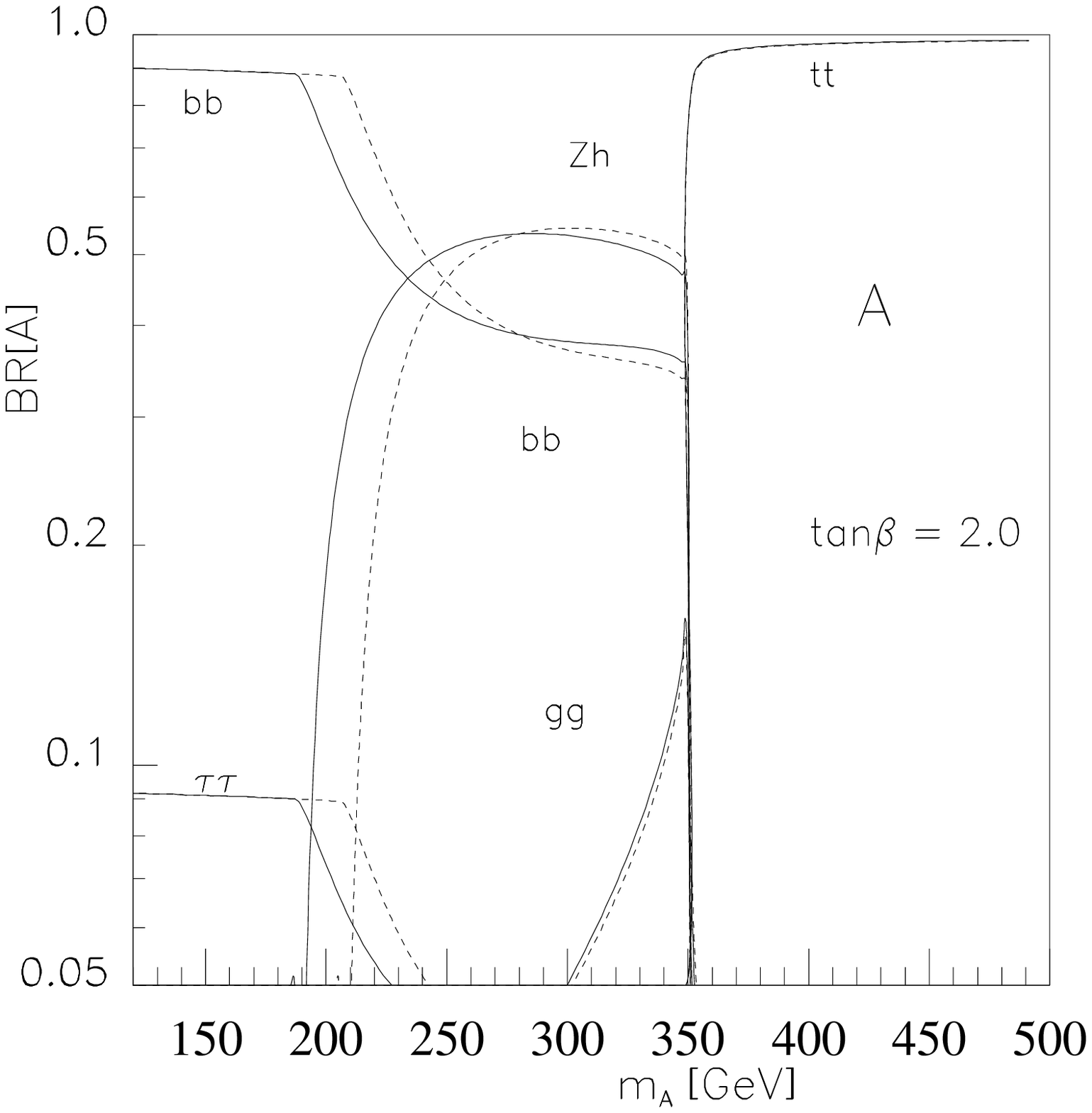}}}
\put(-1,-5)
{\mbox{\epsfysize=9cm\epsffile{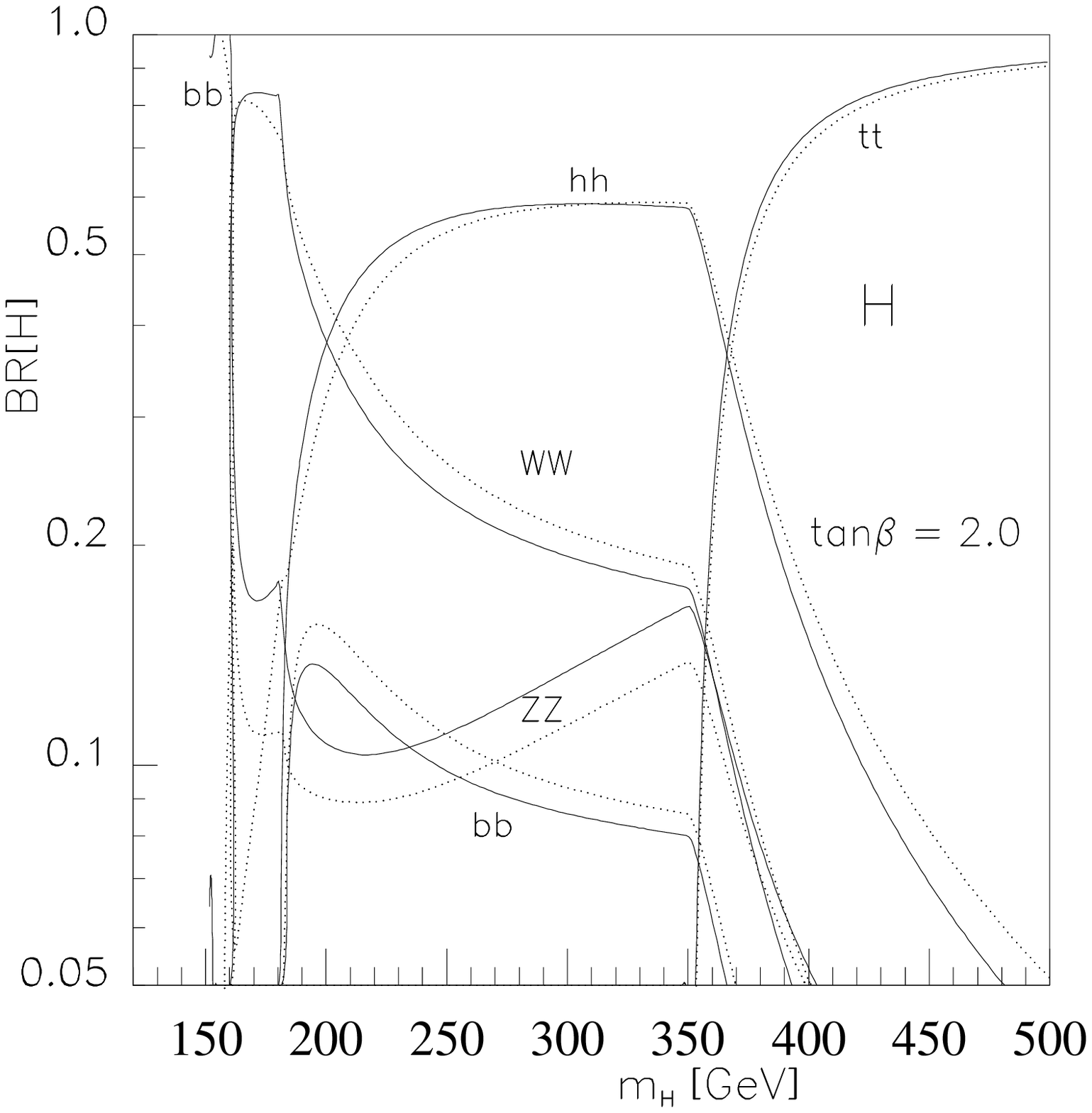}}
 \mbox{\epsfysize=9cm\epsffile{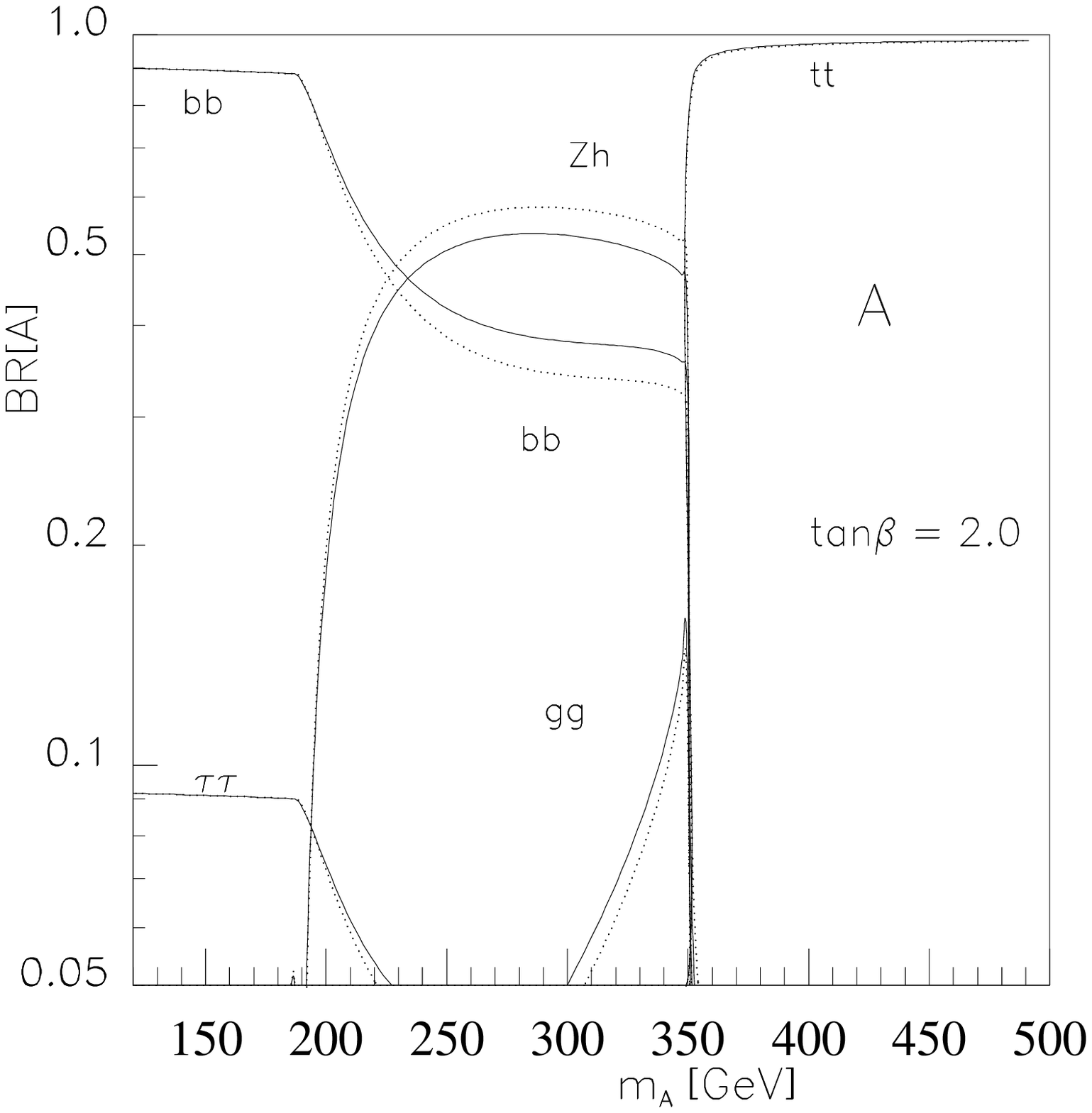}}}
\end{picture}
\vspace*{38mm}
\caption{Branching ratios for the decay modes of the $CP$-even
heavy Higgs boson $H$, and the $CP$-odd Higgs boson $A$ 
for $\tan\beta = 2.0$.
Solid curves are for no mixing, $A=0$, $\mu=0$. 
Dashed and dotted curves refer to mixing:
$A=1.0$~TeV, $\mu=1.0$~TeV (dashed) and 
$A=1.0$~TeV, $\mu=-1.0$~TeV (dotted).}
\end{center}
\end{figure}

\begin{figure}[htb]
\refstepcounter{figure}
\label{Fig:BR-H-A-5}
\addtocounter{figure}{-1}
\begin{center}
\setlength{\unitlength}{1cm}
\begin{picture}(16,9)
\put(-1,4)
{\mbox{\epsfysize=9cm\epsffile{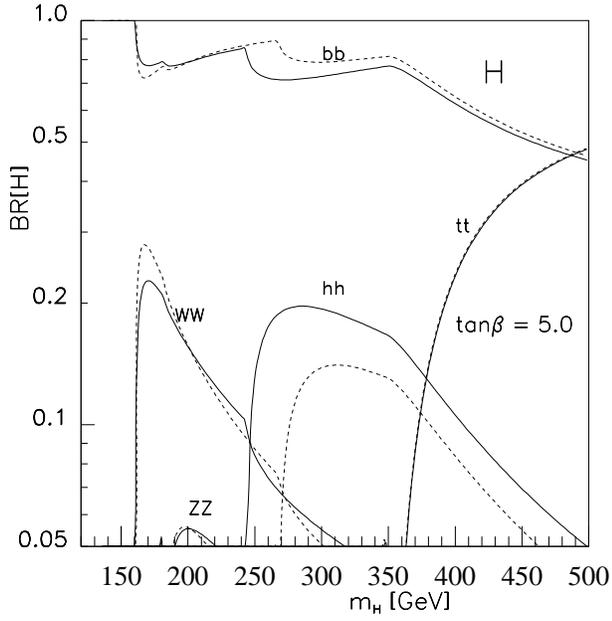}}
 \mbox{\epsfysize=9cm\epsffile{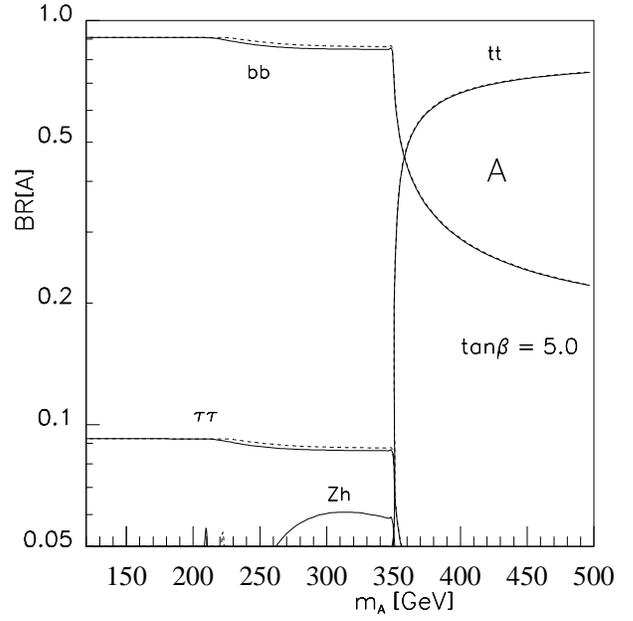}}}
\put(-1,-5)
{\mbox{\epsfysize=9cm\epsffile{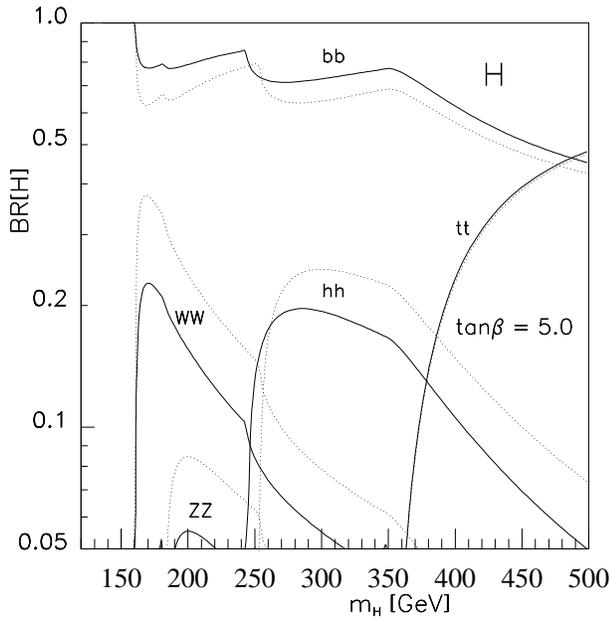}}
 \mbox{\epsfysize=9cm\epsffile{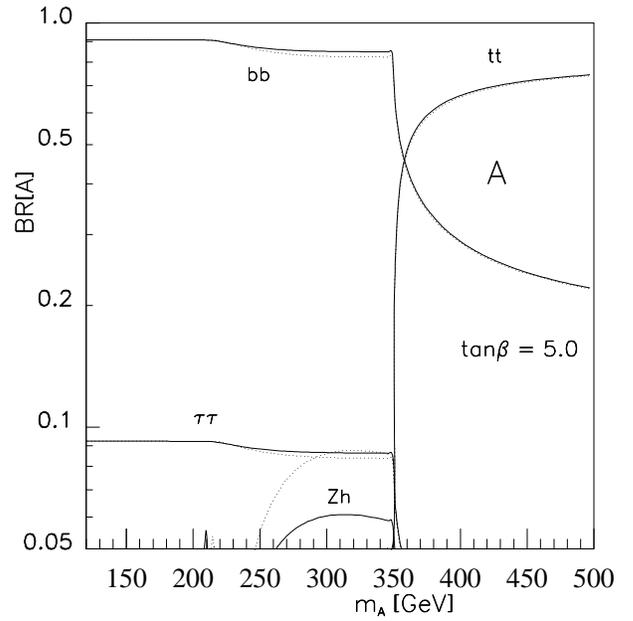}}}
\end{picture}
\vspace*{38mm}
\caption{As in Fig.~\ref{Fig:BR-H-A-2}, but with $\tan\beta = 5.0$. 
Solid curves are for no mixing, $A=0$, $\mu=0$. 
Dashed and dotted curves refer to mixing:
$A=1.0$~TeV, $\mu=1.0$~TeV (dashed) and 
$A=1.0$~TeV, $\mu=-1.0$~TeV (dotted).}
\end{center}
\end{figure}

\begin{figure}[htb]
\refstepcounter{figure}
\label{Fig:hole}
\addtocounter{figure}{-1}
\begin{center}
\setlength{\unitlength}{1cm}
\begin{picture}(16,9)
\put(-1,4)
{\mbox{\epsfysize=9cm\epsffile{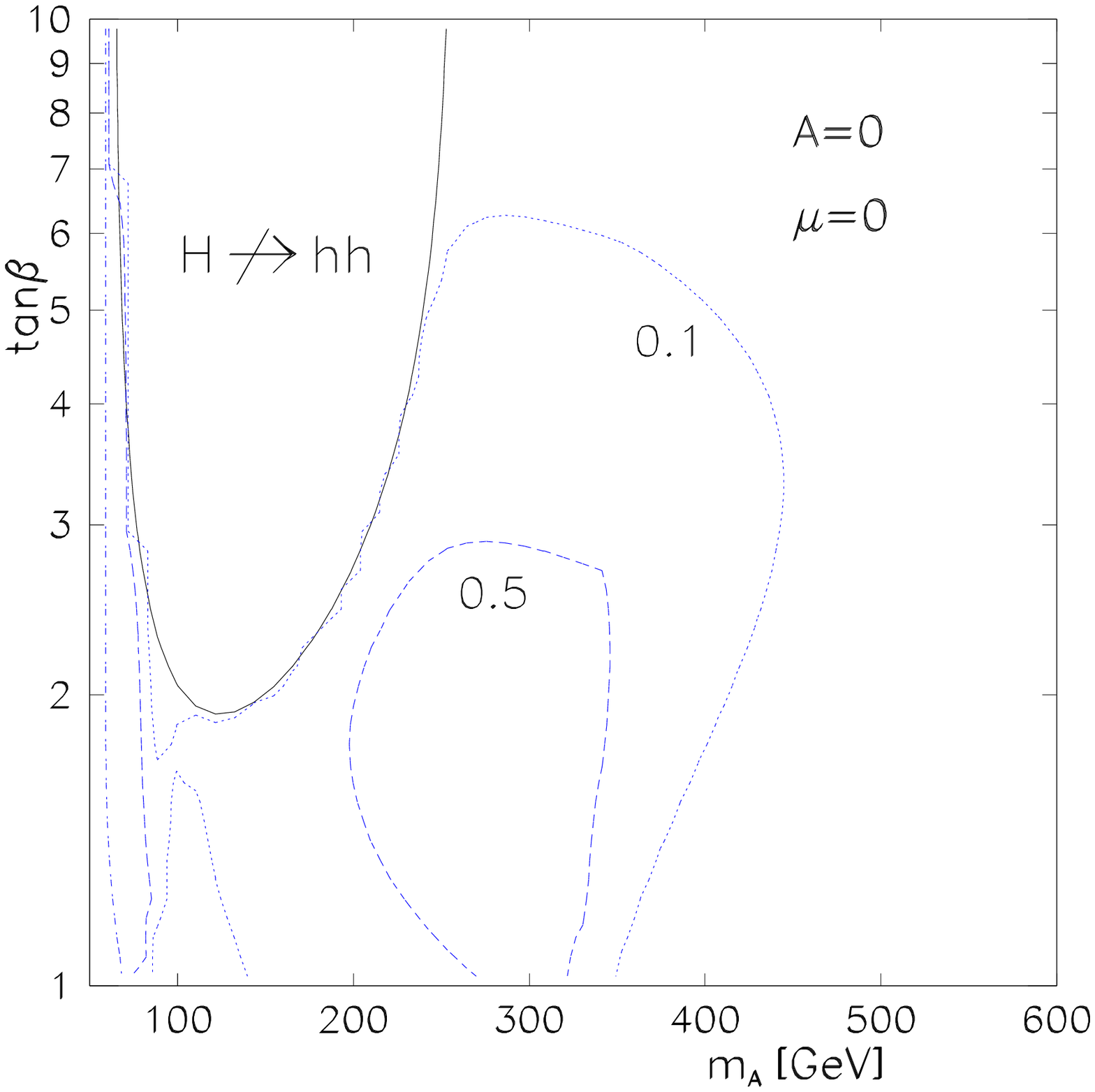}}
 \mbox{\epsfysize=9cm\epsffile{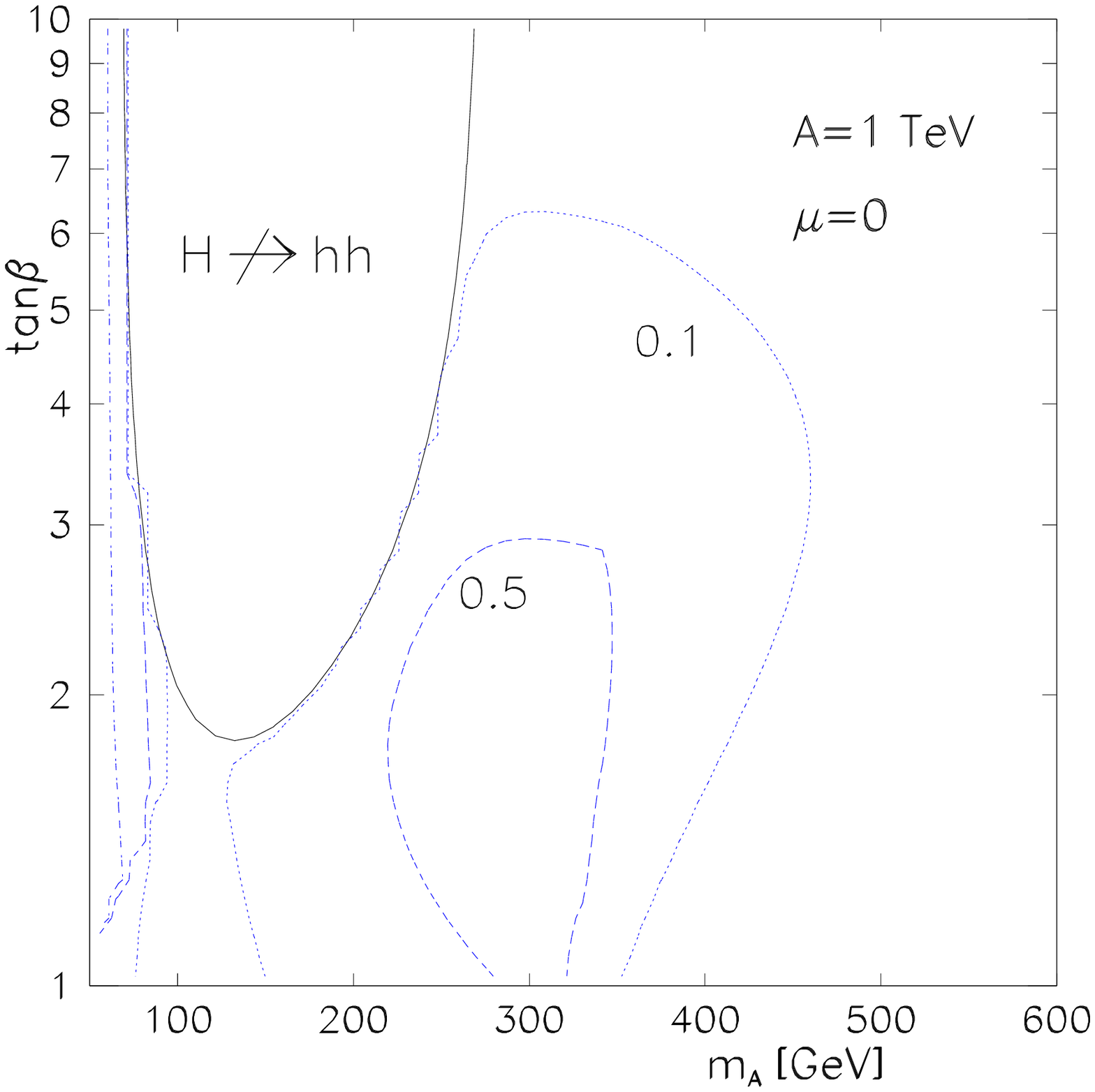}}}
\put(-1,-5)
{\mbox{\epsfysize=9cm\epsffile{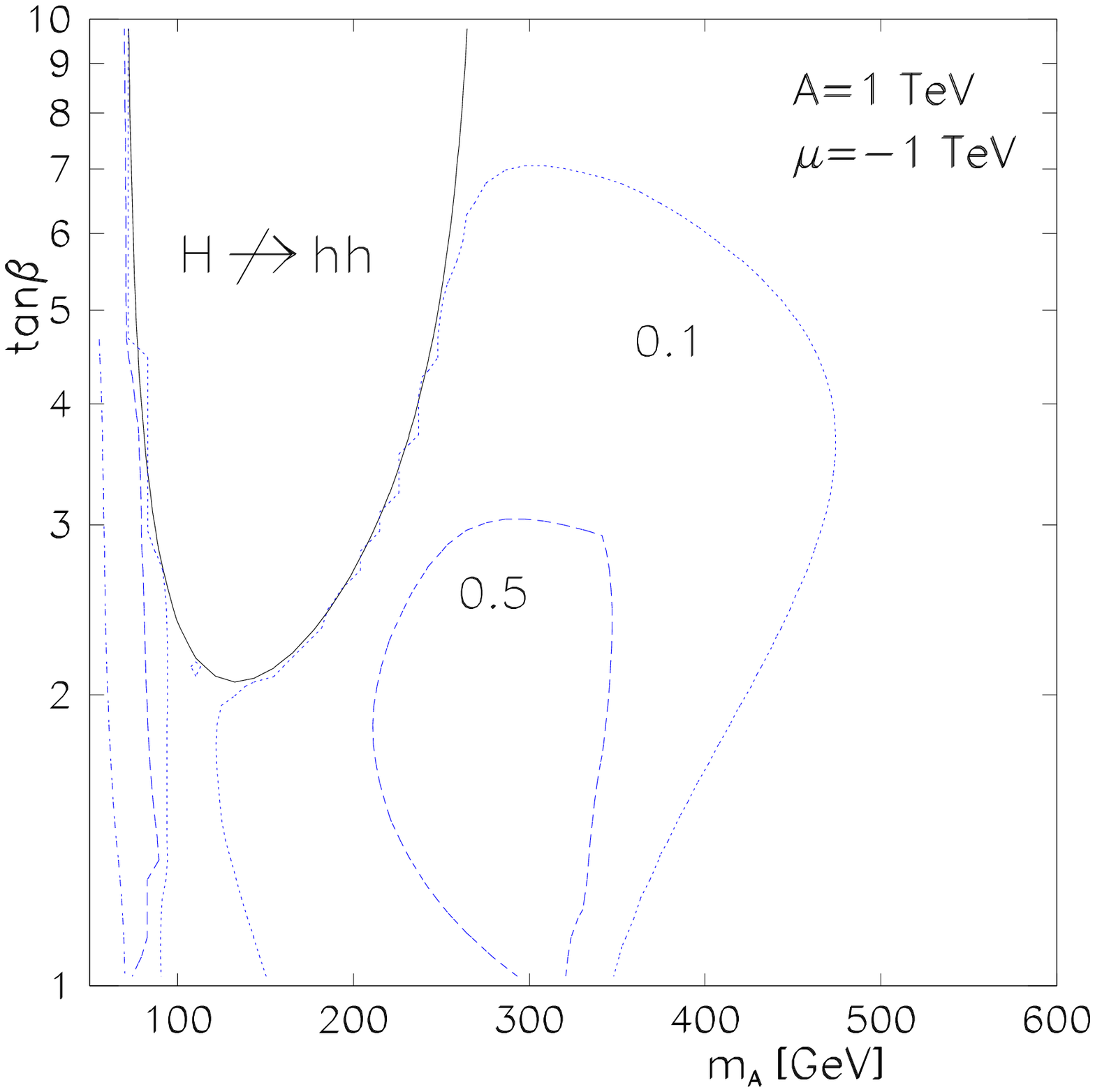}}
 \mbox{\epsfysize=9cm\epsffile{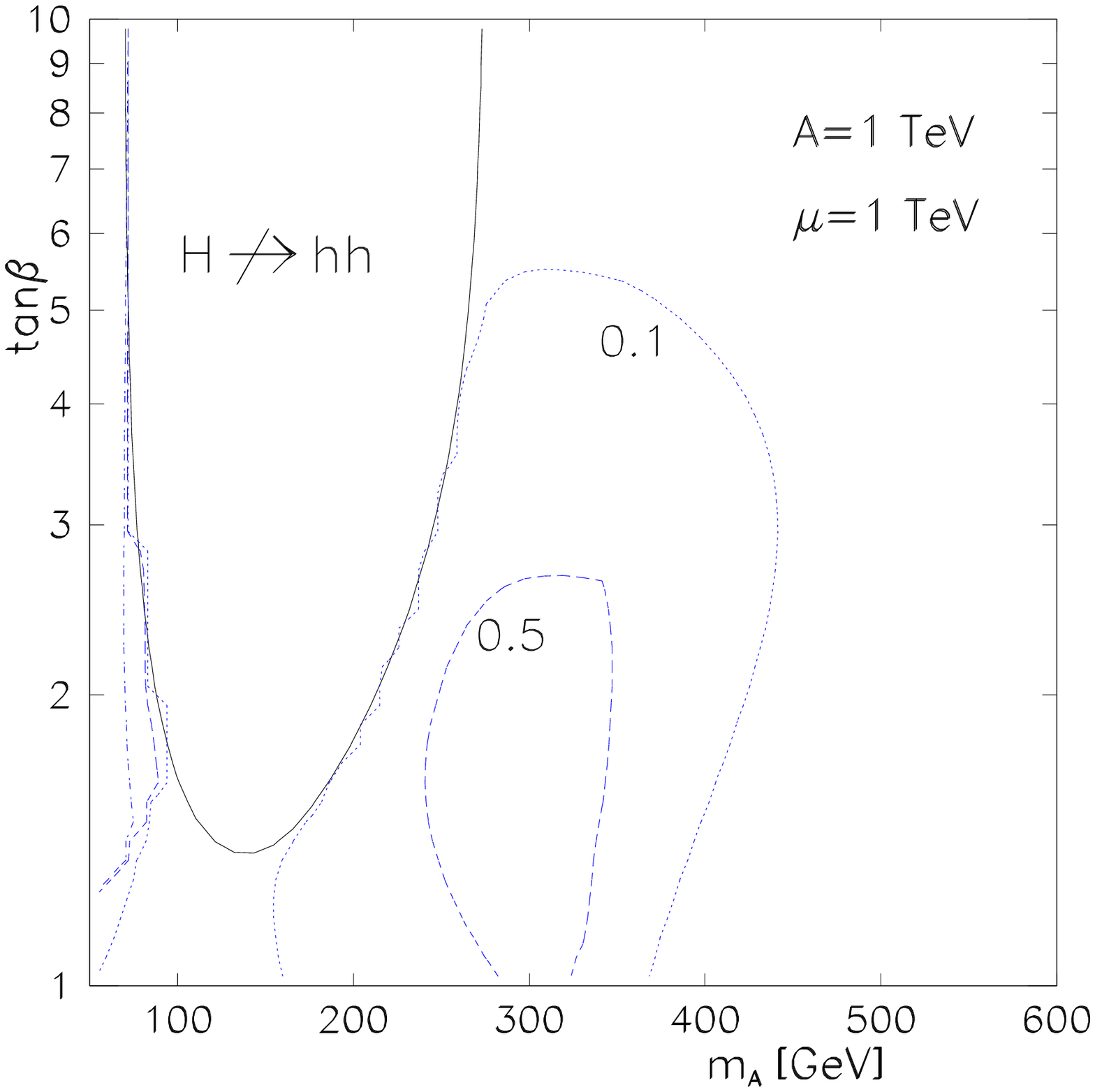}}}
\end{picture}
\vspace*{38mm}
\caption{The region in the $m_A$--$\tan\beta$ plane where the decay
$H\to hh$ is kinematically {\em forbidden} is indicated by a solid 
line contour.
Also given are contours at which the branching ratio equals 0.1 (dotted),
0.5 (dashed) and 0.9 (dash-dotted, at the far left).
Four cases of mixing parameters $A$ and $\mu$ are considered, 
as indicated.}
\end{center}
\end{figure}

\begin{figure}[htb]
\refstepcounter{figure}
\label{Fig:sig-Zll-2}
\addtocounter{figure}{-1}
\begin{center}
\setlength{\unitlength}{1cm}
\begin{picture}(16,9)
\put(-1,4)
{\mbox{\epsfysize=9cm\epsffile{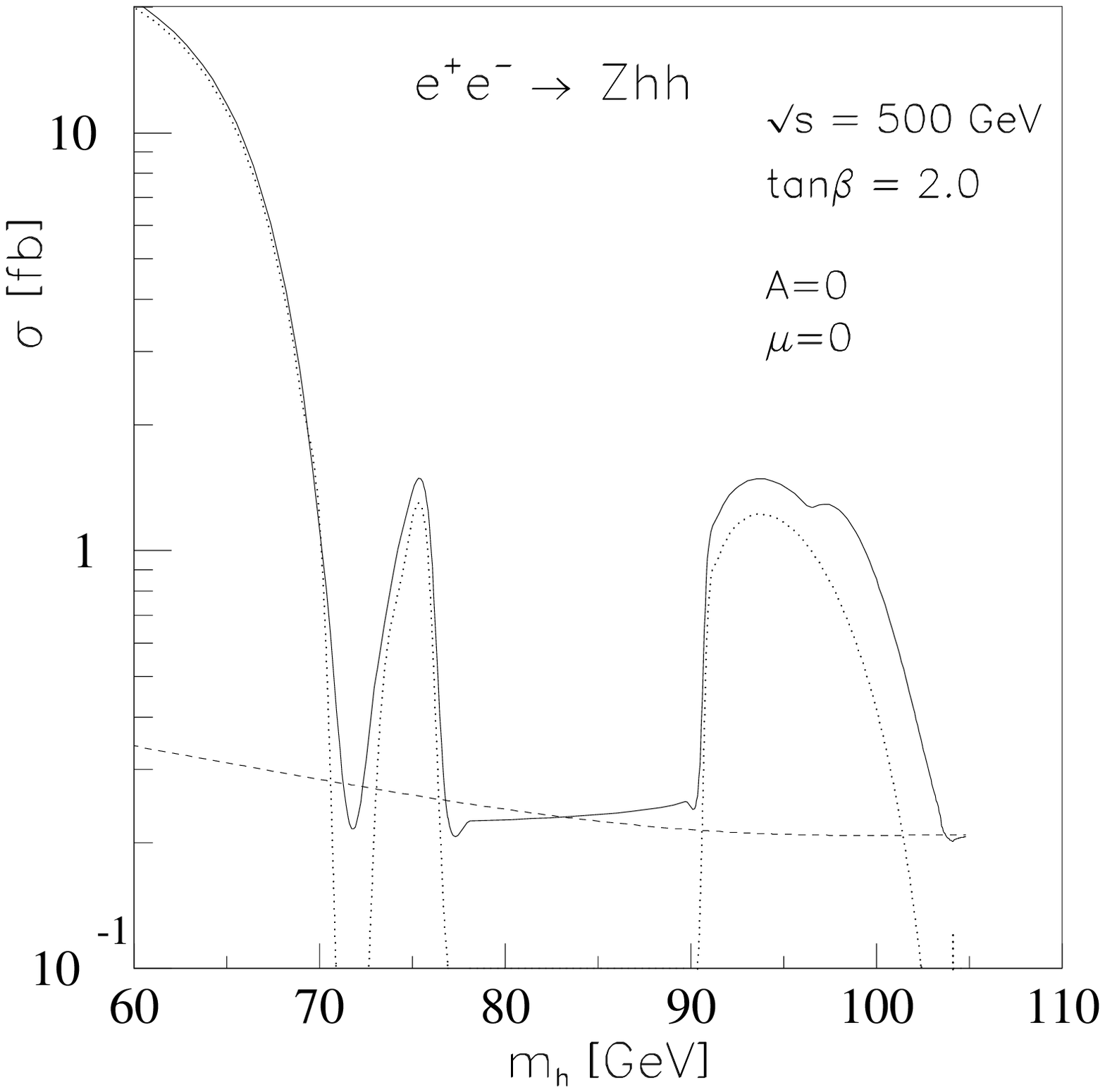}}
 \mbox{\epsfysize=9cm\epsffile{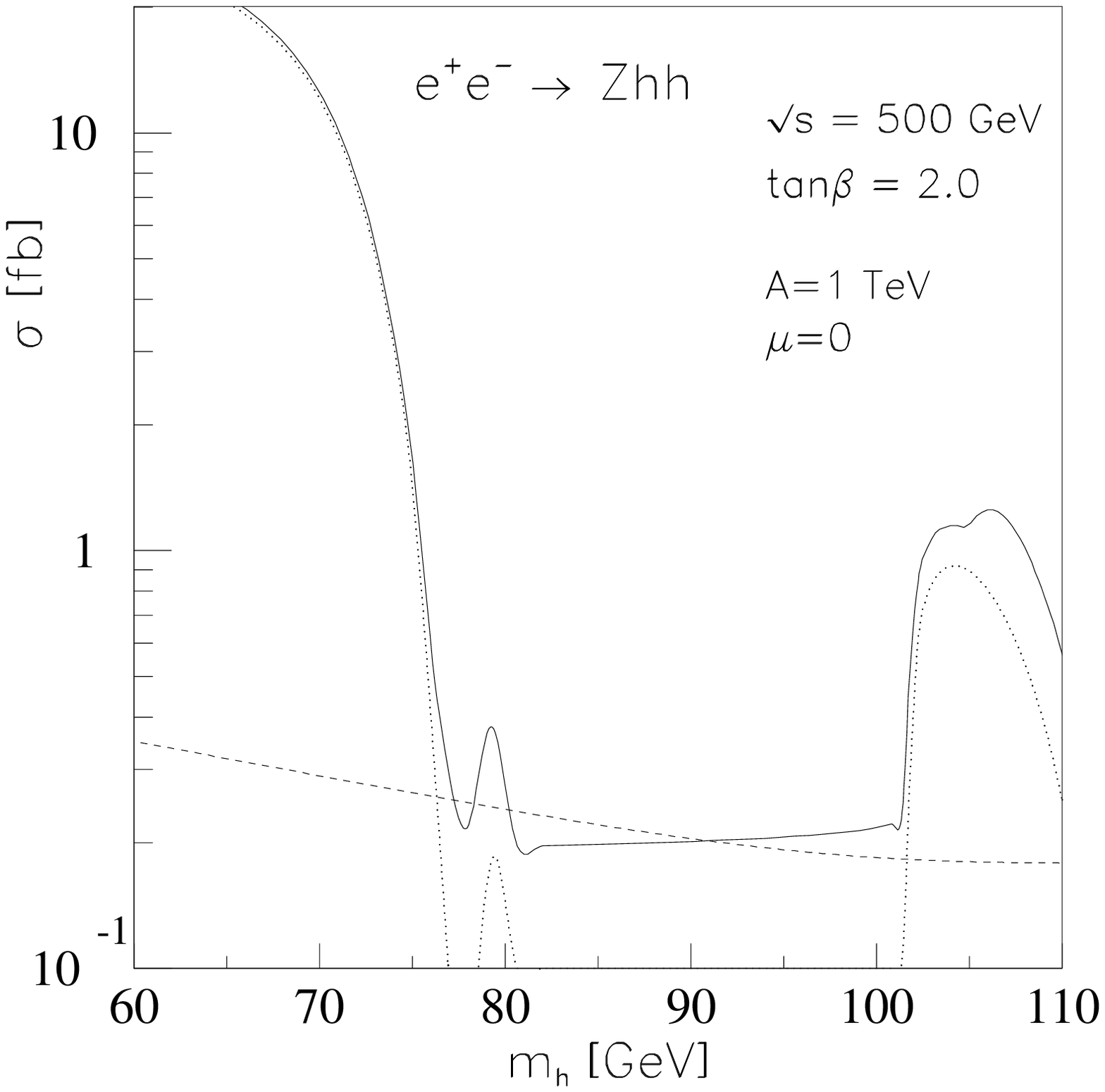}}}
\put(-1,-5)
{\mbox{\epsfysize=9cm\epsffile{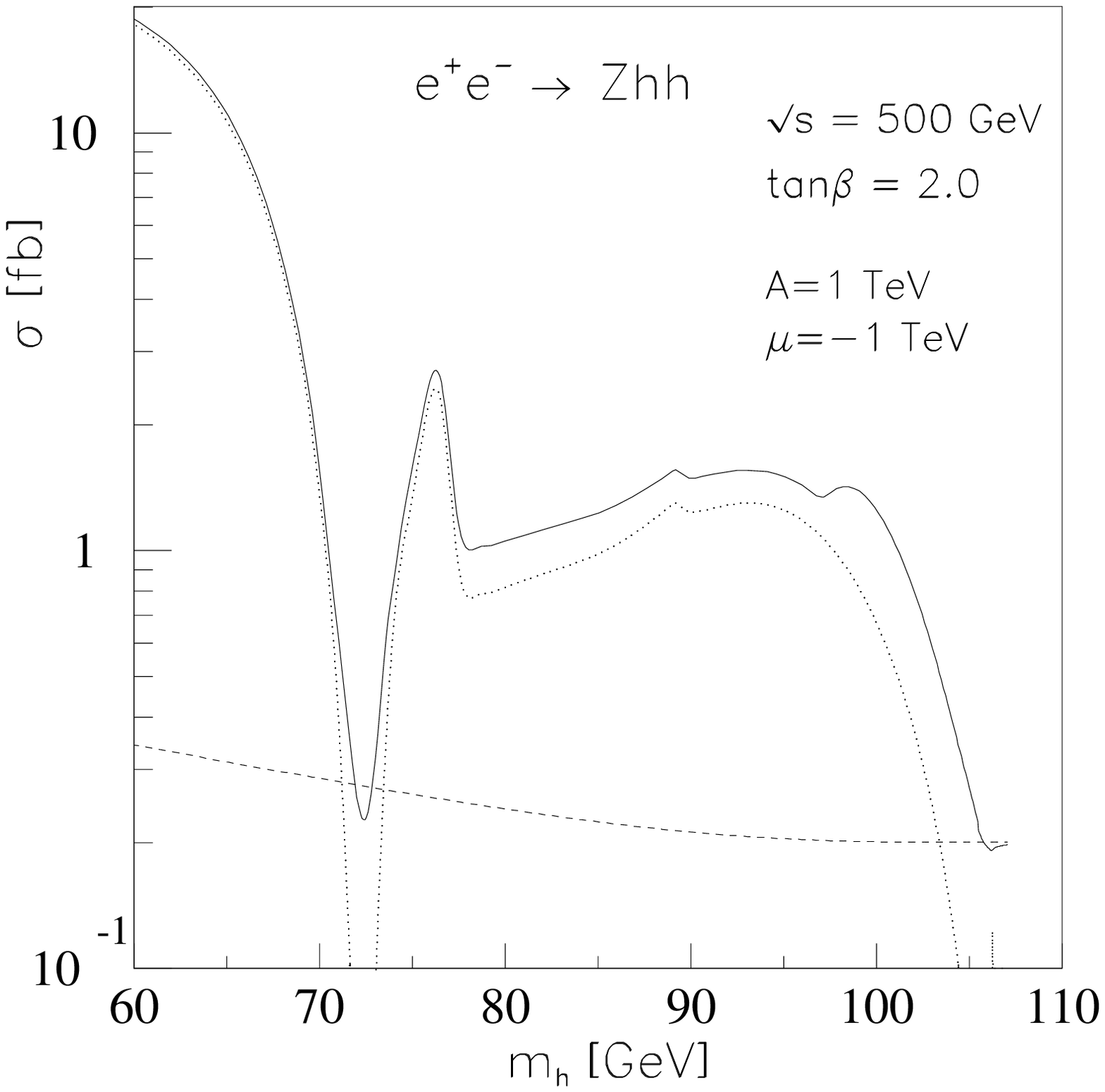}}
 \mbox{\epsfysize=9cm\epsffile{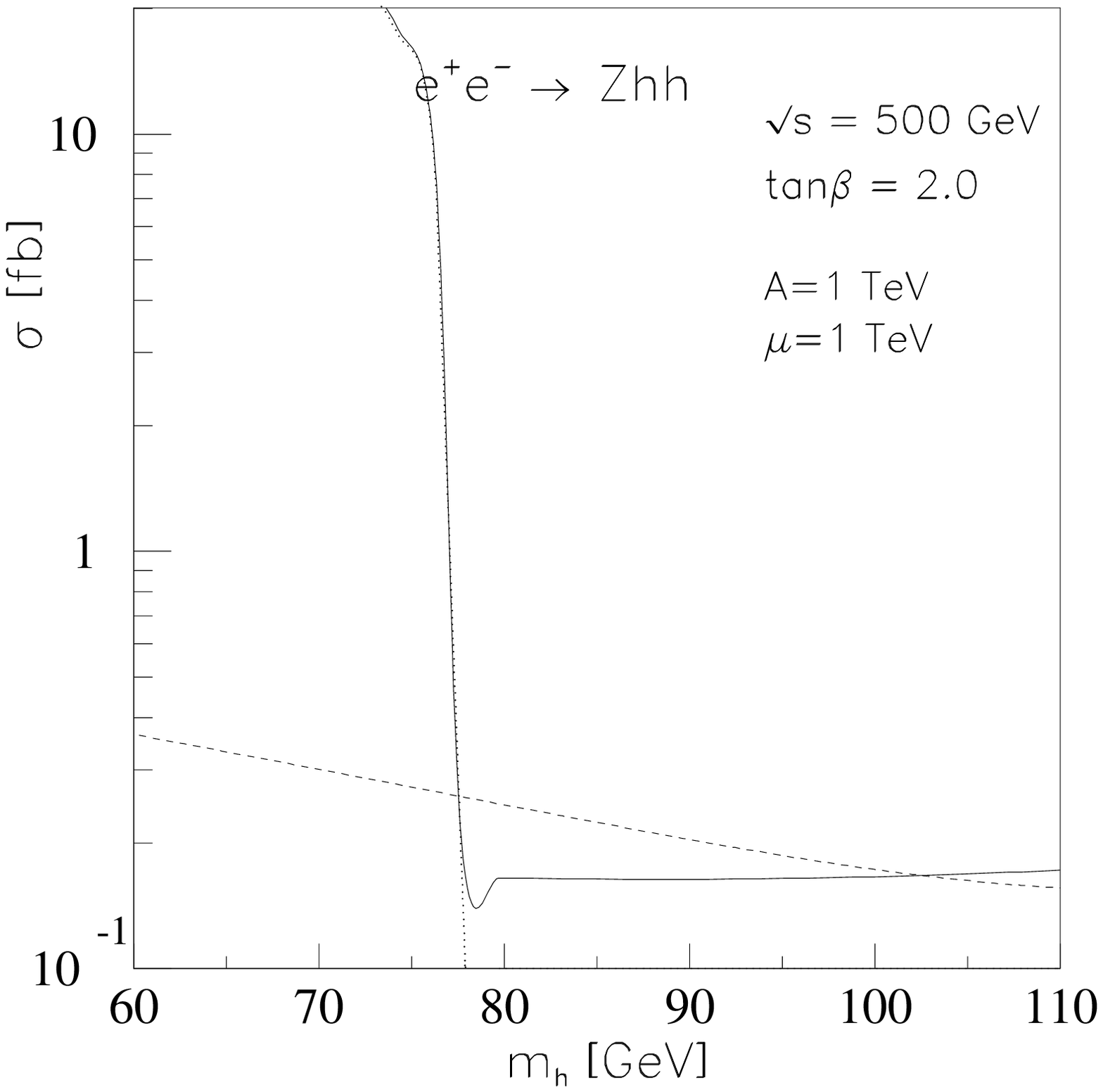}}}
\end{picture}
\vspace*{38mm}
\caption{Cross section $\sigma(e^+e^-\to Zhh)$ as a function of $m_h$,
for four cases: (a) no mixing, 
(b)--(d) $A=1~\TeV$, $\mu=0$, $-1$ and 1~TeV, as indicated.
The dotted curve is the resonant production.
The dashed curve gives the decoupling limit.
}
\end{center}
\end{figure}

\begin{figure}[htb]
\refstepcounter{figure}
\label{Fig:sig-WW-2}
\addtocounter{figure}{-1}
\begin{center}
\setlength{\unitlength}{1cm}
\begin{picture}(16,9)
\put(-1,4)
{\mbox{\epsfysize=9cm\epsffile{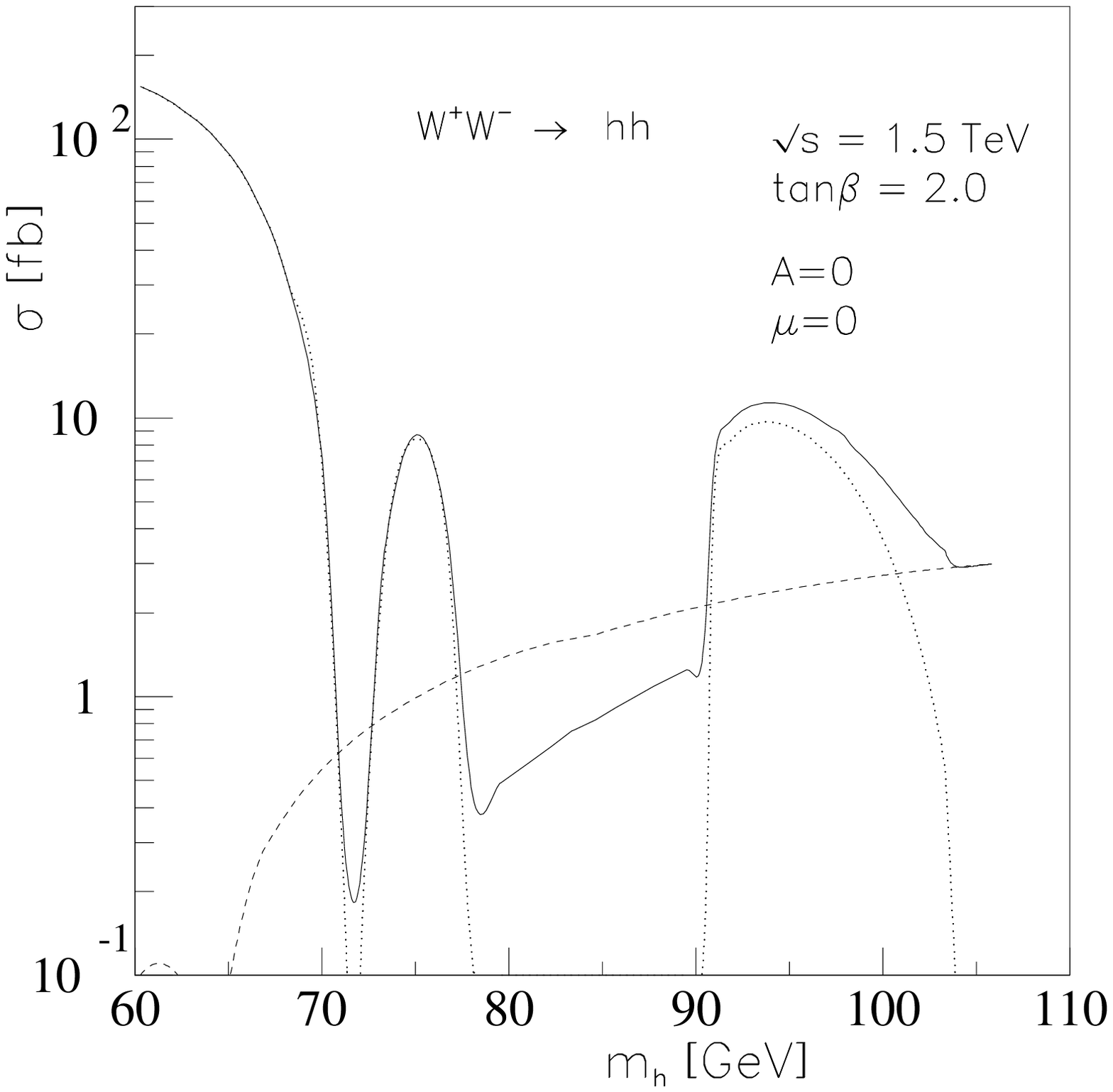}}
 \mbox{\epsfysize=9cm\epsffile{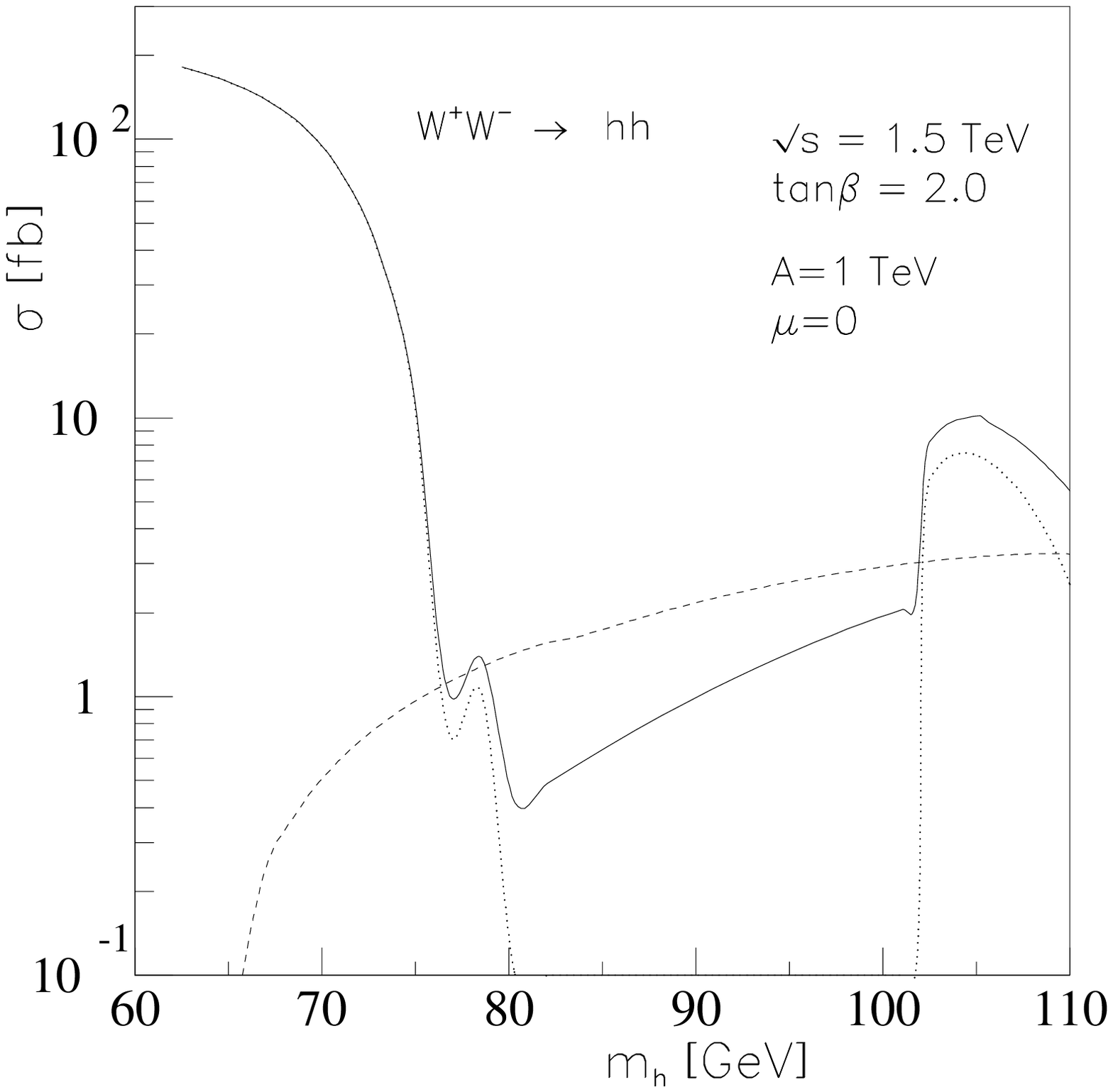}}}
\put(-1,-5)
{\mbox{\epsfysize=9cm\epsffile{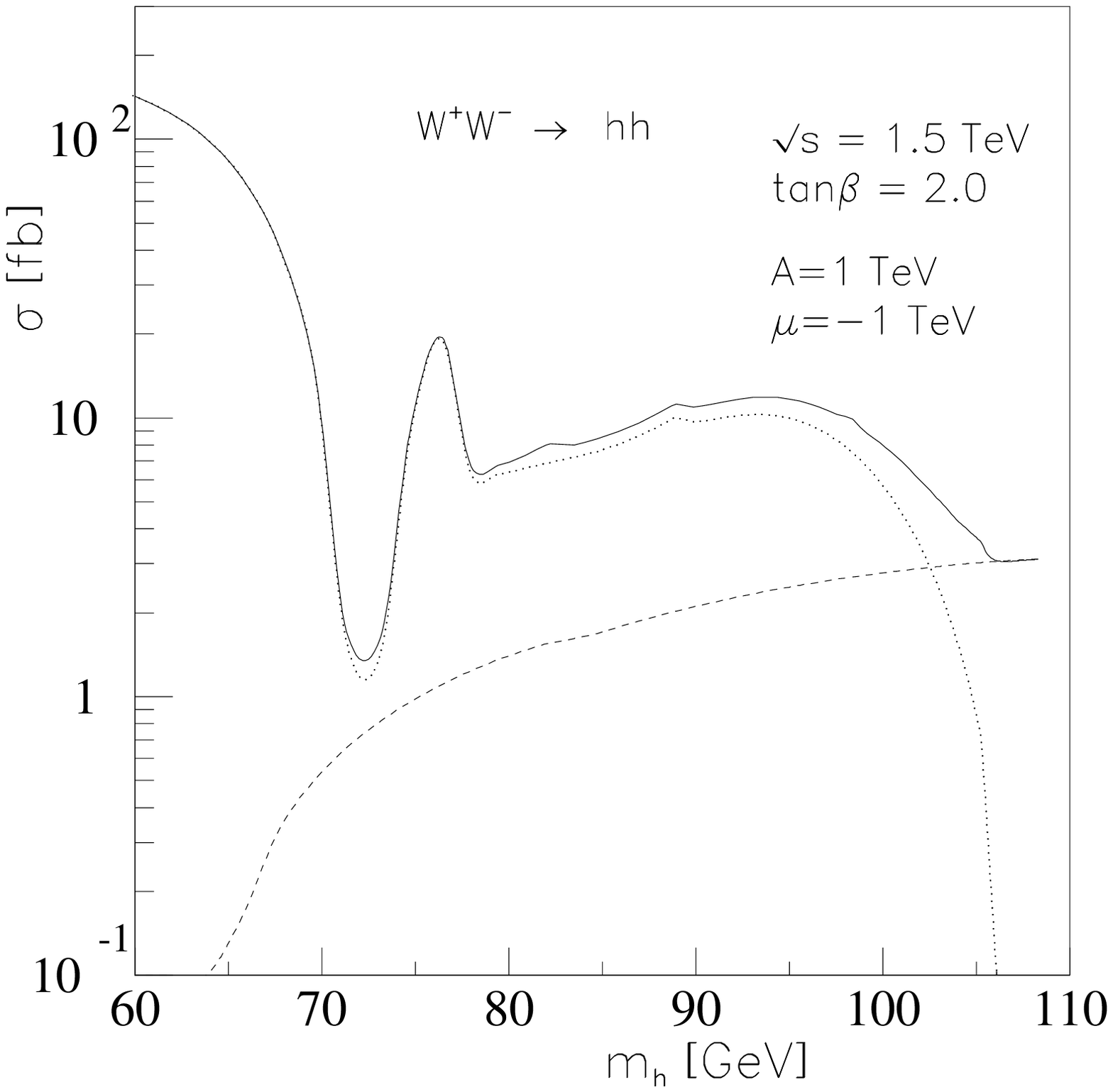}}
 \mbox{\epsfysize=9cm\epsffile{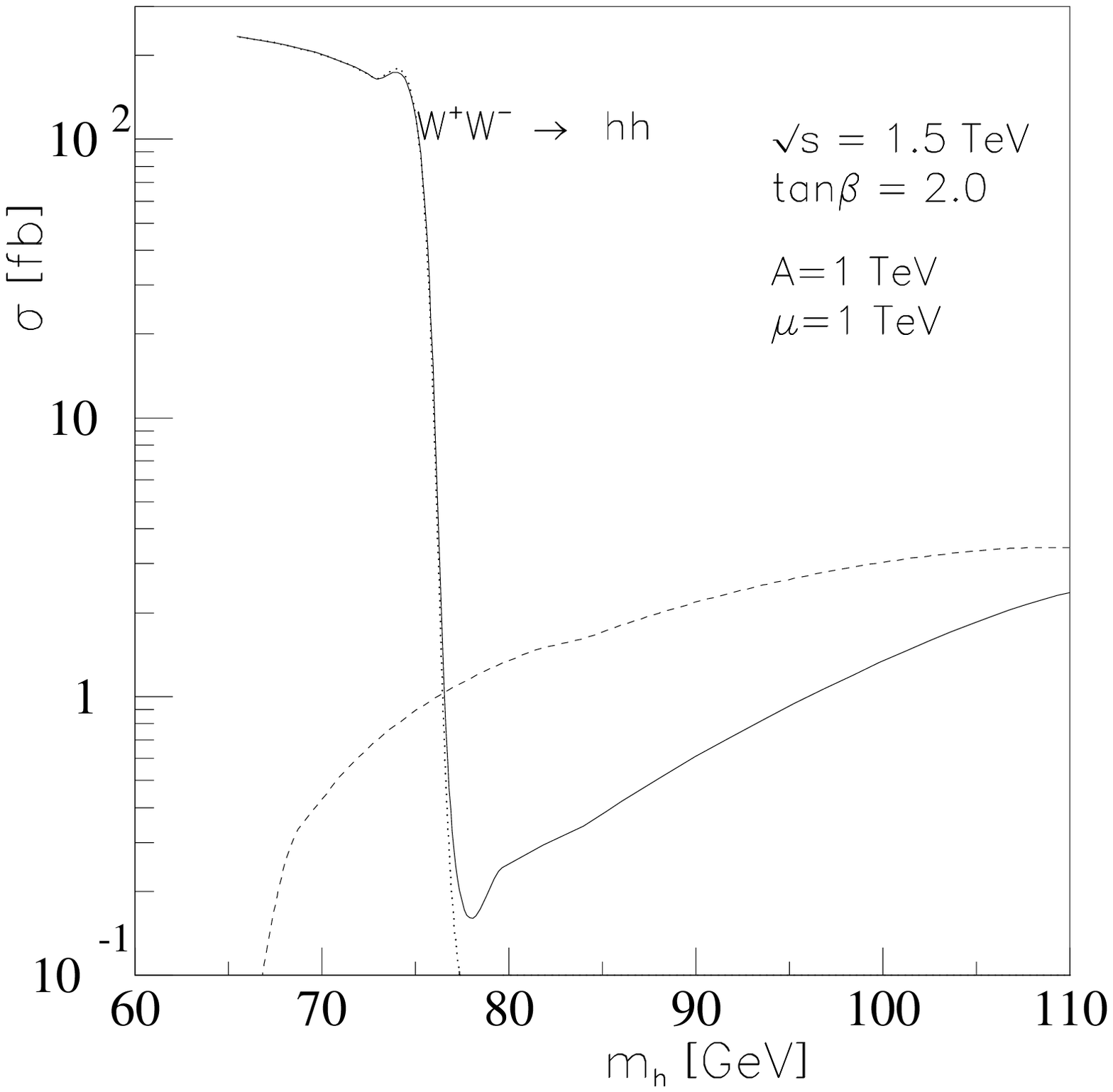}}}
\end{picture}
\vspace*{38mm}
\caption{Cross section $\sigma(e^+e^-\to \nu_e\bar\nu_e hh)$ 
(via $WW$ fusion) as a function of $m_h$,
for four cases: (a) no mixing, 
(b)--(d) $A=1~\TeV$, $\mu=0$, $-1$ and 1~TeV, as indicated.
The dotted curve is the resonant production. 
The dashed curve gives the decoupling limit.
}
\end{center}
\end{figure}

\begin{figure}[htb]
\refstepcounter{figure}
\label{Fig:sensi-500}
\addtocounter{figure}{-1}
\begin{center}
\setlength{\unitlength}{1cm}
\begin{picture}(16,9)
\put(-1,3)
{\mbox{\epsfysize=9cm\epsffile{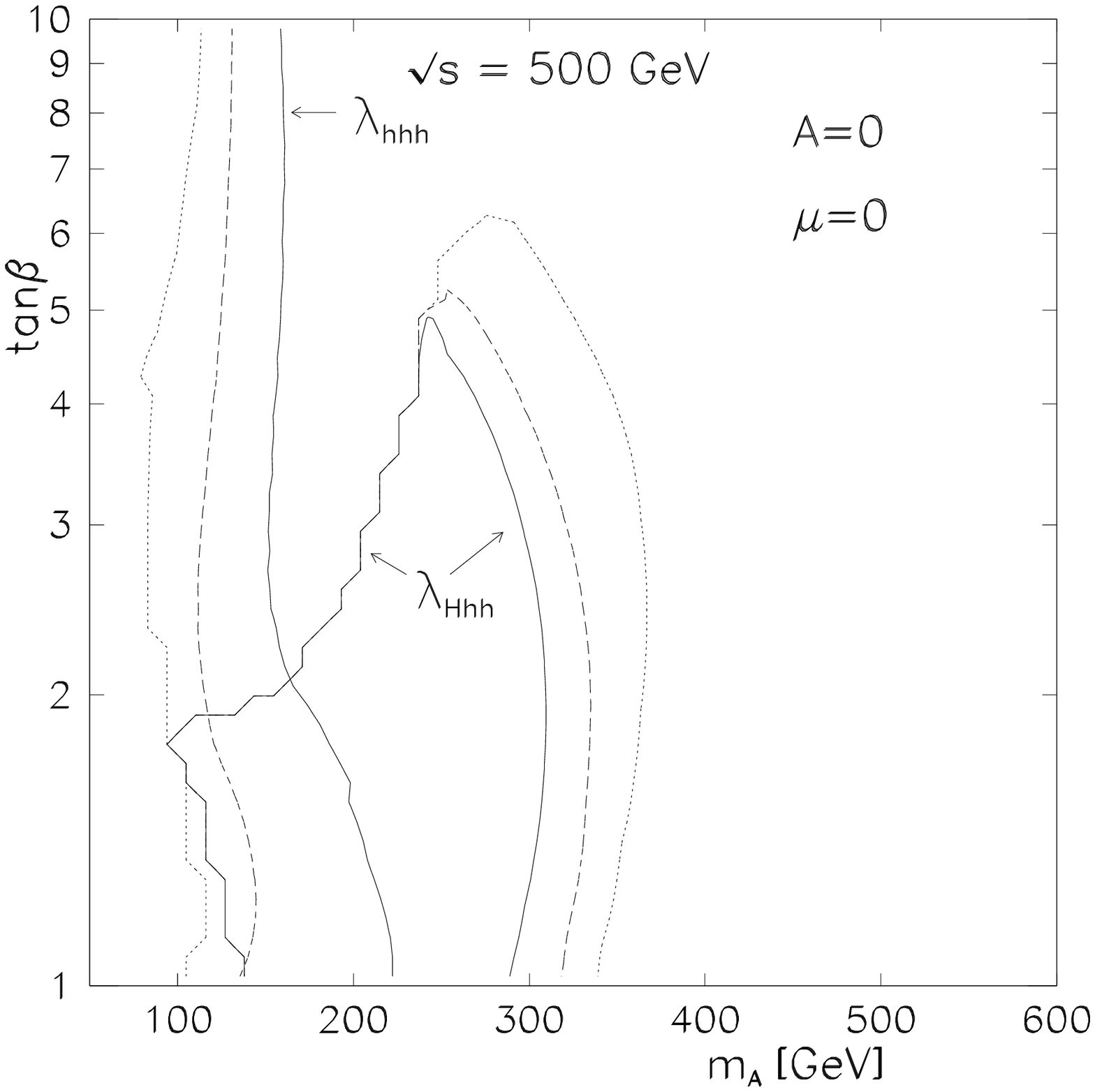}}
 \mbox{\epsfysize=9cm\epsffile{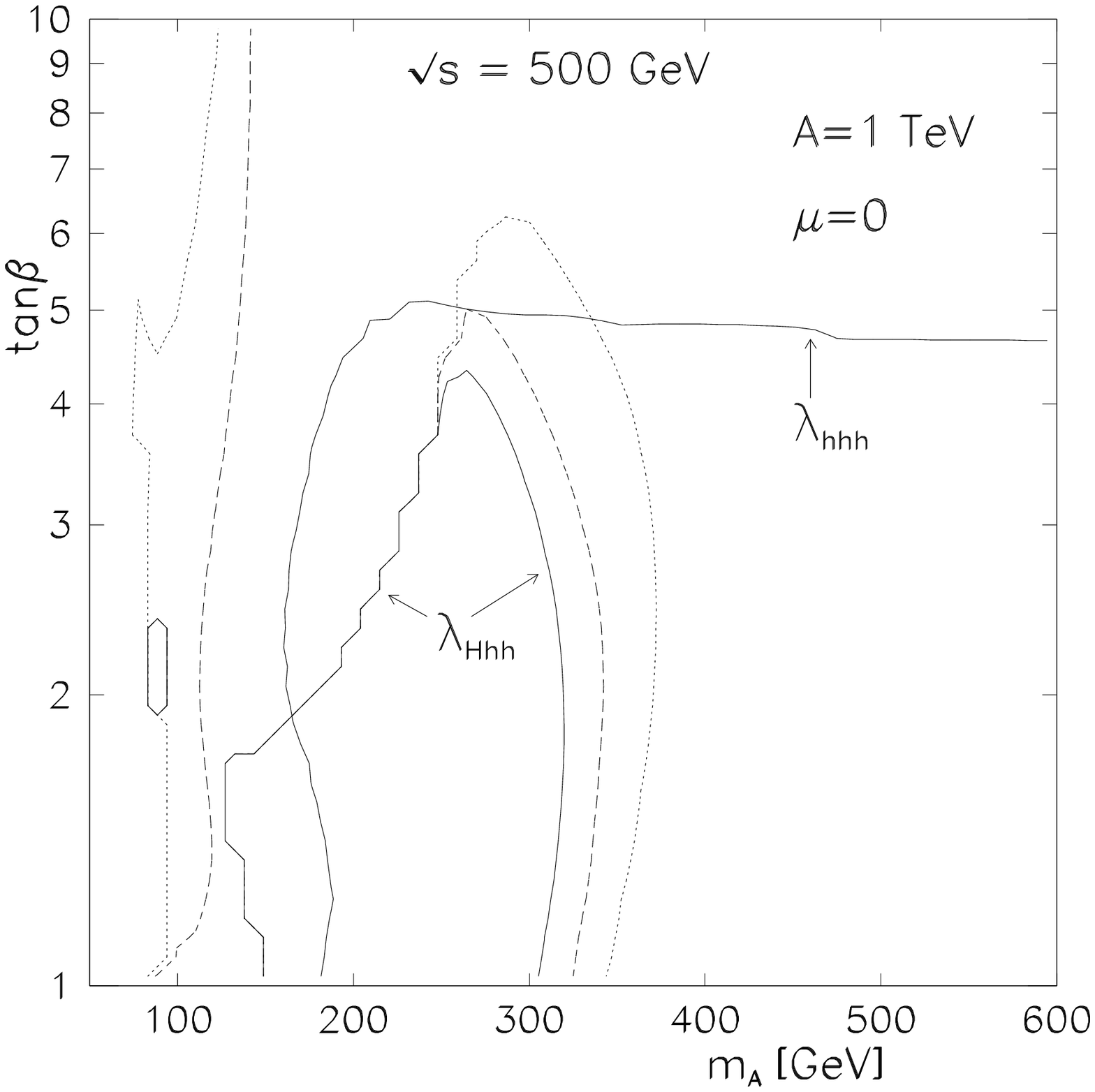}}}
\put(-1,-6)
{\mbox{\epsfysize=9cm\epsffile{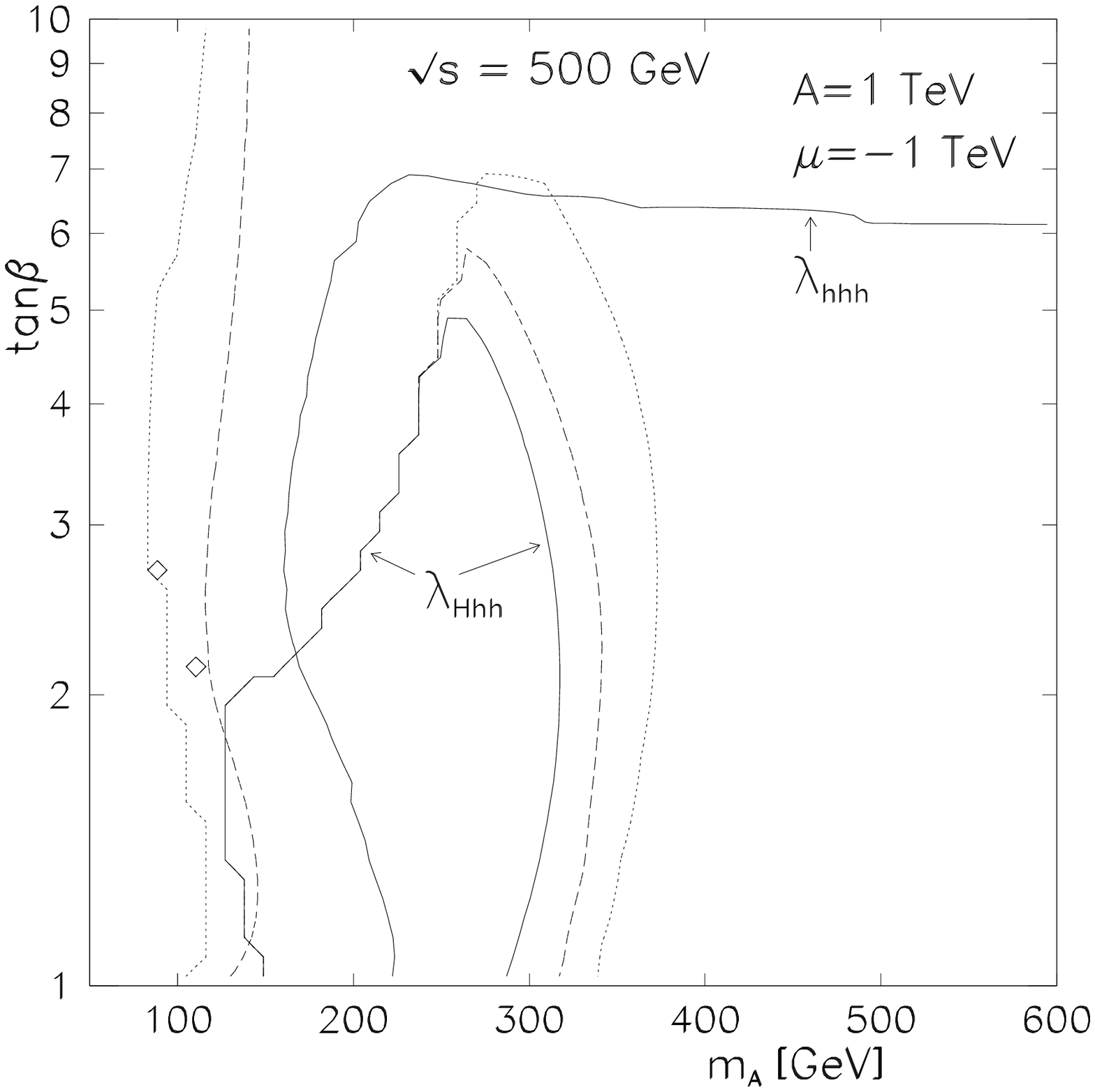}}
 \mbox{\epsfysize=9cm\epsffile{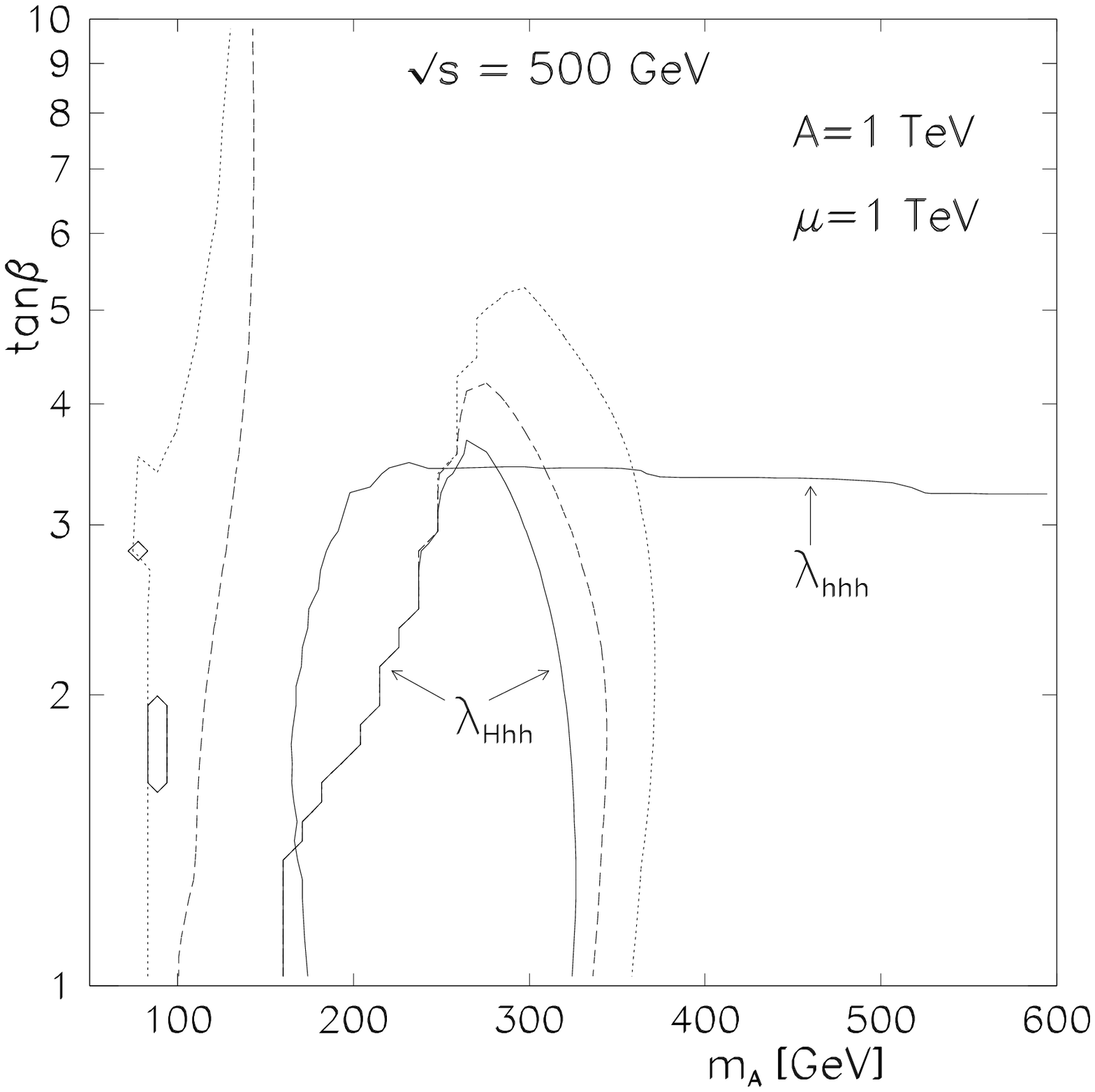}}}
\end{picture}
\vspace*{48mm}
\caption{Regions where trilinear couplings $\lambda_{Hhh}$ and 
$\lambda_{hhh}$ might be measurable at $\sqrt{s}=500$~GeV.
Inside contours labelled $\lambda_{Hhh}$, 
$\sigma(H)\times\mbox{ BR}(H\to hh) > 0.1~\mbox{fb}$ (solid),
while $0.1<\mbox{BR}(H\to hh)<0.9$.
Inside (to the right or below) contour labelled $\lambda_{hhh}$,
the {\it continuum} $WW\to hh$ cross section exceeds 0.1~fb (solid).
Analogous contours are given for 0.05 (dashed) and 0.01~fb (dotted).
Four cases of mixing are considered, with $A=0$ or 1~TeV, and
$\mu=0$ or $\pm1$~TeV, as indicated.
}
\end{center}
\end{figure}

\begin{figure}[htb]
\refstepcounter{figure}
\label{Fig:sensi-1500}
\addtocounter{figure}{-1}
\begin{center}
\setlength{\unitlength}{1cm}
\begin{picture}(16,9)
\put(-1,4)
{\mbox{\epsfysize=9cm\epsffile{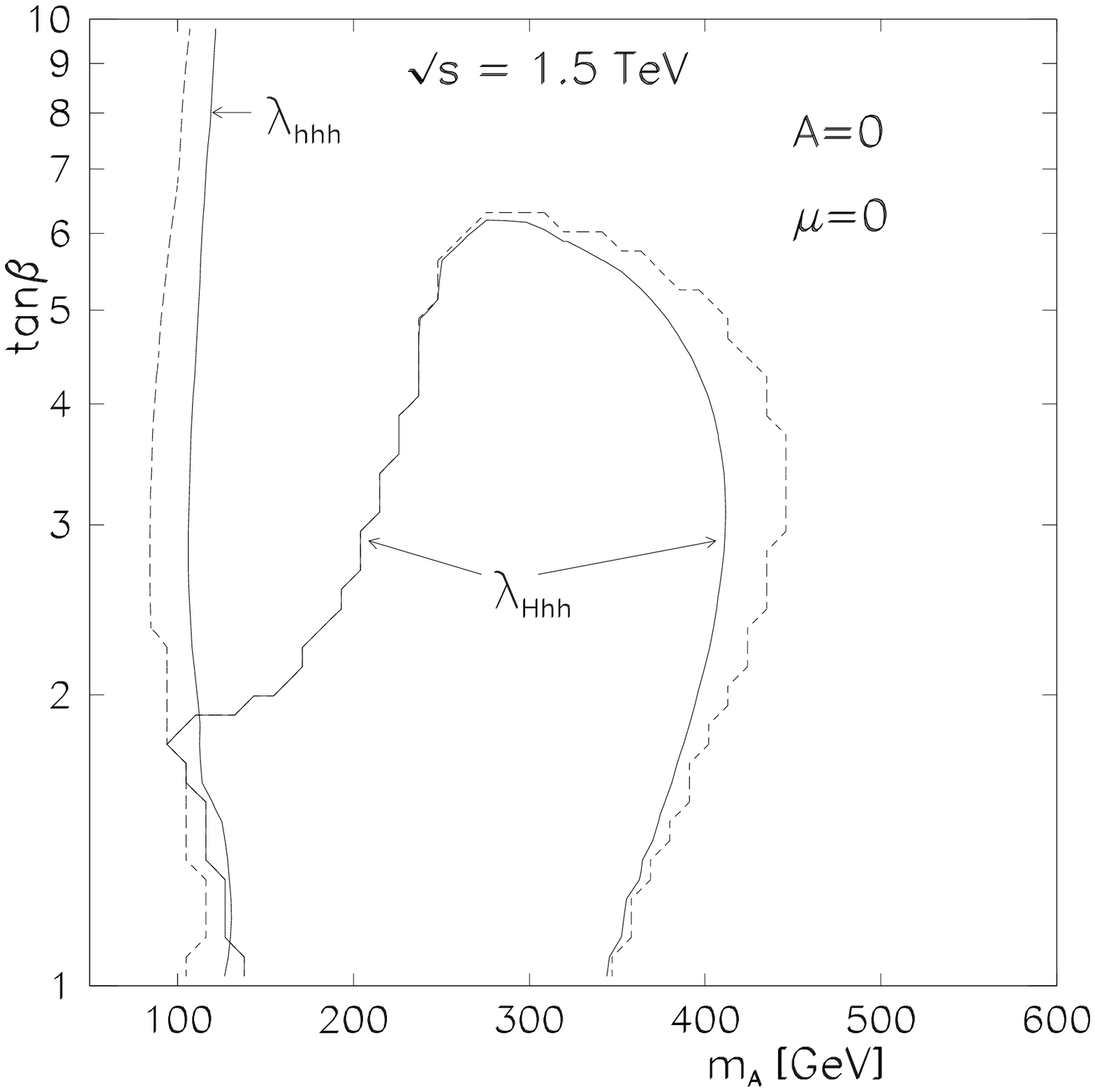}}
 \mbox{\epsfysize=9cm\epsffile{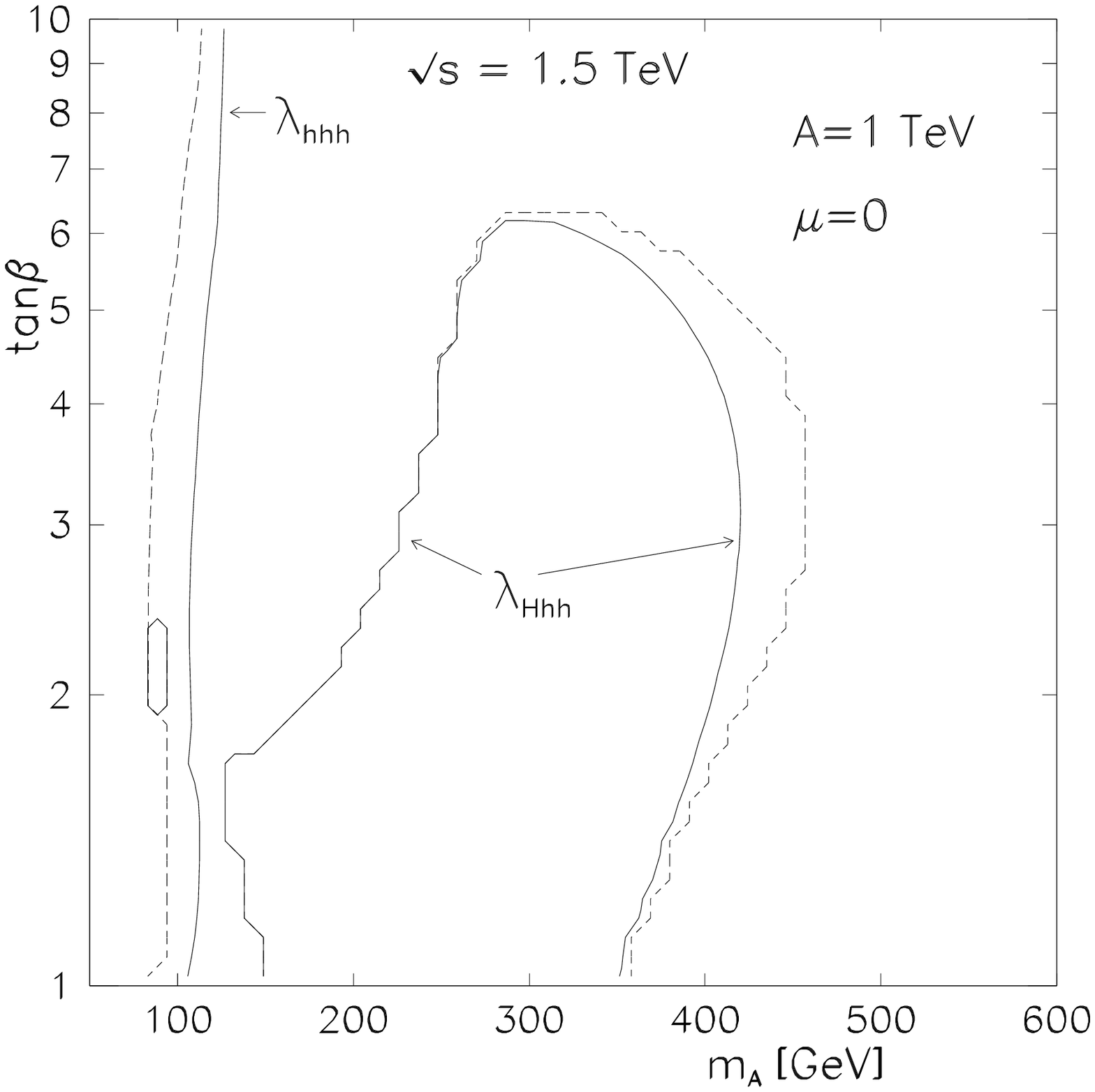}}}
\put(-1,-5)
{\mbox{\epsfysize=9cm\epsffile{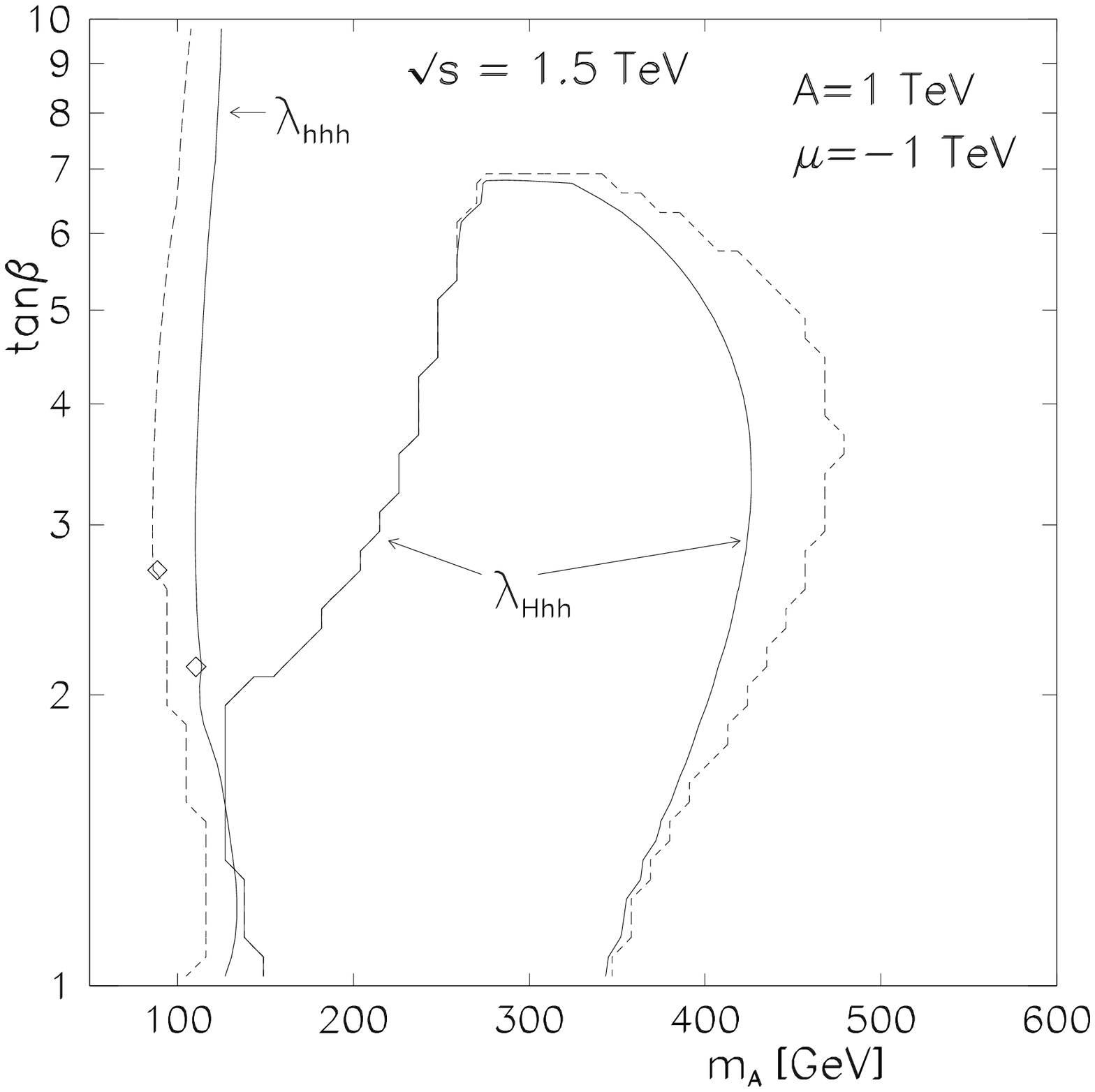}}
 \mbox{\epsfysize=9cm\epsffile{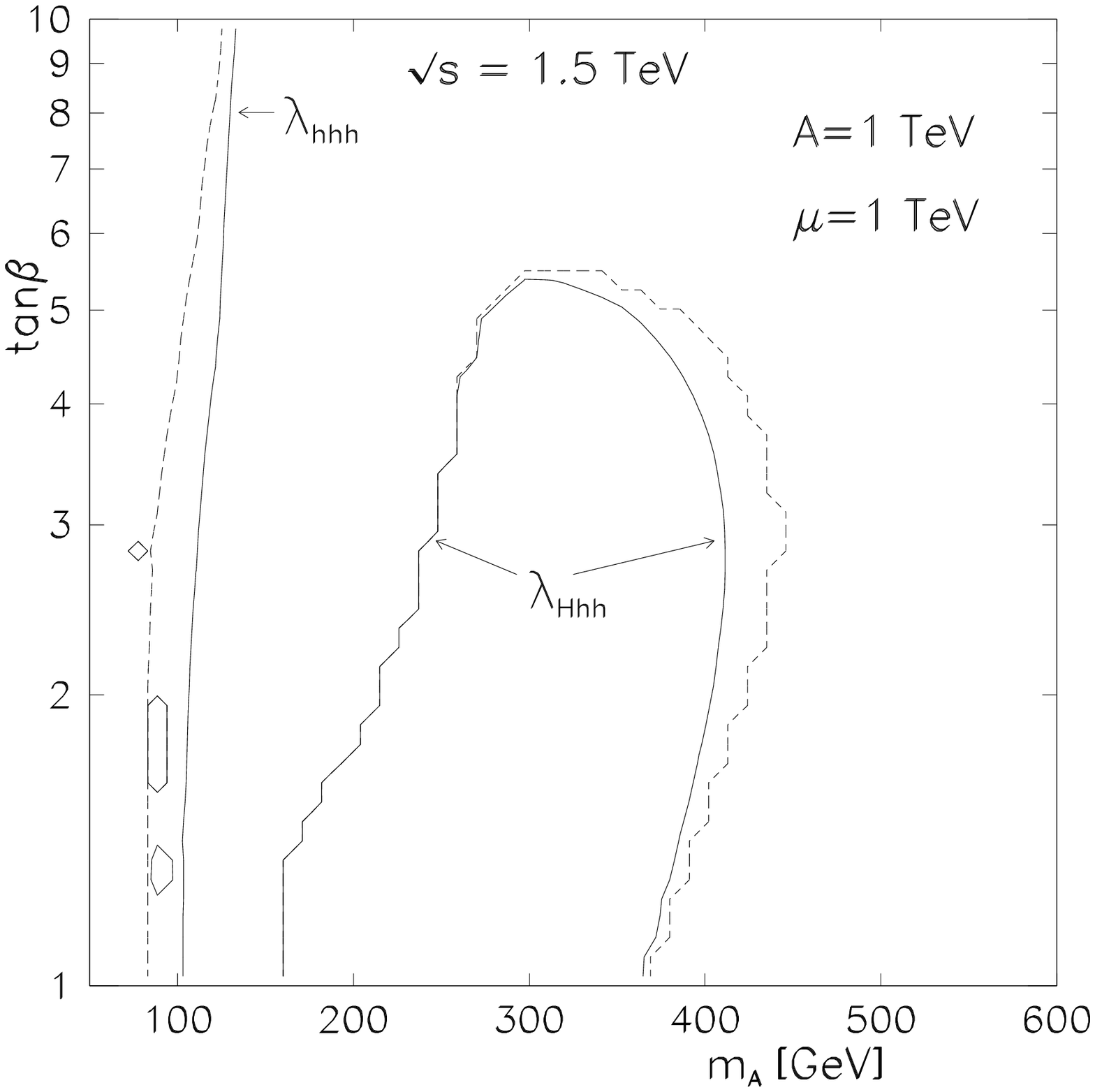}}}
\end{picture}
\vspace*{38mm}
\caption{Regions where trilinear couplings $\lambda_{Hhh}$ and 
$\lambda_{hhh}$ might be measurable at $\sqrt{s}=1.5$~TeV.
The contours correspond to 0.5~fb (solid) and 0.1~fb (dashed).
All other parameters are the same as in Fig.~\ref{Fig:sensi-500}.
}
\end{center}
\end{figure}

\end{document}